\documentclass[prd,twocolumn,reprint,preprintnumbers,nofootinbib,superscriptaddress]{revtex4-1}
\usepackage{eqnarray, mathtools}
\def\vecsign{\mathchar"017E}
\def\dvecsign{\smash{\stackon[-1.95pt]{\vecsign}{\rotatebox{180}{$\vecsign$}}}}
\def\dvec#1{\def\useanchorwidth{T}\stackon[-4.2pt]{#1}{\,\dvecsign}}
\usepackage{stackengine}
\stackMath
\usepackage{graphicx}

\input{header}

\begin{document}
\title{Constraining the top electroweak sector of the SMEFT through $Z$ associated top pair and single top production at the HL-LHC }

\author{Rahool Kumar Barman}\email{rahool.barman@okstate.edu}
\affiliation{Department of Physics, Oklahoma State University, Stillwater, OK, 74078, USA}
\author{Ahmed Ismail}\email{ahmed.ismail.osu@gmail.com}
\affiliation{Department of Physics, Oklahoma State University, Stillwater, OK, 74078, USA}

\begin{abstract}
    We study the processes $pp \to t\bar{t}Z$ and $pp \to tZj$ in the framework of Standard Model Effective Field Theory (SMEFT), employing conventional cut-and-count as well as machine learning techniques to take advantage of kinematic information in complex final states involving multiple leptons and $b$ jets. We explore the projected sensitivity for two SMEFT operators, $\mathcal{O}_{tZ}$ and $\mathcal{O}_{tW}$, that induce electroweak dipole moment interactions for top quarks, through direct searches in these electroweak top production processes at the HL-LHC. New physics modifications to dominant backgrounds are also considered. We show that the new physics sensitivity can be enhanced through a combination of differential distributions for relevant kinematic observables and machine learning techniques. Searches in $t\bar{t}Z$ and $tZj$ production result in stronger constraints on $\mathcal{C}_{tZ}$ and $\mathcal{C}_{tW}$, respectively. At the HL-LHC, $\mathcal{C}_{tZ}$ can be probed up to $-0.41 \lesssim \mathcal{C}_{tZ} \lesssim 0.47$ through searches in the $pp \to t\bar{t}Z + tWZ \to 3\ell + 2b\ + \geq 2j$ channel while $\mathcal{C}_{tW}$ can be probed up to $-0.14 \lesssim \mathcal{C}_{tW} < 0.11$ from searches in the $pp \to tZj + t\bar{t}Z + tWZ \to 3\ell + 1b + 1/2j$ channel, at $95\%$ CL.
\end{abstract}

\maketitle

\section{Introduction}
\label{sec:intro}

The precise measurement of top quark interactions has been a cornerstone of the new physics search program at the Large Hadron Collider~(LHC). New physics effects can manifest themselves in the couplings of the top quark, and can be potentially probed via direct or indirect collider searches. 
While the ability of high energy colliders to probe new physics through strong top quark production has been well-established, electroweak top processes are only now being discovered for the first time. Future runs of the LHC will provide access to not only the cross section, but also the kinematics of such processes. This can give us a better handle on the electroweak couplings of the top quark.

Non-standard top quark couplings have been widely explored in the literature~(c.f.~Refs.~\cite{PECCEI1991305, Zhang:1994fb, Dawson:1995wa, Zhang:2012cd} and references therein). A model-agnostic way to ask about potential deviations from the SM is to use effective field theory (EFT). The Standard Model effective field theory~(SMEFT)~\cite{WEINBERG1979327, BUCHMULLER1986621, Leung:1984ni, Brivio:2017vri} has become a standard tool for assessing the sensitivity of future LHC analyses to new physics, in a way that is independent of assumptions on the details of BSM states. The first basis of non-redundant dimension-6 operators, known as the Warsaw basis, was introduced in Ref.~\cite{Grzadkowski:2010es}. Building upon robust theoretical developments in SMEFT~\cite{Grzadkowski:2010es, Zhang:2013xya, Zhang:2014rja, Lehman:2014jma, Hartmann:2015oia, Ghezzi:2015vva, Gauld:2015lmb, Mimasu:2015nqa, Zhang:2016omx, BessidskaiaBylund:2016jvp, Maltoni:2016yxb, Alioli:2018ljm, Murphy:2020rsh, Li:2020gnx, Li:2020xlh}, a comprehensive exploration of the phenomenological aspects of higher dimension operators has been performed. Non-standard interactions in the Higgs and top sectors encoded by SMEFT operators have been actively scrutinized in the context of current LHC data and with regards to its future sensitivity~\cite{Degrande:2010kt, Degrande:2012gr, Rontsch:2014cca, Degrande:2014tta, Rontsch:2015una, Hartmann:2015aia, Englert:2015hrx,   Azatov:2015oxa, Schulze:2016qas, Degrande:2016dqg, BessidskaiaBylund:2016jvp, Maltoni:2016yxb, Cirigliano:2016nyn, Alioli:2017jdo, Aguilar-Saavedra:2018ksv,  Hartland:2019bjb, Banerjee:2020vtm, Araz:2020zyh}, often in conjunction with measurements at the Tevatron~\cite{D0:2012jgw} and electroweak precision tests at LEP~\cite{ALEPH:2005ab}~(c.f. Refs.~\cite{Ellis:2014dva, Ellis:2014jta, BuarqueFranzosi:2015jrv, Buckley:2016cfg, Cirigliano:2016nyn, deBlas:2016ojx, Biekoetter:2018ypq, Ellis:2018gqa, Ellis:2020unq, Ethier:2021bye}).  

In this work, we focus on the associated production of top quarks with a $Z$ boson at the high luminosity LHC~(HL-LHC: $\sqrt{s}=13~\mathrm{TeV}$, $\mathcal{L}=3~\mathrm{ab^{-1}}$) in the SMEFT framework with dimension 6 operators that affect electroweak top couplings. We consider production of top quark pairs, $pp \to t\bar{t}Z$, as well as single top quark-$Z$ associated production $pp \to tZj$. Both of these channels can be instrumental in probing the neutral current interactions of the top, but have so far remained limited by statistics. The inclusive cross-section for $pp \to t\bar{t}Z$ has been measured by the ATLAS and CMS collaborations using LHC Run-II data collected at $\mathcal{L}\sim 36~\mathrm{fb^{-1}}$ not very long ago~\cite{ATLAS:2019fwo, CMS:2017ugv}, and differential measurements in $t\bar{t}Z$~\cite{CMS:2019too} have only recently begun. Likewise, the CMS collaboration reported the first $tZj$ observation in Ref.~\cite{CMS:2018sgc} using the Run-II dataset corresponding to $\mathcal{L} \sim 77~\mathrm{fb^{-1}}$, and differential cross-section measurements appeared for the first time in Ref.~\cite{CMS:2021ugv}. Differential information for both $t\bar{t}Z$ and $tZj$ is expected to become readily accessible at the HL-LHC. Eventually, the HL-LHC will be an ideal testing ground to explore these rare top electroweak processes. Complementing the rate measurements with differential cross-sections will allow us to probe top electroweak coupings with much better precision than at present~\cite{Degrande:2018fog}. Because the effects of heavy new physics grow with energy, the tails of the distributions are the best places to probe new physics. It is thus crucial that kinematics are maximally leveraged in order to gain the best possible sensitivity to new physics. Furthermore, the $t\bar{t}Z$ and $tZj$ processes feature complicated topologies with many final-state objects. It is thus also interesting to go beyond traditional analyses in order to maximally constrain new physics. We consider both standard cut-and-count techniques as well as novel approaches based on machine learning which allow us to more efficiently optimize our analyses.
Rather than parameterizing the link between amplitudes and experimental data with transfer functions, e.g.~as with the matrix element method, we apply machine learning-based algorithms directly to detector-level events to better approximate a full experimental analysis.

This paper is organized as follows. In Section~\ref{sec:SMEFT}, we discuss the SMEFT framework with a focus on the top electroweak sector. Current constraints on $\mathcal{O}_{tW}$ and $\mathcal{O}_{tZ}$ are also discussed. We review the sensitivity of $pp \to t\bar{t}Z$ to $\mathcal{O}_{tW}$ and $\mathcal{O}_{tZ}$ in Section~\ref{sec:pp_ttz_intro}. Here, we also define relevant kinematic observables and examine their sensitivity to the SMEFT operators. In Section~\ref{sec:pp_tzj_intro}, we perform a detailed collider analysis to extract the projected sensitivity of the HL-LHC in probing $\mathcal{O}_{tW}$ and $\mathcal{O}_{tZ}$ through direct searches in the leptonic $pp \to tZj$ channel. We summarize our results in Section~\ref{sec:conclusions}.

\section{SMEFT Framework}
\label{sec:SMEFT}

Here, we discuss the Standard Model Effective Field Theory~(SMEFT) framework~\cite{WEINBERG1979327, BUCHMULLER1986621, Leung:1984ni} with a main focus on electroweak interactions in the top sector. If new BSM physics is present at a heavy mass scale $\Lambda$ far above the electroweak scale $v = 246~\mathrm{GeV}$, its implications at lower scales can be parameterized through higher dimensional effective operators suppressed by appropriate powers of $\Lambda$. These higher dimensional operators provide a model-independent way of parameterizing deviations from the Standard Model (SM). Considering the SM to be the low energy limit of the full theory, the effective SMEFT Lagrangian can be written by augmenting the SM Lagrangian with these new physics operators, 
\begin{equation}
    \mathcal{L}_{SMEFT} = \mathcal{L}_{SM} + \sum_{i} \frac{\mathcal{C}_{i}}{\Lambda^{2}} \mathcal{O}_{i}^{(6)} + \mathcal{O}\left(\Lambda^{-4}\right).
    \label{eqn:Lag_SMEFT}
\end{equation}

Here $\{\mathcal{O}_{i}^{(6)}\}$ represents the set of operators respecting the symmetries of the SM with mass dimension $d = 6$. New physics effects from operators with mass dimension $d \geq 8$ are subdominant and have been ignored in this work. In Eq.~\ref{eqn:Lag_SMEFT}, $\mathcal{C}_{i}$ are the Wilson coefficients. They are free parameters by definition and are constrained by experimental measurements. Typically, the set of $d = 6$ operators $\{\mathcal{O}_{i}^{(6)}\}$ results in the following modifications to any measured observable $\mathcal{X}$,
\begin{equation}
    \mathcal{X} = \mathcal{X}_{SM} + \sum_{i}\mathcal{X}^{\prime}_{i}\frac{\mathcal{C}_{i}^{(6)}}{\Lambda^{2}} + \sum_{i,j} \mathcal{X}^{\prime\prime}_{i}\frac{\mathcal{C}_{i}^{(6)}\mathcal{C}_{j}^{(6)}}{\Lambda^{4}},
\end{equation}
where the term linear in $\mathcal{C}_{i}$ encodes interference between SM and $\mathcal{O}_{i}^{(6)}$, while the last term represents non-linear pure SMEFT effects. 

There are 59 independent $d = 6$ SMEFT operators for one generation of fermions, assuming baryon number conservation. We will restrict ourselves to operators involving third-generation quarks.
In the Warsaw basis~\cite{Grzadkowski:2010es}, then, 31 of these operators involve the top quark. Restricting to the CP-conserving scenario, 11 such operators can be constructed from four heavy quark fields which include the left-handed quark $SU(2)_{L}$ doublet $Q_{3}$, right-handed top $U_{3}$, and/or right-handed bottom $SU(2)_{L}$ singlet $D_{3}$. These four-heavy-quark operators are mainly constrained by measurements in processes involving $t\bar{t}t\bar{t}$ and $t\bar{t}b\bar{b}$ final states~\cite{DHondt:2018cww}. Furthermore, apart from four-heavy-quark operators, 9 operators involve two heavy quarks along with bosonic fields~\cite{Grzadkowski:2010es,Aguilar-Saavedra:2018ksv,Hartland:2019bjb}. Of these, the top chromomagnetic dipole operator $\mathcal{O}_{tG} = \left( \bar{Q}_3 \sigma^{\mu\nu} T^{A} U_3 \right) \tilde{H} G_{\mu\nu}^a$~\footnote{We adopt the operator notation of Ref.~\cite{Grzadkowski:2010es}.}, modifies the coupling of the top with gluons, and can be constrained by processes such as $t\bar{t}$, $t\bar{t}W$, $t\bar{t}Z$, $t\bar{t}H$, $tZ$, $tW$ and single Higgs production in the gluon fusion channel~$gg \to h$; $\mathcal{O}_{tH} = (H^{\dagger}H)(\bar{Q}_{3}U_{3}\tilde{H})$ modifies the tree-level Higgs-top coupling, and is constrained by $t\bar{t}H$ measurements and $gg \to H$ production; linear combinations of $\mathcal{O}_{HQ}^{(1)} = \left( H^\dagger i \dvec{D}_\mu H \right) \bar{Q}_3 \gamma^\mu Q_3$ and $\mathcal{O}_{HQ}^{(3)} = \left( H^\dagger i \dvec{D}_\mu^{I} H \right) \bar{Q}_3 \tau^{I} \gamma^\mu Q_3$ are constrained by $Zb\bar{b}$ measurements at LEP and electroweak top processes respectively~\cite{Ellis:2020unq}; $\mathcal{O}_{Htb} = i(\tilde{H}^{\dagger}D_{\mu}H)(U_{3}\gamma^{\mu}D_{3})$ can be constrained by measurements of top decay and $h\to b\bar{b}$ measurements~\cite{Alioli:2017ces}; and $\mathcal{O}_{bW} = (\bar{Q}_{3}\sigma^{\mu\nu}D_{3})\tau^{I}HW_{\mu\nu}^{I}$ can be constrained by single top production. Then, each of the operators $\mathcal{O}_{Htb}$ and $\mathcal{O}_{bW}$ mainly contributes at $\mathcal{O}(\Lambda^{-4})$ since the interference of these operators with the SM vanishes in the limit $m_{b} \to 0$~\cite{Buckley:2015lku, Hartland:2019bjb}. The remaining three operators with two quarks and bosonic fields are,
\begin{eqnarray}
    \mathcal{O}_{tW} &=& \left( \bar{Q}_3 \sigma^{\mu\nu} U_3 \right) \tau^a \tilde{H} W_{\mu\nu}^a \\
    \mathcal{O}_{tB} &=& \left( \bar{Q}_3 \sigma^{\mu\nu} U_3 \right) \tilde{H} B_{\mu\nu} \\
    \mathcal{O}_{Ht} &=& \left( H^\dagger i \dvec{D}_\mu H \right) \bar{U}_3 \gamma^\mu U_3
\end{eqnarray}
$\mathcal{O}_{tW}$ modifies the charged current coupling of the top quark, and can be probed through $W$ helicity fraction measurements and electroweak top processes~\cite{Ellis:2020unq}, while $\mathcal{O}_{tb}$ and $\mathcal{O}_{Ht}$ are substantially less constrained. The latter two operators modify the neutral current interactions of the top quark, and can only be constrained by $t\bar{t}Z/\gamma$ and $tZ(j)$. As discussed previously, measurements in these processes have remained statistically limited until now, and differential measurements have started to appear only recently. The upcoming differential data is expected to improve the sensitivity to $\mathcal{O}_{tW}$ and $\mathcal{O}_{tB}$, but not as much for $\mathcal{O}_{Ht}$ since the scattering amplitudes do not exhibit any energy growth with $\mathcal{O}_{Ht}$~\cite{Degrande:2018fog}.

Motivated by future LHC measurements in electroweak top processes, then, our focus in this work is the electroweak top dipole operators $\mathcal{O}_{tW}$ and $\mathcal{O}_{tB}$. The current limits from a global fit of Higgs, electroweak and top data are $-0.12 < \mathcal{C}_{tW} < 0.51$ and $-4.5 < \mathcal{C}_{tB} < 1.2$ at $95\%$ CL individually~\cite{Ellis:2020unq}. To separate out the effects of neutral current interactions, we will work in a basis where our operators of interest are $\mathcal{O}_{tW}$ and the combination
\begin{equation}
    \mathcal{O}_{tZ} = -\sin\theta_{W} \mathcal{O}_{tB} + \cos\theta_{W} \mathcal{O}_{tW}
\end{equation}
where, $\theta_{W}$ is the Weinberg angle. Both $\mathcal{O}_{tW}$ and $\mathcal{O}_{tZ}$ can contribute to $tZj$ and $t\bar{t}Z$ processes at the production level. Our goal is to explore the projected sensitivities for $\mathcal{O}_{tW}$ and $\mathcal{O}_{tZ}$ through searches in $t\bar{t}Z$ and $tZj$ at the HL-LHC using a combination of rate and differential cross-section measurements. With the exception of some recent studies~\cite{Ethier:2021bye,CMS:2019too}, most global fits as well as direct probes for $\mathcal{O}_{tZ}$ have relied on rate measurements alone. The differential cross-sections for $p_{T,Z}$ and $\cos\theta^{\star}$, where $\theta^{\star}$ is the angle between the $Z$ boson and the negatively charged lepton in the center of mass frame of the $Z$ boson, have been measured in the $t\bar{t}Z$ channel for the first time by the CMS experiment using LHC Run-II data collected at $\mathcal{L} \sim 77~\mathrm{fb^{-1}}$~\cite{CMS:2019too}. With the inclusion of differential information, $\mathcal{C}_{tZ}$ has been constrained up to $-1.1 \lesssim \mathcal{C}_{tZ} \lesssim 1.1$ at $95\%$ CL, which is a considerable improvement over the previous CMS~($\mathcal{L} \sim 36~\mathrm{fb^{-1}}$) bound $-2.6 \lesssim \mathcal{C}_{tZ} \lesssim 2.6$ at $95\%$ CL~\cite{CMS:2017ugv}. We now proceed to analyze the prospects for HL-LHC measurements to improve upon these existing limits.

\section{Electroweak top production}
\label{sec:electroweak_top}

As discussed previously, we perform a detailed collider analysis to study the sensitivity of top electroweak processes at the HL-LHC to the SMEFT operators $\mathcal{O}_{tW}$ and $\mathcal{O}_{tZ}$, focusing on $t\bar{t}Z$ and $tZj$ production. These processes allow for the testing of neutral top electroweak couplings that are not accessible through top decay, i.e.~$\mathcal{O}_{tZ}$, while $\mathcal{O}_{tW}$ can affect both production and decay. For $t\bar{t}Z$, we study the $3\ell\ + 2b\ + \geq 2 j$ channel, while for $tZj$, we focus on the $3\ell + 1b + 1/2 j$ final state. Our choice for the aforesaid final states is largely motivated by the absence of major background contributions from QCD processes and non-prompt leptons which are relatively difficult to simulate. These final states also offer sufficient statistics at the HL-LHC to make use of kinematic information.
For each final state, our general approach is to maximize the ability of an HL-LHC search to discriminate SM electroweak top production and backgrounds from SMEFT contributions. We make use of three different methods for each channel: (1) a traditional cut-and-count analysis, where we optimize manually on a selection of kinematic variables; (2) a deep neural network (DNN) approach, with a multi-layer perceptron trained on a larger set of kinematic quantities; and (3) likelihood ratio inference using MadMiner~\cite{Brehmer:2019xox}.

Throughout our analysis, we make use of signal and background events that are simulated at leading order~(LO) with \texttt{MadGraph5\_aMC@NLO}~\cite{Alwall:2014hca} in the 5-flavor scheme with the \texttt{NNPDF2.3QED}~\cite{2013290} parton distribution function. We choose a fixed EFT renormalization scale $\mu_{EFT} \sim \left(m_{t}+m_{Z}\right)/4$~\cite{Degrande:2018fog, Demartin:2015uha} and generate events at center-of-mass energy $\sqrt{s}=13$~TeV. \texttt{Pythia~8}~\cite{Sjostrand:2007gs} is used to simulate showering and hadronization effects and \texttt{Delphes-3.5.0}~\cite{deFavereau:2013fsa} is utilized for fast detector simulation with the default HL-LHC card~\cite{delphes_card}. The pre-selection cuts
\begin{eqnarray}
    p_{T_\ell} > 10~{\rm GeV}, p_{T_b} > 25~{\rm GeV}, p_{T_j} > 25~{\rm GeV}, \nonumber \\
    |\eta_{\ell}| < 4.0, |\eta_{b}| < 4.0, |\eta_{j}| < 4.0  \quad.
    \label{eqn:obj_selection_ttz}
\end{eqnarray}
are applied to all final state objects.

In the cut-and-count and DNN analyses, we maximize the NP signal significance,
\be
\sigma_{s}^{NP} = \frac{|S_{SMEFT} - S_{SM}|}{\sqrt{S_{SM}}}.
\label{eq:cb_significance}
\ee 
Here, $S_{SMEFT}$ represents the yield in the signal region including the SMEFT contributions to the signal processes while $S_{SM}$ represents the number of events expected from SM processes alone. That is, $S_{SMEFT}$ includes pure SM contributions, interference between SMEFT operators and SM, and non-linear pure SMEFT terms. In the MadMiner analyses, we calculate the significance in the $\mathcal{O}_{tW}, \mathcal{O}_{tZ}$ plane from the inferred likelihood ratio directly. We note that the inclusion of $\mathcal{O}(\Lambda^{-4})$ pure SMEFT terms is relatively more important for the $t\bar{t}Z$ channel where the interference term $\mathcal{O}(\Lambda^{-2})$ undergoes accidental suppression due to cancellation between the $gg \to t\bar{t}Z$ and $q\bar{q} \to t\bar{t}Z$ production channels~\cite{BessidskaiaBylund:2016jvp}.

We now turn to the application of these approaches to the final states that are relevant for constraining top neutral current couplings.

\subsection{$pp \to t\bar{t}Z + tWZ \to 3\ell +2b\ + \geq 2j$}
\label{sec:pp_ttz_intro}

We select events with exactly three isolated leptons~($l = e,\mu$), two $b$ tagged jets and at least two light jets~($j$) in the final state satisfying the cuts of Eq.~\ref{eqn:obj_selection_ttz}.
We further impose $p_{T,\ell_{1}} > 40~$GeV and $p_{T,\ell_{2}}>20~$GeV, where $\ell_{1}$ and $\ell_{2}$ are the leading and sub-leading 
leptons. We reconstruct the $Z$ boson by requiring at least one same flavor opposite sign~(SFOS) lepton pair with invariant mass $m_{Z}\pm 10~$GeV. In cases where all three isolated leptons are of the same flavor, two such SFOS pairs could be obtained. In such instances, the pair with invariant mass closest to $m_{Z}$ is associated with the $Z$ boson. We next pursue the reconstruction of the semileptonic $t\bar{t}$ system. The full reconstruction of the semileptonic $t\bar{t}$ system is challenging due to the unknown longitudinal momentum of the neutrino $\nu$ produced from the leptonic top $t_{\ell}$, as well as combinatorial ambiguities between $b$ tagged jets and light jets. The non-SFOS lepton~($\ell_{W}$) is associated with the leptonically decaying top~($t_{\ell}$). We then compute the longitudinal momentum~($\slashed{p}_{z}$) of $\nu$ by constraining the invariant mass of $\ell_{W}$ and $\nu$ with the on-shell $W$ boson mass $m_{W}$. This leads to either two solutions or no solutions for $\slashed{p}_{z}$. We reject events with no solutions. In events with two solutions, we choose the one which minimizes ${(m_{\ell_{W}\nu} - m_{W})}^{2}$. Having identified $\slashed{p}_{z}$, the only missing piece in the reconstruction of $t_{\ell}$ is the choice of the $b$ tagged jet $b_{t_{\ell}}$. Before identifying $b_{t_{\ell}}$, however, we discuss the hadronically decaying top $t_{h}$, which decays via $t_{h} \to (W \to jj) b$. The pair of light jets associated with $t_{h}$ is identified by minimizing ${(m_{jj} - m_{W})}^{2}$. 
We refer to this light jet pair as $\{j_{1t},j_{2t}\}$ where $p_{T,j_{1t}} > p_{T,j_{2t}}$. Finally, we pair the $b$ tagged jets with $t_{\ell}$ and $t_{h}$ by minimizing
\begin{equation}
    \left(m_{l_{t}\nu_{t}b_{i}} - m_{t}\right)^{2} + \left(m_{j_{1t}j_{2t}b_{k}} - m_{t}\right)^{2},
\end{equation}
where, $i,k = 1,2$ with $i\neq k$ and $m_{t}$ is the mass of the top quark $m_{t} = 173.3~$GeV~\cite{Zyla:2020zbs}. We refer to the $b$ tagged jet associated with the hadronic top as $b_{t_{h}}$. 

The dominant source of background is SM $t\bar{t}Z$. Sub-dominant contributions can arise from $WZ+\mathrm{jets}$, $tWZ$, $t\bar{t}h$, $t\bar{t}\gamma$, $t\bar{t}W$, $t\bar{t}VV$~($V=W,Z$) and $t\bar{t}t\bar{t}$. We ignore contributions from $t\bar{t}W$, $t\bar{t}VV$ and $t\bar{t}t\bar{t}$ due to their smaller rates at the HL-LHC. Among the remaining background processes, $tWZ$, $t\bar{t}h$ and $t\bar{t}\gamma$ can be modified by $\mathcal{O}_{tZ}$ as well as $\mathcal{O}_{tW}$. However, the event rates for $t\bar{t}h$ and $t\bar{t}\gamma$ are roughly two orders of magnitude smaller compared to that for $t\bar{t}Z$. Therefore, in the cut-based and multivariate DNN analysis, new physics effects in $t\bar{t}\gamma$ and $t\bar{t}h$ are ignored. We include new physics modifications from $\mathcal{O}_{tW}$ and $\mathcal{O}_{tZ}$ in the $t\bar{t}Z$ and $tWZ$ processes only. 

In order to distinguish the SMEFT signal from SM background, we consider an extensive list of kinematic observables,
\begin{equation}
    \begin{split}
        p_{T,\alpha}, \eta_{\alpha}, \phi_{\alpha} \{\alpha = \alpha_{Z}, \alpha_{t_{\ell}}, \alpha_{t_{h}}, t_{\ell}, t_{h}, Z \}\\
        \{\alpha_{Z} = \ell_{1},\ell_{2};~\alpha_{t_{\ell}} = \ell_{W},\nu_{t},b_{t_{\ell}};~\alpha_{t_{h}}= j_{1t}, j_{2t}, b_{t_{h}}\} \\
        \Delta\phi_{\beta \epsilon}, \Delta \eta_{\beta\epsilon} \{\beta, \epsilon = \alpha_{Z}, \alpha_{t_{\ell}}, \alpha_{t_{h}}, t_{\ell}, t_{h}, Z; \beta \neq \epsilon \} \\
        \theta^{\star t\bar{t}Z}_{\alpha_{Z}}, \theta^{\star t\bar{t}Z}_{t_{\ell}}, \theta^{\star t\bar{t}Z}_{t_{h}}, \theta^{\star t\bar{t}}_{t_{\ell}}, \theta^{\star t\bar{t}}_{t_{h}}, \\
        p_{T, t_{\ell}Z}, p_{T, t_{h}Z}, p_{T, t_{\ell}t_{h}}, p_{T, t_{\ell}t_{h}Z}, H_{T} \\ 
        m_{t_{\ell}}, m_{t_{h}}, m_{Z}, m_{t_{\ell}Z}, m_{t_{h}Z}, m_{t_{\ell}t_{h}}, m_{t_{\ell}t_{h}Z}, m_{T,l_{W}} \\ 
        \Delta R_{\ell\ell}^{\mathrm{min}}, \Delta R_{\ell\ell}^{\mathrm{max}}, \Delta R_{\ell b}^{\mathrm{min}}, \Delta R_{\ell b}^{\mathrm{max}},
    \end{split}
    \label{eqn:ttz_observables}
\end{equation}
where, $\alpha_{Z}$, $\alpha_{t_{\ell}}$, and $\alpha_{t_{h}}$ denote the final state objects that reconstruct the $Z$ boson, $t_{\ell}$, and $t_{h}$, respectively. In Eq.~(\ref{eqn:ttz_observables}), $p_{T,i}$, $\eta_{i}$ and $\phi_{i}$ represent the transverse momentum, pseudorapidity and azimuthal angle of object $i$, respectively; $\Delta \phi_{ij}$ and $\Delta \eta_{ij}$ corresponds to the difference between azimuthal angles and pseudorapidities, respectively, for objects $i$ and $j$; $\theta^{\star m}_{n}$ is the angle between particle $n$ and the beam direction in the center of mass frame of particle $m$; $m_{T,\ell_{W}}$ is the transverse mass of
$\ell_{W}$;
$H_{T}$ is the scalar sum of the transverse momenta of all visible final state objects; $\Delta R_{\ell\ell}^{\mathrm{min(max)}}$ is the minimum~(maximum) $\Delta R = \sqrt{\Delta \eta^{2} + \Delta \phi^{2}}$ separation between any two leptons; and $\Delta R_{\ell b}^{\mathrm{min(max)}}$ is the minimum~(maximum) $\Delta R$ separation between a lepton and $b$ jet. The other notations have their usual meaning. 

\begin{table*}[!htb]
    \centering
    \begin{tabular}{|c||c|c|c|c|} \hline
         \multicolumn{5}{|c|}{$\mathcal{C}_{tZ}=2.0$}\\\hline 
         Optimized & $m_{t_{h}Z}~$ $>$ & $H_{T}~$ $>$ & $\Delta R_{\ell\ell}^{\mathrm{min}}$ $<$ & $\Delta \phi_{\ell_{W}t_{\ell}}$ $<$ \\ 
         cuts & 250~GeV & 300~GeV & 2.75 & 3.1 \\ \hline 
         SMEFT $t\bar{t}Z$ & 2664 & 2611 & 2609 & 2608 \\
         SMEFT $tWZ$ & 151 & 149 & 148 & 148 \\
         $t\bar{t}Z$ & 1853 & 1800 & 1796 & 1795 \\
         $tWZ$ & 118 & 115 & 115 & 115\\
         $WZ$ & 153 & 147 & 147 & 147\\
         $t\bar{t}h$ & 14.1 & 12.8 & 12.8 & 12.8 \\
         $t\bar{t}\gamma$ & 19.7 & 18.9 & 18.9 & 18.9\\ \hline
        Significance & 18.17 & 18.47 & 18.51 & 18.51 \\\hline \hline
        \multicolumn{5}{|c|}{$\mathcal{C}_{tZ}=1.5$}\\\hline 
         Optimized & $m_{t_{h}Z}~$ $>$ & $H_{T}~$ $>$ & $\Delta R_{\ell\ell}^{\mathrm{min}}$ $<$ & $\Delta \phi_{\ell_{W}t_{\ell}}$ $<$ \\ 
         cuts & - & 300~GeV & 2.0 & 2.8 \\ \hline 
         SMEFT $t\bar{t}Z$ & 2369 & 2302 & 2227 & 2189 \\
         SMEFT $tWZ$ & 138 & 135 & 131 & 128 \\
         $t\bar{t}Z$ & 1892 & 1827 & 1756 & 1721 \\
         $tWZ$ & 120 & 116 & 113 & 111\\
         $WZ$ & 153 & 147 & 139 & 139 \\
         $t\bar{t}h$ & 14.7 & 13.1 & 12.8 & 12.6 \\
         $t\bar{t}\gamma$ & 19.9 & 19.1 & 18.5 & 18.2 \\ \hline
        Significance & 10.55 & 10.72 & 10.83 & 10.84 \\ \hline \hline 
        \multicolumn{5}{|c|}{$\mathcal{C}_{tZ}=1.0$}\\\hline 
         Optimized & $m_{t_{h}Z}~$ $>$ & $H_{T}~$ $>$ & $\Delta R_{\ell\ell}^{\mathrm{min}}$ $<$ & $\Delta \phi_{\ell_{W}t_{\ell}}$ $<$ \\ 
         cuts & 500~GeV & 350~GeV & 2.5 & - \\ \hline 
         SMEFT $t\bar{t}Z$ & 947 & 912 & 906 & 906 \\
         SMEFT $tWZ$ & 69.7 & 67.2 & 66.9 & 66.9 \\
         $t\bar{t}Z$ & 781 & 748 & 742 & 742 \\
         $tWZ$ & 65.1 & 62.6 & 62.2 & 62.2\\
         $WZ$ & 117 & 106 & 104 & 104 \\
         $t\bar{t}h$ & 3.2 & 2.7 & 2.7 & 2.7 \\
         $t\bar{t}\gamma$ & 8.1 & 7.7 & 7.7 & 7.7 \\ \hline
        Significance & 5.47 & 5.54 & 5.57 & 5.57\\ \hline \hline
        \multicolumn{5}{|c|}{$\mathcal{C}_{tZ}=0.5$}\\\hline 
         Optimized & $m_{t_{h}Z}~$ $>$ & $H_{T}~$ $>$ & $\Delta R_{\ell\ell}^{\mathrm{min}}$ $<$ & $\Delta \phi_{\ell_{W}t_{\ell}}$ $<$ \\ 
         cuts & 500~GeV & 350~GeV & 0.75 & 1.9 \\ \hline 
         SMEFT $t\bar{t}Z$ & 830 & 794 & 368 & 325 \\
         SMEFT $tWZ$ & 66.4 & 64.1 & 27.6 & 24.5 \\
         $t\bar{t}Z$ & 781 & 748 & 337 & 289 \\
         $tWZ$ & 65 & 63 & 26 & 23 \\
         $WZ$ & 117 & 106 & 43.8 & 38.4 \\
         $t\bar{t}h$ & 3.2 & 2.7 & 1.0 & 0.8 \\
         $t\bar{t}\gamma$ & 8.1 & 7.7 & 3.8 & 3.4 \\ \hline
        Significance & 1.61 & 1.55 & 1.61 & 1.99 \\ \hline \hline 
    \end{tabular}
    \begin{tabular}{|c||c|c|c|c|} \hline
         \multicolumn{5}{|c|}{$\mathcal{C}_{tZ}=-2.0$}\\\hline 
         Optimized & $m_{t_{h}Z}~$ $>$  & $H_{T}~$ $>$ & $\Delta R_{\ell\ell}^{\mathrm{min}}$ $<$ & $\Delta \phi_{\ell_{W}t_{\ell}}$ $<$ \\ 
         cuts & - & 300~GeV & 2.0 & - \\ \hline 
         SMEFT $t\bar{t}Z$ & 2781 & 2710 & 2622 & 2622 \\
         SMEFT $tWZ$ & 148 & 146 & 142 & 142 \\
         $t\bar{t}Z$ & 1892 & 1827 & 1756 & 1756 \\
         $tWZ$ & 120 & 116 & 113 & 113\\
         $WZ$ & 153 & 147 & 139 & 139\\
         $t\bar{t}h$ & 14.8 & 13.1 & 12.8 & 12.8 \\
         $t\bar{t}\gamma$ & 19.9 & 19.1 & 18.5 & 18.5\\ \hline
        Significance & 19.55 & 19.82 & 19.82 & 19.82 \\ \hline \hline
        \multicolumn{5}{|c|}{$\mathcal{C}_{tZ}=-1.5$}\\\hline 
         Optimized & $m_{t_{h}Z}~$ $>$ & $H_{T}~$ $>$ & $\Delta R_{\ell\ell}^{\mathrm{min}}$ $<$ & $\Delta \phi_{\ell_{W}t_{\ell}}$ $<$ \\ 
         cuts & 250~GeV & 300~GeV & - & - \\ \hline 
         SMEFT $t\bar{t}Z$ & 2350 & 2292 & 2292 & 2292 \\
         SMEFT $tWZ$ & 135 & 133 & 132 & 132 \\
         $t\bar{t}Z$ & 1853 & 1800 & 1800 & 1800 \\
         $tWZ$ & 118 & 115 & 115 & 115 \\
         $WZ$ & 153 & 147 & 147 & 147\\
         $t\bar{t}h$ & 14.1 & 12.8 & 12.8 & 12.8 \\
         $t\bar{t}\gamma$ & 19.7 & 18.9 & 18.9 & 18.9 \\ \hline
        Significance & 11.06 & 11.12 & 11.12 & 11.12 \\ \hline \hline 
        \multicolumn{5}{|c|}{$\mathcal{C}_{tZ}=-1.0$}\\\hline 
         Optimized & $m_{t_{h}Z}~$ $>$ & $H_{T}~$ $>$ & $\Delta R_{\ell\ell}^{\mathrm{min}}$ $<$ & $\Delta \phi_{\ell_{W}t_{\ell}}$ $<$ \\ 
         cuts & 400~GeV & 500~GeV & 2.25 & - \\ \hline 
         SMEFT $t\bar{t}Z$ & 1382 & 920 & 909 & 908 \\
         SMEFT $tWZ$ & 95.0 & 68.2 & 67.3 & 67.3 \\
         $t\bar{t}Z$ & 1195 & 770 & 757 & 757 \\
         $tWZ$ & 89.4 & 63.7 & 63.0 & 63.0 \\
         $WZ$ & 136 & 76.1 & 73.4 & 73.4 \\
         $t\bar{t}h$ & 5.8 & 2.6 & 2.6 & 2.6 \\
         $t\bar{t}\gamma$ & 12.3 & 7.4 & 7.3 & 7.3 \\ \hline
        Significance & 5.08 & 5.09 & 5.17 & 5.17 \\ \hline \hline 
        \multicolumn{5}{|c|}{$\mathcal{C}_{tZ}=-0.5$}\\\hline 
         Optimized & $m_{t_{h}Z}~$ $>$ & $H_{T}~$ $>$ & $\Delta R_{\ell\ell}^{\mathrm{min}}$ $<$ & $\Delta \phi_{\ell_{W}t_{\ell}}$ $<$ \\ 
         cuts & 350~GeV & 650~GeV & 2 & 0.4 \\ \hline 
         SMEFT $t\bar{t}Z$ & 1489 & 419 & 411 & 215 \\
         SMEFT $tWZ$ & 103 & 38.1 & 37.1 & 17.5 \\
         $t\bar{t}Z$ & 1442 & 392 & 381 & 185 \\
         $tWZ$ & 101.8 & 36.5 & 35.7 & 16.9 \\
         $WZ$ & 150 & 32.7 & 32.7 & 21.8 \\
         $t\bar{t}h$ &  8.4 & 0.82 & 0.81 & 0.26 \\
         $t\bar{t}\gamma$ & 15.1 & 3.7 & 3.6 & 1.6\\ \hline
        Significance & 1.16 & 1.32 & 1.47 & 2.04 \\\hline \hline 
    \end{tabular}
    \caption{Optimized selection cuts on $m_{t_{h}Z}$, $H_{T}$, $\Delta R_{\ell\ell}^{\mathrm{min}}$ and $\Delta \phi_{\ell_{W}t_{\ell}}$, applied successively, to maximize the NP signal significance $\sigma_{S}^{NP}$ of cut-based collider analysis in the $pp \to t\bar{t}Z + tWZ \to 3\ell + 2b\ + \geq 2j $ channel to explore the projected sensitivity to $\mathcal{O}_{tZ}$ at $\sqrt{s}=13~$TeV LHC with $\mathcal{L}=3~{\rm ab^{-1}}$. The optimized cuts, signal and background yields, and $\sigma_{s}$ values are shown for $\{\mathcal{C}_{tZ} = \pm 2.0, \pm 1.5, \pm 1.0, \pm 0.5\}$. No cuts are applied on $m_{t_{h}Z}$ and $\Delta \phi_{\ell_{W}t_{\ell}}$ in the signal regions that are optimized for $\mathcal{C}_{tZ}=1.5,-2.0$, and $\mathcal{C}_{tZ}=1.0, -1.0, -1.5$, respectively.} \label{tab:ttz_OtZ_cut_flow}
\end{table*}

\begin{figure*}[!t]
    \centering
    \includegraphics[scale=0.28]{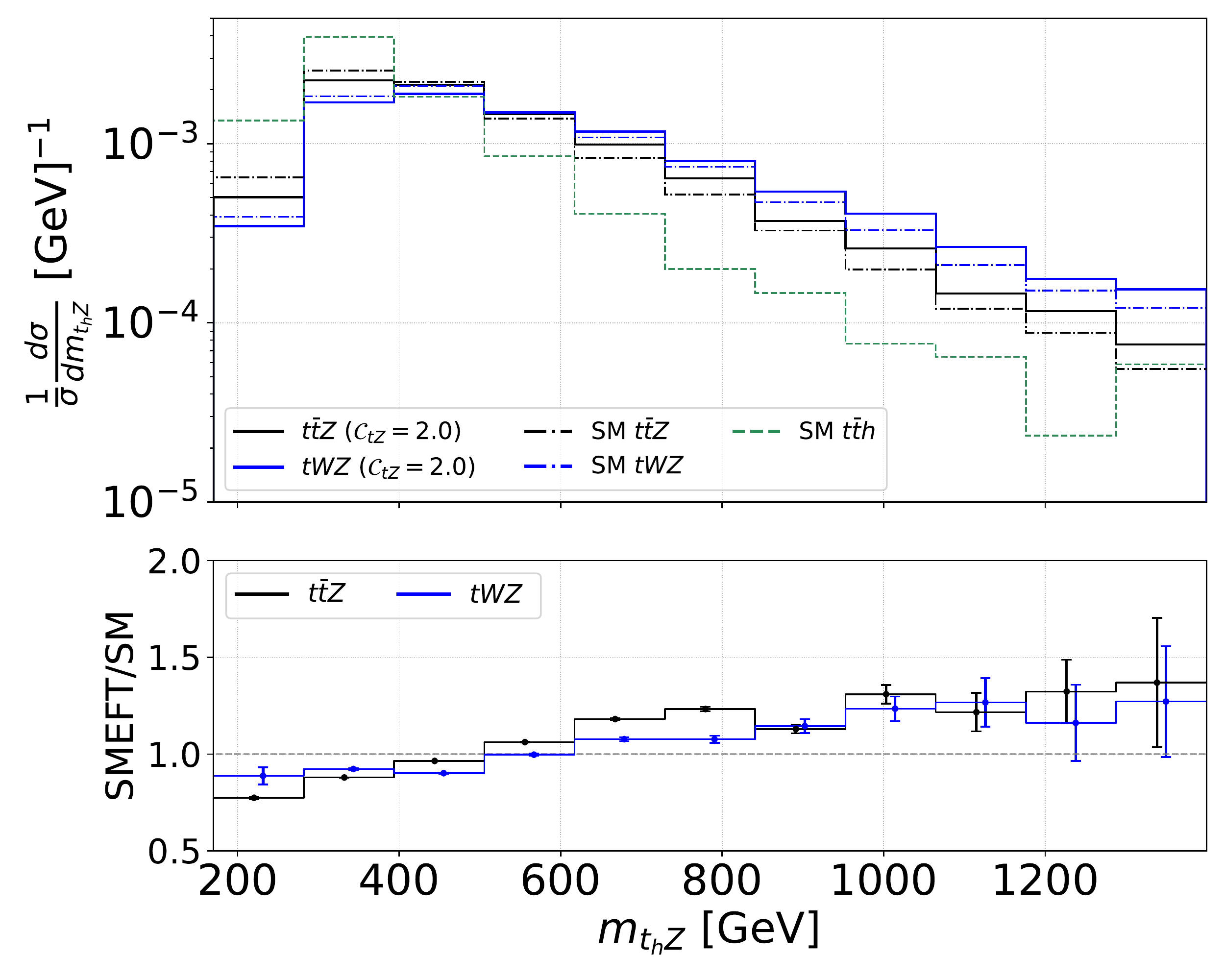}\hspace{2.0cm}\includegraphics[scale=0.28]{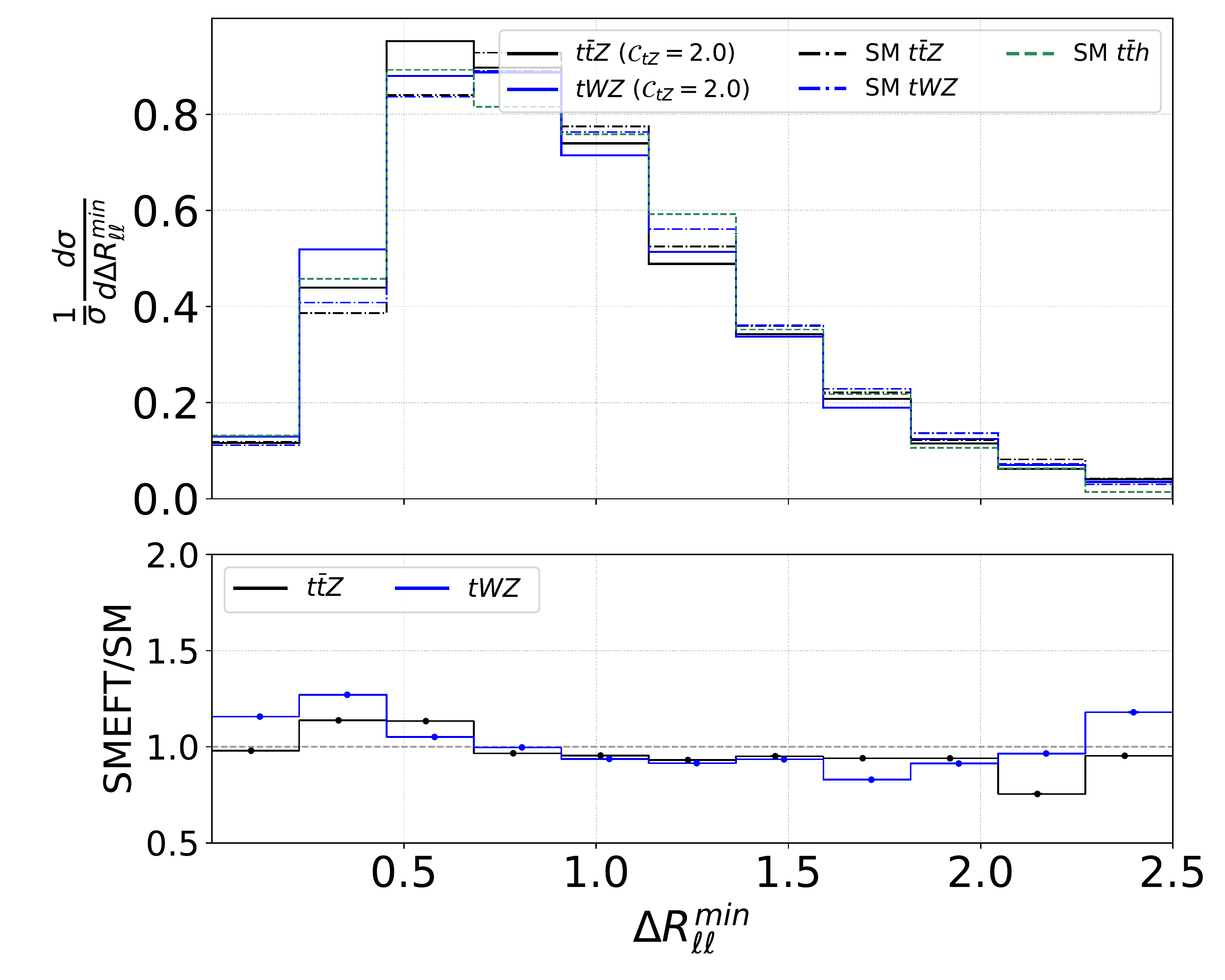}\\
     \includegraphics[scale=0.28]{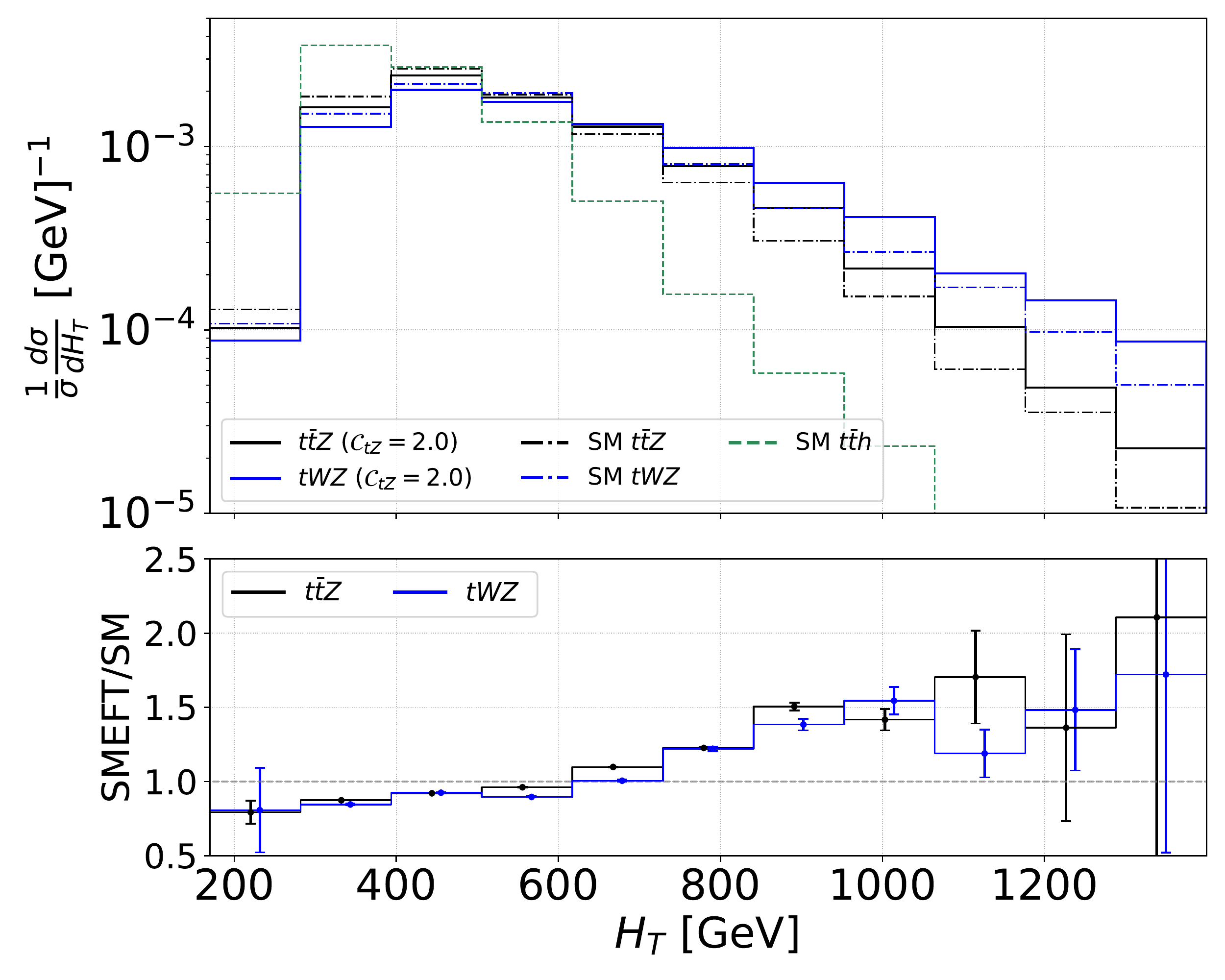}\hspace{2.0cm}\includegraphics[scale=0.28]{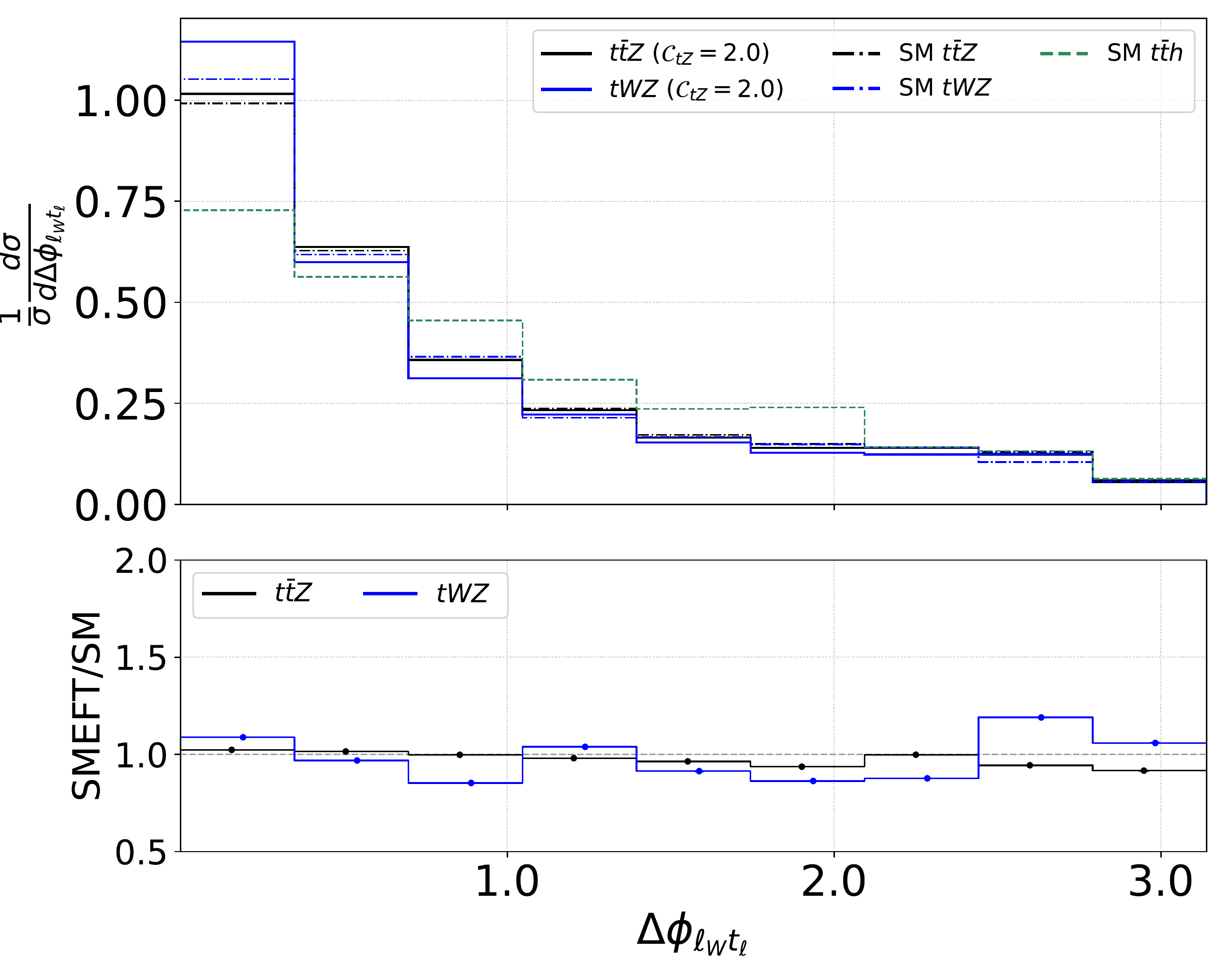}
    \caption{\textit{Top panels:} Distributions for the invariant masses of the hadronically decaying top and $Z$ boson $m_{t_{h}Z}$~(left), and minimum $\Delta R$ separation between a pair of leptons $\Delta R_{\ell\ell}^{\mathrm{min}}$~(right). \textit{Bottom panels:} Distributions for the scalar sum of the transverse momenta of all visible final state objects $H_{T}$~(left), and the azimuthal angle difference $\Delta \phi_{\ell_{W}t_{\ell}}$ between the leptonically decaying top and $\ell_{W}$~(right). The distributions correspond to SMEFT $t\bar{t}Z$~(black solid) and $tWZ$~(blue solid) with $\mathcal{C}_{tZ}=2.0$, SM $t\bar{t}Z$~(black dashed), $tWZ$~(blue dashed) and $t\bar{t}h$~(red dashed). The results are shown at detector level for the LHC with $\sqrt{s}=13~$TeV.}
    \label{fig:ttz_OtZ_distributions}
\end{figure*}

Before turning to the cut-based analysis to estimate the projected sensitivity for $\mathcal{O}_{tZ}$ at the HL-LHC, we discuss some of the distributions of these kinematic variables. In Fig.~\ref{fig:ttz_OtZ_distributions}, we present the distributions for $m_{t_{h}Z}$, $H_{T}$, $\Delta R_{\ell\ell}^{\mathrm{min}}$ and $\Delta \phi_{\ell_{W}t_{\ell}}$ at the detector level, for SMEFT $t\bar{t}Z$ and $tWZ$ with $\mathcal{C}_{tZ}=2.0$, and their pure SM counterparts. The subset of these four observables resulted in the maximal value of $\sigma_{S}^{NP}$ among several other combinations of observables from Eq.~(\ref{eqn:ttz_observables}) considered for the cut-based analysis.
In the bottom panel of the respective figures, we show the ratio of new physics to SM scenario SMEFT/SM. We observe that the ratio SMEFT/SM exceeds $\gtrsim 1$ in the tails of the $m_{t_{h}Z}$ and $H_{T}$ distributions, in accordance with the general expectation that the effects of higher dimension operators should grow with energy. At $H_{T} \sim 1~$TeV, the NP contributions from $\mathcal{C}_{tZ}=2.0$ in $t\bar{t}Z$ and $tWZ$ can be larger than their SM counterparts by $\mathcal{O}(50\%)$. On the other hand, the ratio SMEFT/SM is mostly above 1 for both $t\bar{t}Z$ and $tWZ$ at lower values of $\Delta \phi_{\ell_{W}t_{\ell}}$ and $\Delta R_{\ell\ell}^{\mathrm{min}}$, owing to the inverse relationship between the opening angles and boosts of the intermediate-state particles. We particularly highlight the distribution of $\Delta R_{\ell\ell}^{\mathrm{min}}$ due to its negative correlation with the transverse momentum of the $Z$ boson $p_{T,Z}$ which is one of the most sensitive observables to constrain $\mathcal{C}_{tZ}$~\cite{Rontsch:2014cca, CMS-PAS-FTR-18-036}. The ratio SMEFT/SM increases with larger $p_{T,Z}$. At relatively large $p_{T,Z}$, the leptons from $Z$ decay are highly collimated leading to small $\Delta R$ separation, and thus dominantly constitute the lower bins of the $\Delta R_{\ell\ell}^{\mathrm{min}}$ distribution. 

\begin{figure*}[!htb]
    \centering
    \includegraphics[scale=0.24]{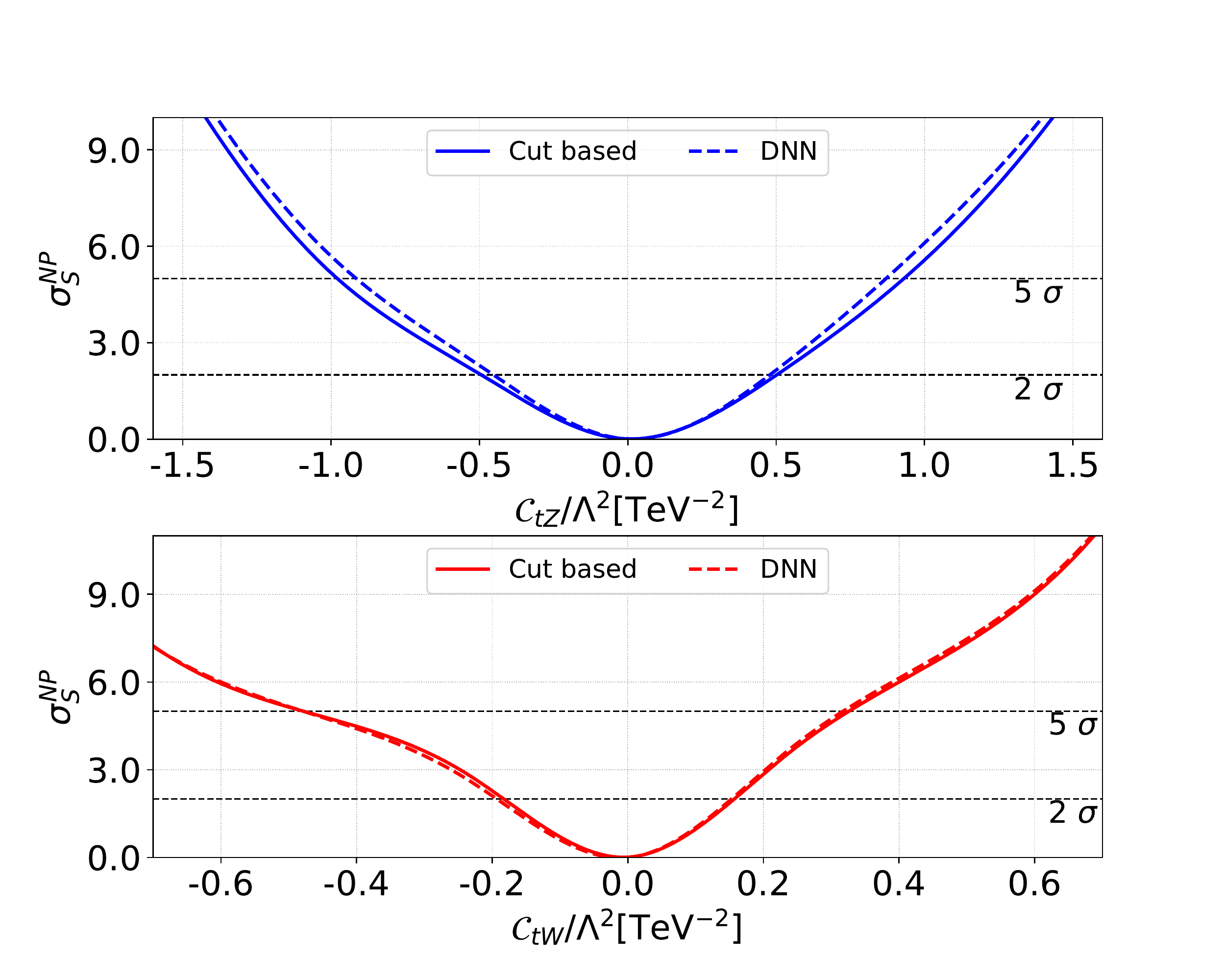}\includegraphics[scale=0.24]{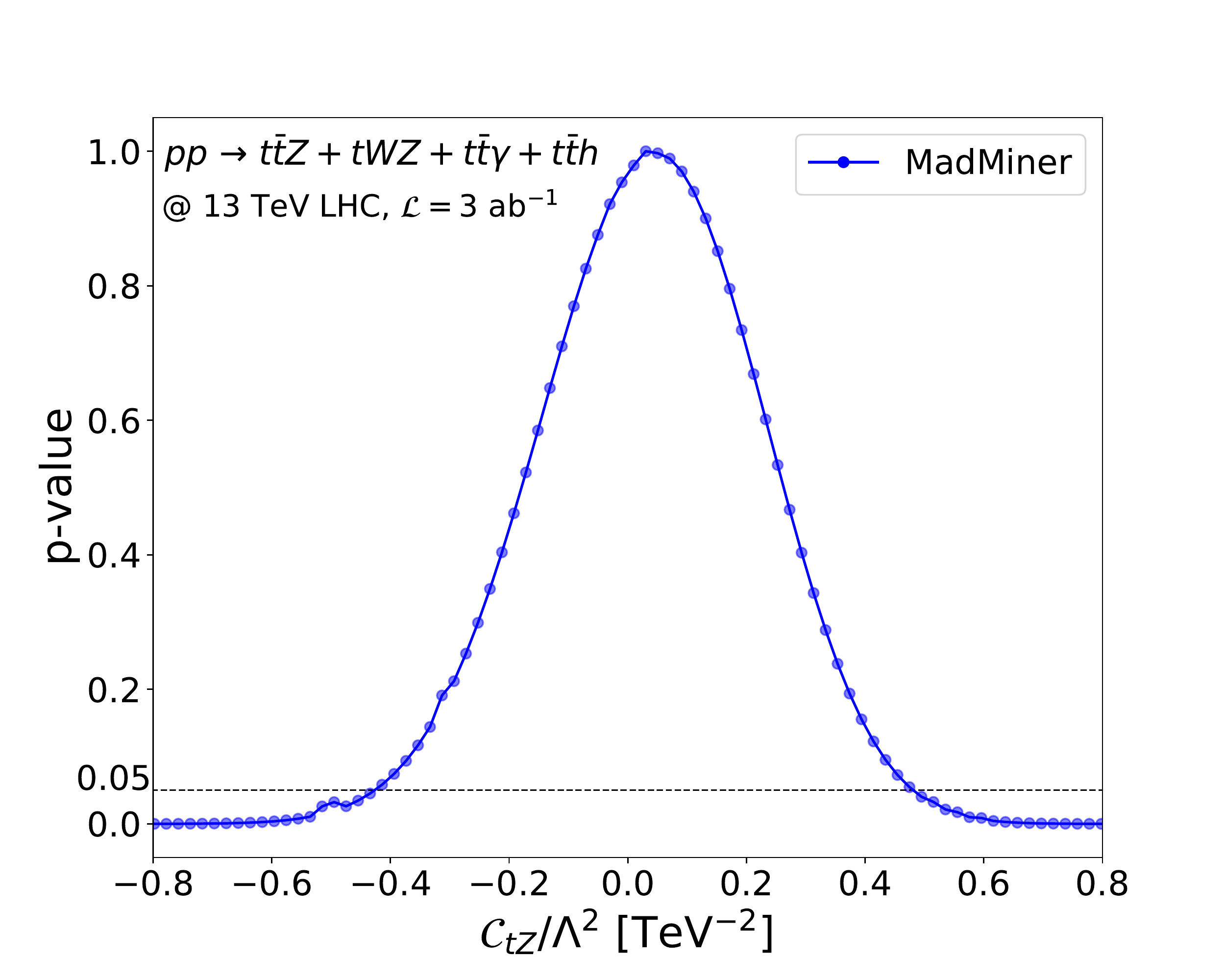}\includegraphics[scale=0.24]{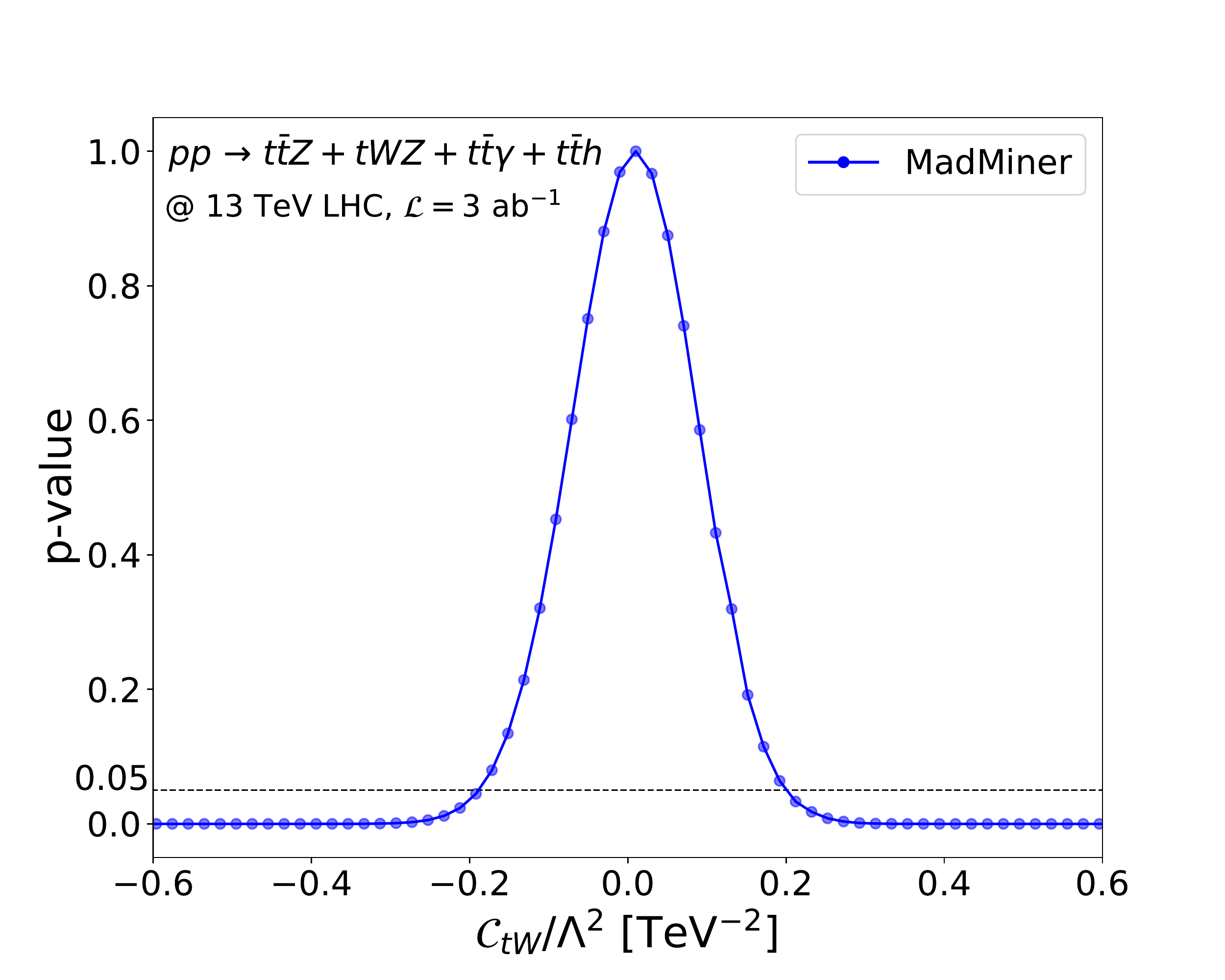}
    \caption{\textit{Left panel:} Projected sensitivity for $\mathcal{O}_{tZ}$~(top panel) and $\mathcal{O}_{tW}$~(bottom panel) from searches in $pp \to t\bar{t}Z + tWZ \to 3\ell + 2b\ + \geq 2j$~(red) channels at the HL-LHC. The solid and dashed lines represent the projections from cut-based and DNN analysis, respectively. \textit{Central panel:} Projected sensitivity for $\mathcal{O}_{tZ}$ through searches in $pp \to t\bar{t}Z + tWZ \to 3\ell + 2b\ + \geq 2j$~(blue) channel using MadMiner. The blue and red solid lines denote the variation of p-values as a function of $\mathcal{C}_{tZ}$ when new physics effects are included at both production and decay level, and only at production, respectively. \textit{Right panel:} Similar to central panel but with $\mathcal{C}_{tW}$ instead of $\mathcal{C}_{tZ}$. The results are presented for $\sqrt{s}=13~$TeV LHC with $\mathcal{L}=3~{\rm ab^{-1}}$.}
    \label{fig:ttz_limits}
\end{figure*}

We proceed to make selection cuts on the aforesaid observables, $m_{t_{h}Z}$, $H_{T}$, $\Delta R_{\ell\ell}^{\mathrm{min}}$ and $\Delta \phi_{\ell_{W}t_{\ell}}$, which maximize $\sigma_{S}^{NP}$ in Eq.~(\ref{eq:cb_significance}).
This cut-based optimization is performed separately for each of 8 signal benchmarks corresponding to different values of $\{\mathcal{C}_{tZ} = \pm 2.0, \pm 1.5, \pm 1.0$, $\pm 0.5\}$. The optimized selection cuts, cut-flow of signal and background rates, and signal significance values $\sigma_{s}^{NP}$, are presented in Table~\ref{tab:ttz_OtZ_cut_flow}. We observe that the optimized signal regions prefer strong cuts on $m_{t_{h}Z}$ and $H_{T}$ $viz$ $m_{t_{h}Z} > 500~$GeV~($400~$GeV) and $H_{T} > 350~$GeV~(500~GeV) at $\mathcal{C}_{tZ}$=1.0~(-1.0), which concurs with the observations in Fig.~\ref{fig:ttz_OtZ_distributions} where we observe an enhancement in the ratio SMEFT/SM at large values of these observables. Similarly, many of the optimized signal regions prefer lower $\Delta R_{\ell\ell}^{\mathrm{min}}$ and $\Delta \phi_{\ell_{W}t_{\ell}}$. This aspect is more apparent at relatively smaller values of $\mathcal{C}_{tZ}=+0.5, -0.5$ where the large $H_{T}$ and $m_{t_{h}Z}$ regions feature a weaker SMEFT-induced enhancement. For $\mathcal{C}_{tZ} = 0.5$, we obtain a signal significance of $1.99$ which increases to $5.57$ at $\mathcal{C}_{tZ}=1.0$. For negative values of $\mathcal{C}_{tZ}$, $\sigma_{S}^{NP}$ improves from 2.04 at $\mathcal{C}_{tZ}=-0.5$ to 5.17 at $\mathcal{C}_{tZ}=-1.0$. This variation of $\sigma_{S}^{NP}$ with $\mathcal{C}_{tZ}$ is summarized in the left panel of Fig.~\ref{fig:ttz_limits} as blue solid line. We observe that $\mathcal{O}_{tZ}$ can be probed up to $-0.49 \lesssim \mathcal{C}_{tZ} \lesssim 0.51$ at the HL-LHC at the $2\sigma$ level through searches in the $pp \to t\bar{t}Z + tWZ \to 3\ell + 2b\ + \geq 2j$ channel.

\begin{table}[!htb]
    \centering\scalebox{0.7}{
    \begin{tabular}{|c||c|c|c|} \hline
          & \multicolumn{3}{|c|}{$\mathcal{C}_{tW}=0.72$}\\\hline 
         Optimized & $H_{T}$ & $\Delta R_{\ell b}^{\mathrm{min}}$ $<$ & $\Delta \phi_{\ell_{W}t_{\ell}}$ $<$ \\ 
         cuts & - & 3.5 & 3.1 \\ \hline 
         SMEFT $t\bar{t}Z$ & 2430 & 2429 & 2428 \\
         SMEFT $tWZ$ & 144 & 144 & 144  \\
         $t\bar{t}Z$ & 1893 & 1892 & 1892 \\
         $tWZ$ & 123 & 122 & 122 \\
         $WZ$ & 153 & 150 & 150 \\
         $t\bar{t}h$ & 14.8 & 14.8 & 14.8 \\
         $t\bar{t}\gamma$ & 19.9 & 19.9 & 19.9\\ \hline
        Significance & 11.90 & 11.91 & 11.92 \\\hline \hline
         & \multicolumn{3}{|c|}{$\mathcal{C}_{tW}=0.48$}\\\hline 
         Optimized & $H_{T} >$ & $\Delta R_{\ell b}^{\mathrm{min}}$ $<$ & $\Delta \phi_{\ell_{W}t_{\ell}}$ $<$ \\ 
         cuts & $250~$GeV & 2.75 & - \\ \hline 
         SMEFT $t\bar{t}Z$ & 2210 & 2201 & 2201 \\
         SMEFT $tWZ$ & 131 & 129 & 129  \\
         $t\bar{t}Z$ & 1889 & 1881 & 1881 \\
         $tWZ$ & 122 & 121 & 121 \\
         $WZ$ & 150 & 136 & 136 \\
         $t\bar{t}h$ & 14.6 & 14.6 & 14.6 \\
         $t\bar{t}\gamma$ & 19.8 & 19.8 & 19.8\\ \hline
        Significance & 7.04 & 7.06 & 7.06 \\\hline \hline
         & \multicolumn{3}{|c|}{$\mathcal{C}_{tW}=0.24$}\\\hline 
         Optimized & $H_{T}$ $>$ & $\Delta R_{\ell b}^{\mathrm{min}}$ $<$ & $\Delta \phi_{\ell_{W}t_{\ell}}$ $<$ \\ 
         cuts & 250~GeV & - & 1.6 \\ \hline 
         SMEFT $t\bar{t}Z$ & 2050 & 2050 & 1678 \\
         SMEFT $tWZ$ & 124 & 124 & 101 \\
         $t\bar{t}Z$ & 1889 & 1889 & 1528 \\
         $tWZ$ & 122 & 122 & 99.5 \\
         $WZ$ & 150 & 150 & 117 \\
         $t\bar{t}h$ & 14.6 & 14.6 & 11.3 \\
         $t\bar{t}\gamma$ & 19.8 & 19.8 & 15.9\\ \hline
        Significance & 3.48 & 3.48 & 3.62\\\hline \hline
        \end{tabular}
        \begin{tabular}{|c|c|c|} \hline
         \multicolumn{3}{|c|}{$\mathcal{C}_{tW}=-0.72$}\\\hline 
          $H_{T}$ & $\Delta R_{\ell b}^{\mathrm{min}}$ $<$ & $\Delta \phi_{\ell_{W}t_{\ell}}$ $<$ \\ 
          - & 2.5 & 3.1 \\ \hline 
          1530 & 1509 & 1508 \\
          133 & 129 & 128  \\
          1893 & 1871 & 1870 \\
          123 & 119 & 119 \\
          153 & 139 & 139 \\
          14.8 & 14.7 & 14.7 \\
          19.9 & 19.7 & 19.7\\ \hline
          7.52 & 7.57 & 7.58\\\hline \hline
        \multicolumn{3}{|c|}{$\mathcal{C}_{tW}=-0.48$}\\\hline 
           $H_{T}$ $>$ & $\Delta R_{\ell b}^{\mathrm{min}}$ $<$ & $\Delta \phi_{\ell_{W}t_{\ell}}$ $<$  \\ 
          250~GeV & 3.0 & - \\ \hline 
          1650 & 1647 & 1647 \\
          128 & 127 & 127  \\
          1889 & 1887 & 1887 \\
          122 & 122 & 122 \\
          150 & 150 & 142 \\
          14.6 & 14.6 & 14.6 \\
          19.8 & 19.8 & 19.8\\ \hline
          4.98 & 5.01 & 5.01\\\hline \hline
        \multicolumn{3}{|c|}{$\mathcal{C}_{tW}=-0.24$}\\\hline 
         $H_{T}$ & $\Delta R_{\ell b}^{\mathrm{min}}$ $<$ & $\Delta \phi_{\ell_{W}t_{\ell}}$ $<$ \\ 
         - & 2.25 & 2.8 \\ \hline 
         1767 & 1710 & 1675 \\
         120 & 112 & 110\\
         1893 & 1838 & 1802  \\
         123 & 115 & 113 \\
         153 & 131 & 131 \\
         14.7 & 14.6 & 14.3 \\
         19.9 & 19.5 & 19.2 \\ \hline
        2.77 & 2.88 & 2.90 \\\hline \hline
        \end{tabular}}
        \caption{Optimized selection cuts on $H_{T}$, $\Delta R_{\ell b}^{\mathrm{min}}$ and $\Delta \phi_{\ell_{W}t_{\ell}}$, applied successively, to maximize the signal significance $\sigma_{S}^{NP}$ from cut-based collider analysis in the $pp \to t\bar{t}Z + tWZ \to 3\ell + 2b\ + \geq 2j $ channel to estimate the projected sensitivity for $\mathcal{O}_{tW}$ at $\sqrt{s}=13~$TeV LHC with $\mathcal{L}=3~{\rm ab^{-1}}$. The optimized cuts, signal and background yields, and $\sigma_{S}^{NP}$ values are shown for $\{\mathcal{C}_{tW} = \pm 0.72, \pm 0.48, \pm 0.24\}$. No cuts are applied on $H_{T}$ for $\mathcal{C}_{tW}=0.72, -0.72, -0.24$, on $\Delta R_{\ell b}^{min}$ for $\mathcal{C}_{tW} = 0.24$, and on $\Delta \phi_{\ell_{W}t_{\ell}}$ for $\mathcal{C}_{tW} = 0.48, -0.48$.}
    \label{tab:ttz_OtW_cut_flow}
\end{table}

\begin{figure*}[!htb]
    \centering
    \includegraphics[scale=0.23]{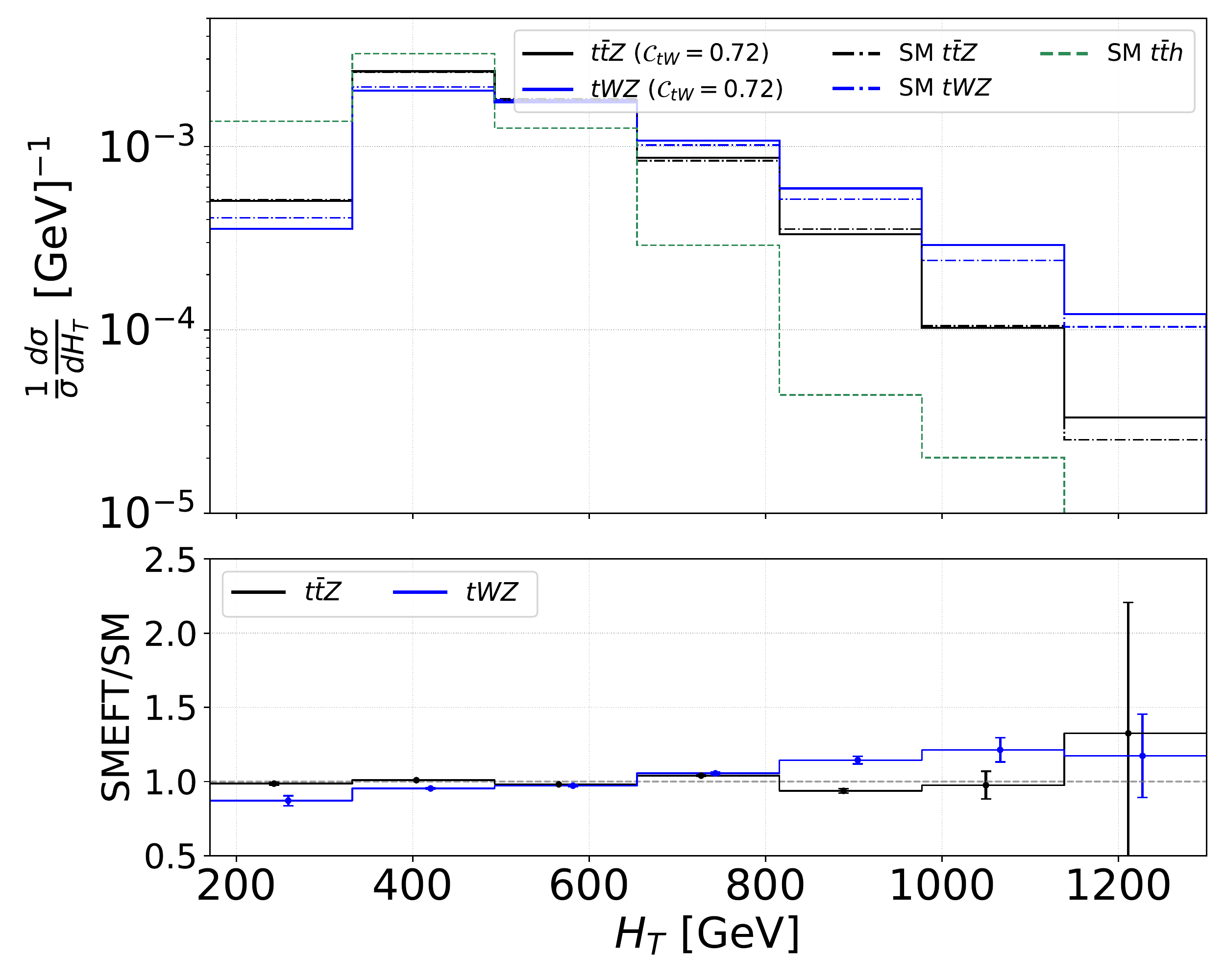}\includegraphics[scale=0.23]{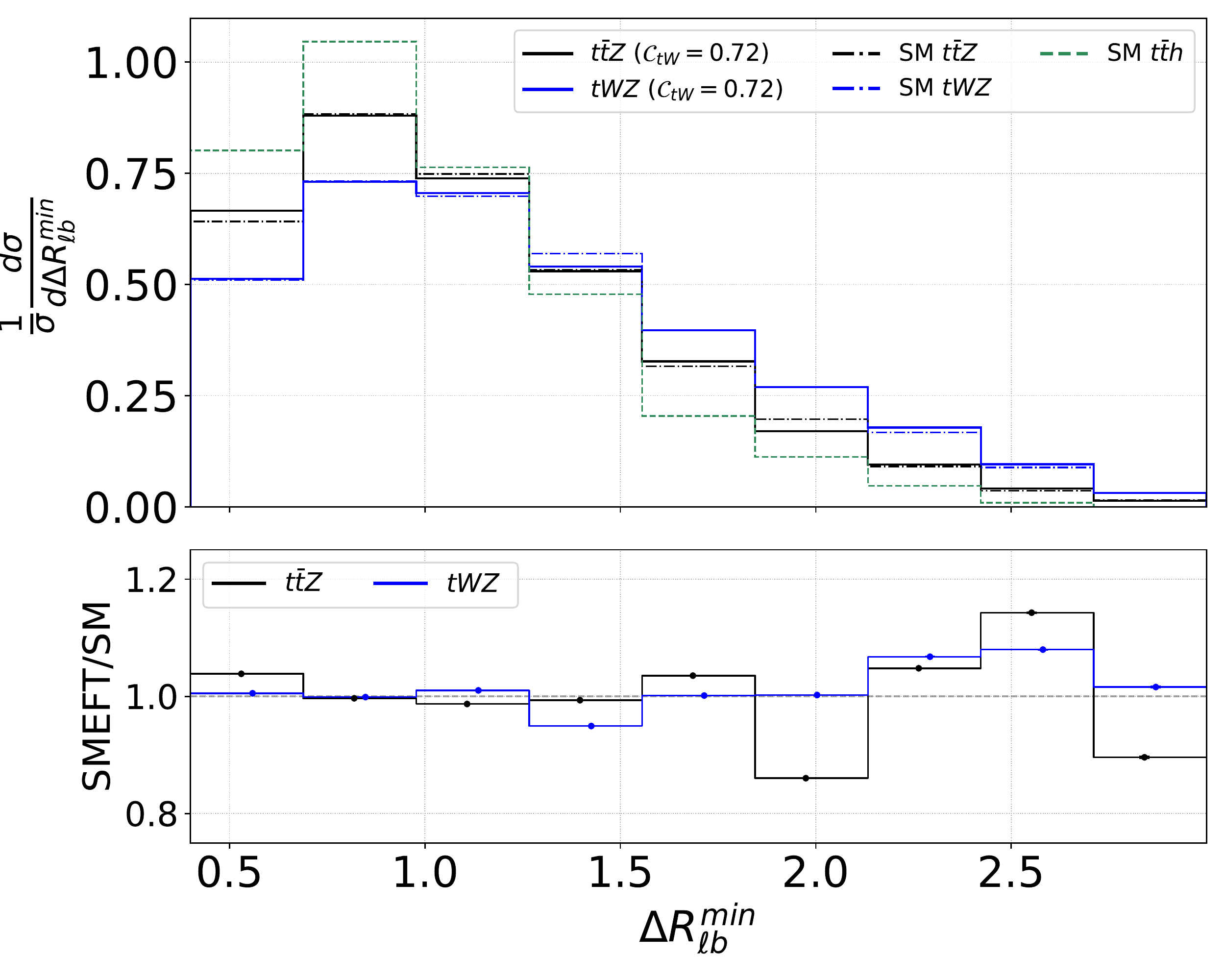}\includegraphics[scale=0.23]{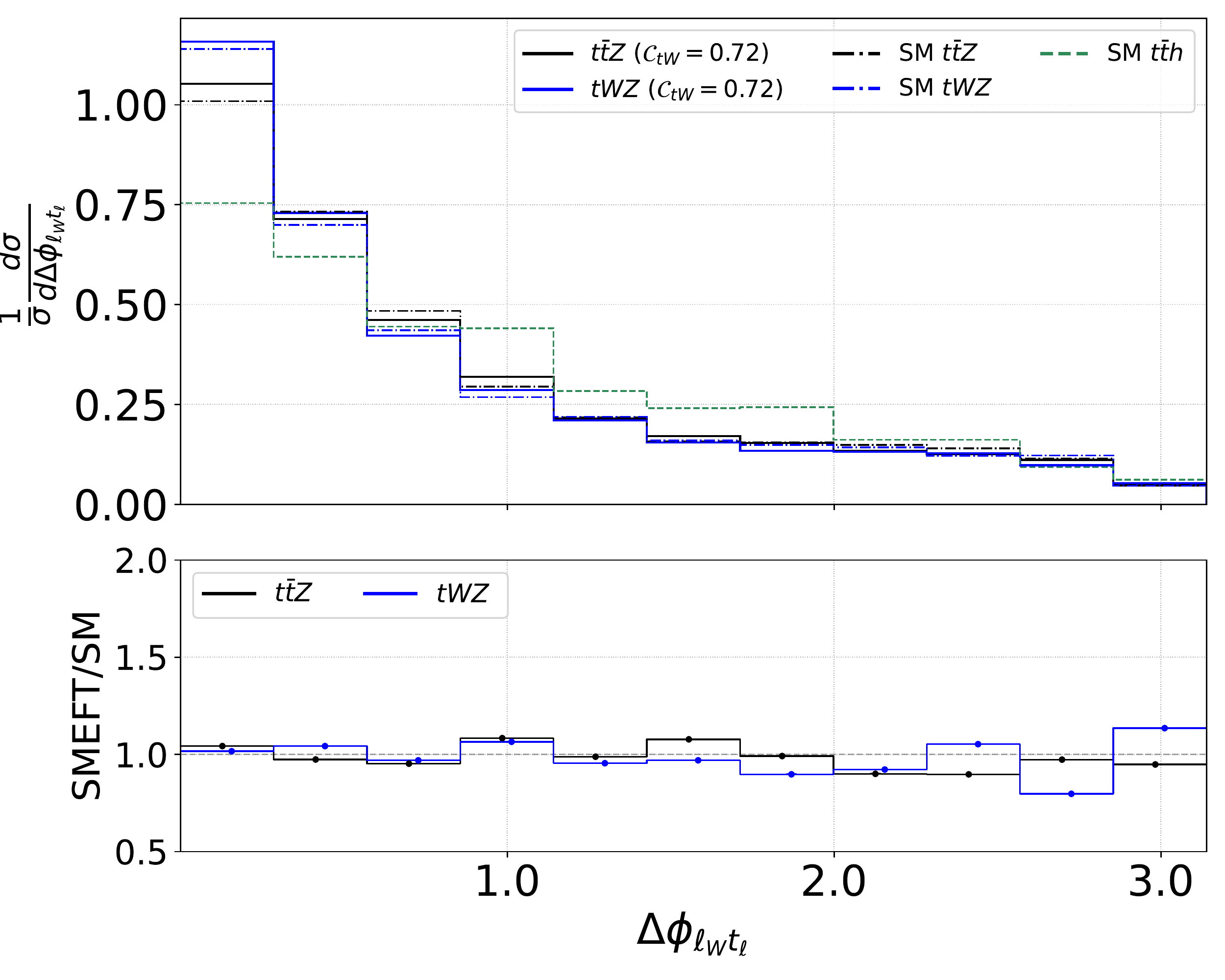}
    \caption{Distributions for the scalar sum of the transverse momenta of all visible final state objects $H_{T}$~(left), minimum $\Delta R$ separation between a lepton and $b$ jet pair $\Delta R_{\ell b}^{\mathrm{min}}$~(center), and difference of azimuthal angles for the leptonically decaying top and $\ell_{W}$ $\Delta \phi_{\ell_{W}t_{\ell}}$~(right), for SMEFT $t\bar{t}Z$~(black solid) and $tWZ$~(blue solid) with $\mathcal{C}_{tW}=0.72$, SM $t\bar{t}Z$~(black dashed), $tWZ$~(blue dashed) and $t\bar{t}h$~(red dashed). The results are presented at the detector level for the LHC with $\sqrt{s}=13~$TeV.}
    \label{fig:ttz_OtW_distributions}
\end{figure*}

While $\mathcal{O}_{tZ}$ leads to new physics contributions only at the production level, $\mathcal{O}_{tW}$ can induce modifications of both top production and decay by virtue of its modification to the $tWb$ vertex. We perform a separate cut-based analysis to estimate the projected sensitivity for $\mathcal{O}_{tW}$ at the HL-LHC. For this analysis, we consider several subsets of observables from Eqn.~(\ref{eqn:ttz_observables}) for cut-based optimization. Among them, the subset of $\{H_{T},~\Delta R_{\ell b}^{\mathrm{min}},~\Delta \phi_{\ell_{W}t_{\ell}}\}$, leads to the strongest sensitivity. In Fig.~\ref{fig:ttz_OtW_distributions}, we present their distributions, at the detector level,
for SMEFT $t\bar{t}Z$ and $tWZ$ at $\mathcal{C}_{tW}=0.72$, and SM $t\bar{t}Z$, $tWZ$ and $t\bar{t}h$. Unlike the $\mathcal{C}_{tZ}$ scenario, the ratio SMEFT/SM for $t\bar{t}Z$ in Fig.~\ref{fig:ttz_OtW_distributions} remains roughly close to $1$, demonstrating a reduced sensitivity to $\mathcal{C}_{tW}$, except in the highly boosted $H_{T}$ regime, $H_{T} > 1100~\mathrm{GeV}$, which is marred by large statistical uncertainty. We consider 6 different signal benchmarks corresponding to $\{\mathcal{C}_{tW}=\pm 0.72, \pm 0.48, \pm 0.24\}$. The optimized cuts on $H_{T}$, $\Delta R_{\ell b}^{\mathrm{min}}$ and $\Delta \phi_{\ell_{W}t_{\ell}}$, signal and background yields, and $\sigma_{S}^{NP}$ values are presented in Table~\ref{tab:ttz_OtW_cut_flow}. We obtain $\sigma_{S}^{NP} = 11.92~(7.62)$ for $\mathcal{C}_{tW} = 0.72~(-0.72)$, which decreases to 3.64~(2.90) at $\mathcal{C}_{tW}=0.24~(-0.24)$. From the cut flows in Table~\ref{tab:ttz_OtW_cut_flow}, we see that the kinematic cuts do not significantly increase the significance. This follows from the reduced dependence of the kinematic distributions in Fig.~\ref{fig:ttz_OtW_distributions} on the EFT operator, and we do not expect large gains beyond a rate-only measurement. Using the results from Table~\ref{tab:ttz_OtW_cut_flow}, we interpolate the variation of $\sigma_{S}^{NP}$ as a function of $\mathcal{C}_{tW}$, as illustrated in the left panel of Fig.~\ref{fig:ttz_limits} as red solid line. We observe that the HL-LHC would be able to probe $\mathcal{C}_{tW}$ up to $-0.19 \lesssim \mathcal{C}_{tW} \lesssim 0.16$ at $2\sigma$ uncertainty through searches in the $pp \to t\bar{t}Z + tWZ \to 3\ell + 2b\ + \geq 2j $ channel. We also find other subsets of observables that lead to roughly comparable sensitivity $viz.$ $\{H_{T},p_{T,Z},p_{T,W},\Delta R_{\ell b}^{\mathrm{min}}\}$, $\{p_{T,Z}/p_{T,W}, m_{t_{h}t_{\ell}Z}, \Delta R_{\ell b}^{\mathrm{min}},\Delta \phi_{\ell_{W}t_{\ell}}\}$.

\begin{table}[!htb]
\centering
\begin{tabular}{|c|c|c|c|c|c|c|c|c|c|} \hline 
    \multirow{2}{*}{$\mathcal{C}_{tZ}$} & \multicolumn{2}{c|}{SMEFT} & \multicolumn{5}{c|}{Background} & \multirow{2}{*}{$\alpha$}  & \multirow{2}{*}{$\sigma_{S}^{NP}$} \\ \cline{2-8}
    & $t\bar{t}Z$ & $tWZ$ & $t\bar{t}Z$ & $tWZ$ & $WZ$ & $t\bar{t}h$ & $t\bar{t}\gamma$ &  & \\ \hline
    2.0 & 1557 & 84.5 & 942 & 58.6 & 73.6 & 1.9 & 7.7 & 0.60 & 19.23\\   
    1.5 & 979 & 56.9 & 673 & 44.5 & 51.8 & 0.9 & 5.9 & 0.64 & 11.22 \\   
    1.0 & 1185 & 68.8 & 984 & 63.6 & 81.8 & 2.3 & 8.7 & 0.60 & 6.10\\   
    0.5 & 640 & 40.3 & 582 & 41.9 & 57.2 & 3.5 & 5.5 & 0.60 & 2.15 \\   
    -0.5 & 1038 & 63.7 & 963 & 61.8 & 81.8 & 7.0 & 9.8 & 0.56 & 2.3 \\   
    -1.0 & 906 & 56.3 & 743 & 51.2 & 68.2 & 1.5 & 6.2 & 0.62 & 5.69 \\   
    -1.5 & 1594 & 93.7 & 1179 & 77.6 & 111.8 & 3.6 & 11.0 & 0.56 & 11.61 \\   
    -2.0 & 2016 & 111 & 1258 & 82.9 & 114.5 & 4.3 & 11.8 & 0.55 & 20.18 \\ \hline   
    \end{tabular}
    \caption{Signal significance $\sigma_{S}$ from DNN analysis in $pp \to t\bar{t}Z + tWZ \to 3\ell + 2b\ + \geq 2j $ channel for $\{\mathcal{C}_{tZ} = \pm 2.0, \pm 1.5, \pm 1.0, \pm 0.5\}$ at $\sqrt{s}=13~$TeV LHC with $\mathcal{L}=3~{\rm ab^{-1}}$. The signal rates for SMEFT $t\bar{t}Z$ and $tWZ$ processes, background rates for SM $t\bar{t}Z$, $tWZ$, $t\bar{t}\gamma$, $t\bar{t}h$ and $WZ+\mathrm{jets}$ are presented. The optimal DNN score $\alpha$ and corresponding signal significance~$\sigma_{S}^{NP}$ are also shown.}
    \label{tab:ttz_OtZ_DNN}
\end{table}

\begin{table}[!htb]
\centering
\begin{tabular}{|c|c|c|c|c|c|c|c|c|c|} \hline 
    \multirow{2}{*}{$\mathcal{C}_{tW}$} & \multicolumn{2}{c|}{SMEFT} & \multicolumn{5}{c|}{Background} & \multirow{2}{*}{$\alpha$}  & \multirow{2}{*}{$\sigma_{S}^{NP}$} \\ \cline{2-8}
      & $t\bar{t}Z$ & $tWZ$ & $t\bar{t}Z$ & $tWZ$ & $WZ$ & $t\bar{t}h$ & $t\bar{t}\gamma$ &  & \\ \hline
     0.72 & 2348 & 132 & 1818 & 111 & 142 & 14.0 & 18.6 & 0.23 & 12.0\\   
     0.48 & 2088 & 115 & 1771 & 106 & 142 & 13.8 & 18.2 & 0.26 & 7.19\\   
     0.24 & 1957 & 110 & 1790 & 108 & 141 & 13.9 & 18.3 & 0.25 & 3.73\\   
     -0.24 & 1763 & 120 & 1888 & 122 & 150 & 14.7 & 19.9 & 0.20 & 2.72\\   
     -0.48 & 1651 & 126 & 1890 & 122 & 153 & 14.7 & 19.9 & 0.11 & 5.01\\   
     -0.72 & 1527 & 132 & 1890 & 122 & 153 & 14.7 & 19.9 & 0.06 & 7.52\\  \hline  
    \end{tabular}
    \caption{Signal significance $\sigma_{S}$ from DNN analysis in $pp \to t\bar{t}Z + tWZ \to 3\ell + 2b\ + \geq 2j $ channel for $\{\mathcal{C}_{tW} = \pm 0.72, \pm 0.48, \pm 0.24\}$ at $\sqrt{s}=13~$TeV LHC with $\mathcal{L}=3~{\rm ab^{-1}}$. The signal rates for SMEFT $t\bar{t}Z$ and $tWZ$, background rates for SM $t\bar{t}Z$, $tWZ$, $t\bar{t}\gamma$, $t\bar{t}h$ and $WZ+\mathrm{jets}$ are presented. The optimal DNN score $\alpha$ and corresponding signal significance~$\sigma_{S}^{NP}$ are also shown.}
    \label{tab:ttz_OtW_DNN}
\end{table}

While the cut-based approach is simple to apply and easily interpretable using the individual features of the selected observables, it is less effective in exploring any correlations which might exist among the observables. Furthermore, it becomes progressively more cumbersome as the dimensionality of input features is increased. Therefore, in order to comprehensively explore the sensitivity for $\mathcal{O}_{tZ}$ and $\mathcal{O}_{tW}$, we also perform a machine-learning based multivariate analysis using a Deep Neural Network~(DNN), using the same signal benchmarks as for the cut-based approach.
While the cut-based analysis only takes into account the differences in the shape of a few selected distributions, the neural networks can exploit the shape information of a much larger number of input features while also taking into account the NP deviations in their correlations.
For each signal benchmark, we construct a fully connected DNN using Keras, which takes as input the 150 observables of Eq.~(\ref{eqn:ttz_observables}). Each DNN has between 4 and 8 hidden layers; the number of layers and the number of nodes in each layer are optimized for each benchmark.
We use the Rectified Linear Unit~(ReLU) activation function in each layer except for the final one, where the Sigmoid activation function is used instead in order to provide an output classifying an event as SM-like (0) or SMEFT-like (1). Training is performed on a subset of our event sample using Adam optimization to minimize binary cross-entropy loss over 200 epochs, with a learning rate of $10^{-5}$ and a batch size of 64. In order to avoid overtraining, we apply early stopping using a validation set. 

The training data for the DNN is comprised of $t\bar{t}Z$ and $tWZ$ events with at least one SMEFT vertex, and SM $t\bar{t}Z$, $tWZ$ and $WZ+\mathrm{jets}$ events. The network is trained to distinguish the pure EFT $t\bar{t}Z$ and $tWZ$ events from the SM processes. The test data is comprised of SMEFT $t\bar{t}Z$ and $tWZ$ events~(sensitive to pure SM, interference terms, and NP squared terms), and SM $t\bar{t}Z$, $tWZ$, $t\bar{t}\gamma$, $t\bar{t}h$ and $WZ+\mathrm{jets}$ events. SM $t\bar{t}h$ and $t\bar{t}\gamma$ events are not included in the training data due to their relatively lower cross sections.

We identify events with DNN output values above a cutoff $\alpha$ as signal-like, and those with output scores below $\alpha$ as background. After each network is trained, we choose the value of $\alpha$ that maximizes $\sigma_{S}^{NP}$. The resulting signal and background yields are listed in Tables~\ref{tab:ttz_OtZ_DNN} and \ref{tab:ttz_OtW_DNN}. In the case of $\mathcal{O}_{tZ}$, the multivariate DNN analysis improves the projected sensitivity by $\sim \mathcal{O}(5\text{-}10)\%$ compared to cut-based optimization. For example, $\sigma_{S}^{NP}$ for $\mathcal{C}_{tZ} = 0.5$~(-0.5) improves from 1.9~(2.0) with the cut-based analysis to 2.1~(2.3) with the DNN. For $\mathcal{O}_{tW}$, the differences between the cut-based and DNN results are smaller. We interpolate $\sigma_{S}^{NP}$ as a function of $\mathcal{C}_{tZ}$~($\mathcal{C}_{tW}$) using the results in Table~\ref{tab:ttz_OtZ_DNN}~(Table~\ref{tab:ttz_OtW_DNN}), and present them in the left panel of Fig.~\ref{fig:ttz_limits} as blue~(red) solid lines. The projected sensitivity for $\mathcal{O}_{tZ}$ reaches up to $-0.45 \lesssim \mathcal{C}_{tZ} \lesssim 0.48$ at $2\sigma$ uncertainty, thus, registering $\mathcal{O}(7\%)$ improvement over the cut-based results. In the case of $\mathcal{O}_{tW}$, the projected sensitivity reaches up to $-0.19 \lesssim \mathcal{C}_{tW} \lesssim 0.15$ which is almost comparable to the results from the cut-based analysis. Next, we assess the most important input observables in the dataset using the Python-based \texttt{ELI5} tool~\cite{eli5}.
Specifically, we calculate the permutation feature importance by measuring the decrements in model score when the data for each feature is randomly shuffled among events. We compute the permutation importance scores of input observables in all of the DNN models trained on our signal benchmarks.
Although the relative weight of observables exhibits variation across different signal benchmarks, the subset of the most sensitive observables remains almost unchanged. We list 35 such observables which typically feature in the list of leading permutation scores for all signal benchmarks:
\begin{equation}
\begin{split}    
 &m_{t_{\ell}t_{h}}, m_{t_{\ell/h}Z}, m_{t_{\ell}t_{h}Z}, H_{T}, p_{T,\ell_{1/2}}, p_{T,\ell_{W}}, p_{T,b_{h}}, \\
 &p_{T,Z}, p_{T,t_{\ell/h}}, p_{T,t_{\ell/h}Z}, p_{T,t_{\ell}t_{h}}, \Delta R_{\ell\ell}^{\mathrm{max}}, \Delta R_{\ell\ell}^{\mathrm{min}}, \Delta R_{\ell b}^{\mathrm{min}}, \\
 &\phi_{\ell_{W}}, \phi_{t_{\ell}}, \Delta \phi_{\ell_{W}t_{\ell}}, \Delta \phi_{\ell_{W}\ell_{1}}, \Delta \phi_{\nu\ell_{1}}, \Delta \phi_{\nu b_{2}},\\ 
 &\Delta \phi_{b_{\ell}\ell_{1}}, \Delta \phi_{b_{\ell}\ell_{2}}, \Delta \phi_{j_{2}\ell_{1}}, \Delta \phi_{b_{h}\ell_{2}}, \Delta \phi_{\ell_{1}t_{\ell}},\\
 &\eta_{\ell_{1}}, \Delta \eta_{t_{\ell}Z}, \Delta \eta_{\nu Z}, \Delta \eta_{\ell_{W}\ell_{2}}, \Delta \eta_{b_{\ell}\ell_{2}}, \Delta \eta_{\ell_{2}Z}.
 \end{split}
 \end{equation}
Training DNN models with only these 35 input observables results in sensitivities that are comparable to those from models trained using all 150 observables listed in Eq.~(\ref{eqn:ttz_observables}).

We note that pure EFT $t\bar{t}Z$ and $tWZ$ events used in the training dataset have been generated in the \texttt{MG5\_aMC@NLO} framework by setting the squared coupling order $NP^{2} >0$, and particle decay chains cannot be specified together with squared coupling orders. As a result, the pure EFT events used for training the DNN model include NP effects at the production level only and lack spin-correlation effects in top decays. However, in the scenario with non-zero $\mathcal{C}_{tW}$, the SMEFT $t\bar{t}Z$ and $tWZ$ events in the test dataset include NP contributions both for production and top decay, along with top spin-correlation effects in the final state objects. The consequences of this disparity are expected to be more pronounced for $t\bar{t}Z$~(compared with $tWZ$) which also happens to be the most dominant signal. This difference between the training and test datasets could be responsible for the meagre improvement in the projected sensitivity from a multivariate DNN analysis over the traditional cut-based approach. On the other hand, $\mathcal{O}_{tZ}$ affects $t\bar{t}Z$ and $tWZ$ at the production level only. Therefore, when we examine the projected sensitivity to $\mathcal{O}_{tZ}$, the only source of disparity between the training and test datasets is top spin correlation effects. Overall, the inconsistency between the training and test datasets is expected to be less severe in the case of $\mathcal{O}_{tZ}$ than for $\mathcal{O}_{tW}$.

In order to ensure that we are not losing sensitivity to NP operators in the DNN analysis from the missing correlation effects in the training data, we also estimate the projected sensitivities through a likelihood-based approach which takes into account new physics effects both at production and decay along with $t\bar{t}$ spin-correlation effects. To achieve this, we use the \texttt{MadMiner} tool which employs machine-learning based event information extraction techniques to obtain the event likelihood ratio as a function of the SMEFT parameters~\cite{Brehmer:2018hga, Brehmer:2019xox}. The event likelihood ratio $r(x|\theta,\theta_{SM}) = p(x|\theta)/p(x|\theta_{SM})$, where $p(x|\theta)$ is the probability of observables $x$ given theory parameters $\theta = \{\mathcal{C}_{tZ}, \mathcal{C}_{tW}\}$~($\theta_{SM}=\{0,0\}$), is the most powerful test statistic to discriminate the hypothesis $\theta$ from $\theta_{SM}$~\cite{Brehmer:2018hga}. However, at the detector level, $r(x|\theta,\theta_{SM})$ is an intractable function due to conditioning from several latent variables $z$ such as parton showering, hadronization and detector response.
On the other hand, the joint likelihood ratio $r(x,z|\theta,\theta_{SM})$ can be computed for every Monte Carlo~(MC) simulated event at the detector level~\cite{Brehmer:2018hga, Brehmer:2019xox}. In addition, the joint score $t(x,z|\theta_{0}) = \triangledown_{\theta}(p(x,z|\theta))\big|_{\theta_{0}}$, the gradient of the joint likelihood ratio at reference positions $\theta_0$ in theory parameter space, can also be computed from MC simulation and used to help estimate the true likelihood ratio $r(x|\theta,\theta_{SM})$. \texttt{MadMiner} uses matrix element information from MC event samples and shape information in reconstructed observables to train a neural network using an appropriate loss functional that depends on $r(x,z|\theta,\theta_{SM})$ and/or $t(x,z|\theta_{0})$. The loss function is defined such that its minimizing function is the intractable event likelihood ratio $r(x|\theta)$ and the trained NN is an estimator of $r(x|\theta)$~\cite{Brehmer:2018hga}. This estimated likelihood ratio is sensitive to both linear and non-linear NP effects. In the presence of SMEFT operators $\theta_{i}$, the matrix squared element at the parton level $|\mathcal{M}|^{2}$ is given by,
\begin{equation}
\begin{split}
    |\mathcal{M}|^{2} = &~1 \cdot |\mathcal{M}|_{SM}^{2}(x,\theta_{SM}) + \sum_{i} \theta_{i}^{2} \cdot |\mathcal{M}|_{BSM}^{2}(x,\theta_{i}) \\ & + \sum_{i} 2~\theta_{i} \cdot \mathrm{Re} |\mathcal{M}|_{SM}^{\dagger}(x,\theta_{SM})~|\mathcal{M}|_{BSM}(x,\theta_{i}) \\ & + \sum_{i,j}^{i\neq j} 2~\theta_{i}~\theta_{j}\cdot \mathrm{Re}|\mathcal{M}|_{BSM}^{\dagger}(x,\theta_{i})~|\mathcal{M}|_{BSM}(x,\theta_{j}),
    \label{eq:matrix_squared_element}
\end{split}
\end{equation}
where, $|\mathcal{M}|_{SM}^{2}(x,\theta_{SM})$ represents the matrix squared element for the SM, while $|\mathcal{M}|_{BSM}^{2}(x,\theta_{i})$ represents the matrix squared element for pure SMEFT interactions corresponding to $\theta_{i}$. Eq.~(\ref{eq:matrix_squared_element}) can be factorized through a morphing technique into the product of an analytic function $w_{c}(\theta)$ that is exclusively dependent on $\theta$ and a phase space dependent function $f_{c}(x)$, summed over $c$ components,~\cite{Brehmer:2018hga,Brehmer:2019xox}
\begin{equation}
\begin{split}
    |\mathcal{M}|^{2} = & \sum_{c} ~w_{c}(\theta) \cdot f_{c}(x).
\end{split}
\end{equation}
Here, the $f_{c}(x)$ are not necessarily positive or normalized distributions. The number of components $c$ is equal to the number of elements in Eq.~(\ref{eq:matrix_squared_element}), which also defines the number of signal benchmarks that form the morphing basis. Once the parton level event weights~(or matrix squared elements) are computed at these $c$ signal benchmarks, the ``morphing setup" can evaluate the event weights at any given $\theta$. In the present study, $\theta_{i}$ has two components, $\mathcal{C}_{tZ}$ and $\mathcal{C}_{tW}$. Thereby, the morphing basis would include 6 components~(or signal benchmarks with different $\theta$) if new physics effects are considered at the production level only. Since $\mathcal{C}_{tW}$ also affects top decay, the morphing basis consists of 12 benchmarks. Accordingly, we generate event samples for $t\bar{t}h$, $tWZ$, $t\bar{t}\gamma$ and $t\bar{t}h$ processes at 12 different benchmark values of $\theta$. \texttt{MadMiner} utilizes the event weights for these 12 benchmarks to interpolate the event weights in the $\{\mathcal{C}_{tZ},\mathcal{C}_{tW}\}$ plane through the morphing setup. We consider squared and quartic ansatz for $\mathcal{C}_{tZ}$ and $\mathcal{C}_{tW}$, respectively, in the morphing technique. The squared ansatz for $\mathcal{C}_{tZ}$ is prompted by its contribution at the production level only while $\mathcal{C}_{tW}$ contributes both at production and decay level actuating the quartic ansatz. We generate $10^{6}$ events for each of the aforesaid processes and reconstruct all of the observables in Eq.~(\ref{eqn:ttz_observables}). We consider a fully connected neural network with 3 hidden layers each containing 100 nodes. Training is performed using the \texttt{ALICES} algorithm~\cite{Stoye:2018ovl} over 120 epochs. The ALICES loss functional depends on both the joint likelihood ratio $r(x,z|\theta,\theta_{SM})$ and the joint score $t(x,z|\theta_{0})$ to maximize the inclusion of information that can be obtained from the MC event samples simulated at the detector level. The relative weights of the terms in the \texttt{ALICES} loss functional that depend on the joint score and the joint likelihood ratio is parametrized by the hyperparameter $\alpha$, which we set equal to 1. We employ a batch size of 128, the tanh activation function, and Adam optimization, with a learning rate that exponentially decays from $10^{-4}$ to $10^{-5}$.
\begin{figure}[!t]
    \centering
    \includegraphics[scale=0.30]{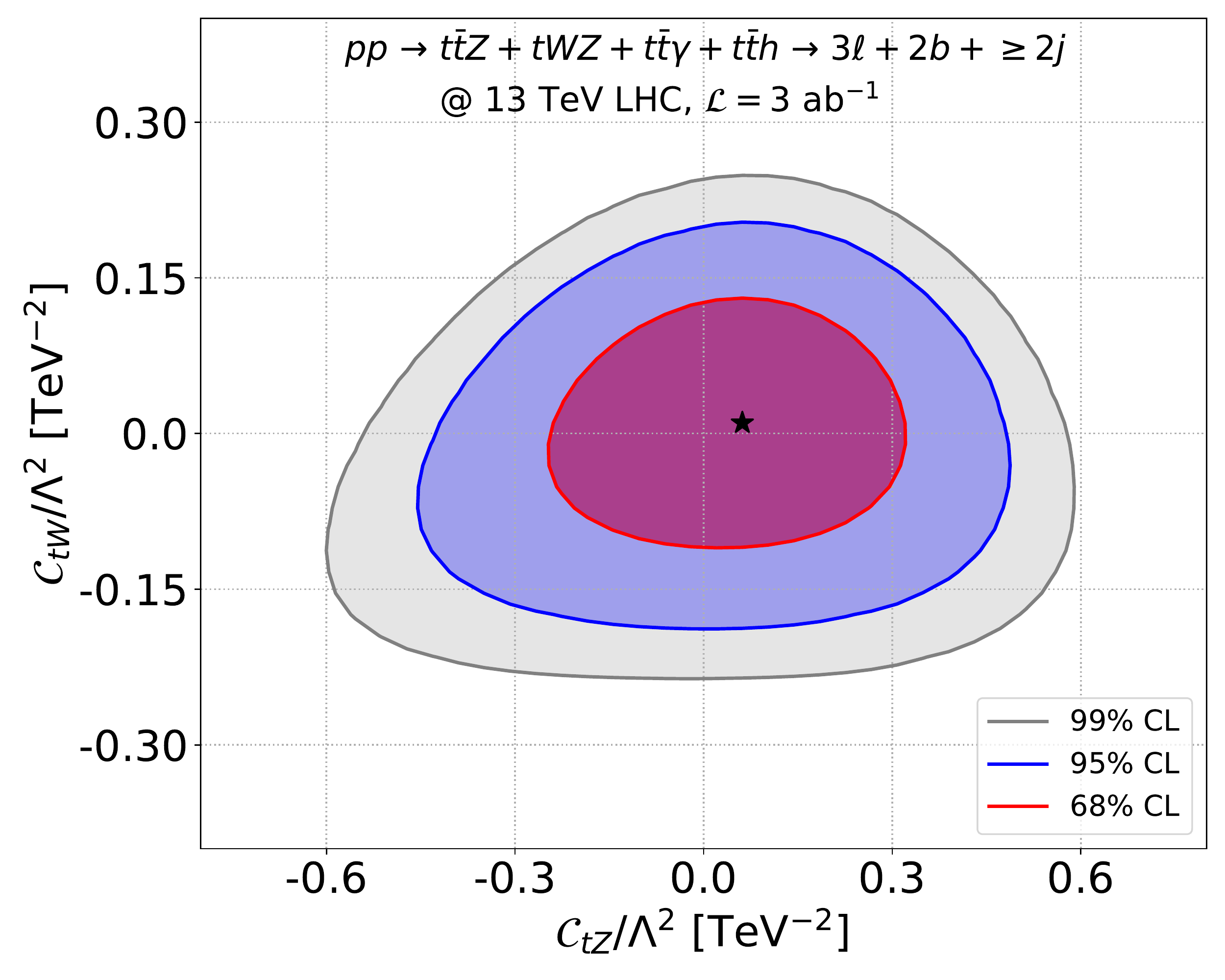}
    \caption{Projected sensitivity using MadMiner in the $\{\mathcal{C}_{tZ},\mathcal{C}_{tW}\}$ plane from searches in the $pp \to t\bar{t}Z + tWZ \to 3\ell + 2b\ + \geq 2j $ channel at the 13~TeV LHC with $\mathcal{L}=3~{\rm ab^{-1}}$.}
    \label{fig:ttz_madminer_projection}
\end{figure}

In Fig.~\ref{fig:ttz_madminer_projection}, we present the projection contours in the $\{\mathcal{C}_{tZ},\mathcal{C}_{tW}\}$ plane from searches in the $pp \to t\bar{t}Z + tWZ  \to 3\ell + 2b\ + \geq 2j$ channel at the HL-LHC using \texttt{MadMiner}. The estimated likelihood ratio is used to draw the 2d contour in Fig.~\ref{fig:ttz_madminer_projection} as a function of $\theta = \{\mathcal{C}_{tZ},\mathcal{C}_{tW}\}$. In order to set 1d limits along the direction of $\mathcal{C}_{tZ}$ or $\mathcal{C}_{tW}$, we profile the estimated event likelihood ratio over the other theory parameter. We present the 1d projection limits for $\mathcal{C}_{tZ}$ and $\mathcal{C}_{tW}$ in the central and right panels of Fig.~\ref{fig:ttz_limits}, respectively. The projected sensitivity for $\mathcal{C}_{tZ}$ reaches up to $-0.41\lesssim \mathcal{C}_{tZ} \lesssim 0.47$ at $95\%$ CL. We note that the \texttt{MadMiner} analysis showcases a marginal improvement over the projected limits from the DNN analysis~($-0.45\lesssim \mathcal{C}_{tZ} \lesssim 0.48$ at $2\sigma$) and roughly $(5-10)\%$ improvement over the cut-based results~($-0.49\lesssim \mathcal{C}_{tZ} \lesssim 0.51$ at $2\sigma$).

As discussed previously, the differential distributions in the $pp \to t\bar{t}Z + tWZ$ channel at the HL-LHC display a sizeable sensitivity to $\mathcal{C}_{tZ}$. Correspondingly, their inclusion in addition to the rate measurements help in boosting new physics sensitivity. The cut-and-count analysis leads to an improvement of roughly $1.5\%$ and $25\%$ in $\sigma_s^{NP}$ for $\mathcal{C}_{tZ} = 2.0$ and 0.5, respectively, compared to pure rate measurements. Typically, we would expect the machine learning techniques to be more efficient in unveiling BSM effects in correlated multi-dimensional feature space than conventional cut-and-count methods.
For example, the DNN  methodology leads to roughly $5\%$~($35\%$) improvement in the projected sensitivity over rate measurements for $\mathcal{C}_{tZ}=2.0~(0.5)$. On the contrary, the cut-and-count optimization and machine learning techniques lead to only $\lesssim 5\%$ enhancement in signal significance over rate-only measurements for the various $\mathcal{C}_{tW}$ benchmarks. This behavior is expected since the differential measurements in the $pp \to t\bar{t}Z+tWZ \to 3\ell + 2b\ + \geq 2j$ channel displays only minuscule sensitivity to $\mathcal{C}_{tW}$. Furthermore, in the $\mathcal{C}_{tW}$ scenario, all three analysis techniques considered in the present study lead to similar sensitivities, \texttt{MadMiner:} $-0.19 \lesssim \mathcal{C}_{tW} \lesssim 0.16$ at $95\%$ CL, DNN: $-0.19 \lesssim \mathcal{C}_{tW} \lesssim 0.15$ at $2\sigma$, and cut-based: $-0.18 \lesssim \mathcal{C}_{tW} \lesssim 0.18$ at $2\sigma$.

\subsection{$pp \to tZj + t\bar{t}Z + tWZ \to 3\ell + 1b + 1/2j$}
\label{sec:pp_tzj_intro}

\begin{figure*}[!ht]
    \centering
    \includegraphics[scale=0.23]{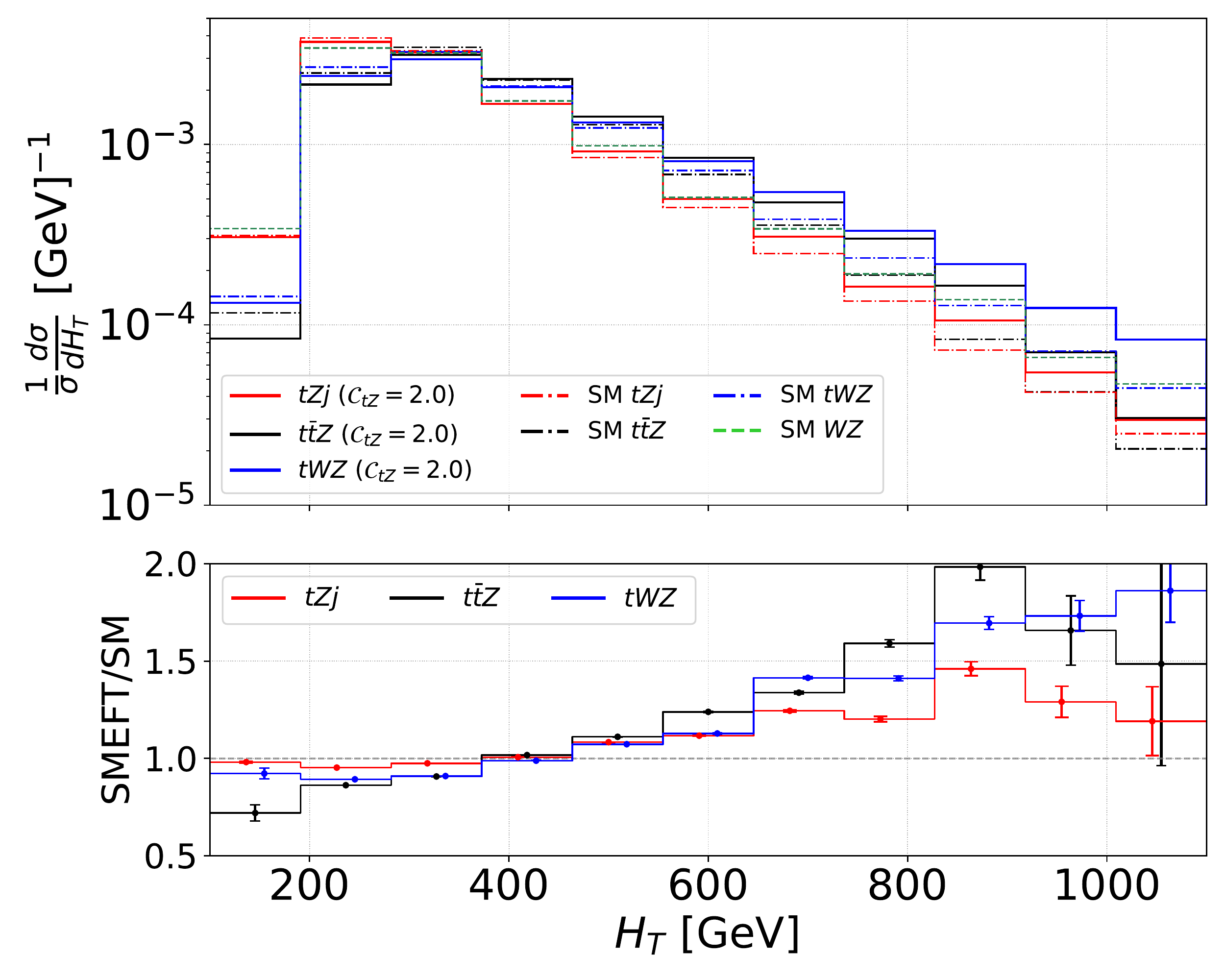}\includegraphics[scale=0.23]{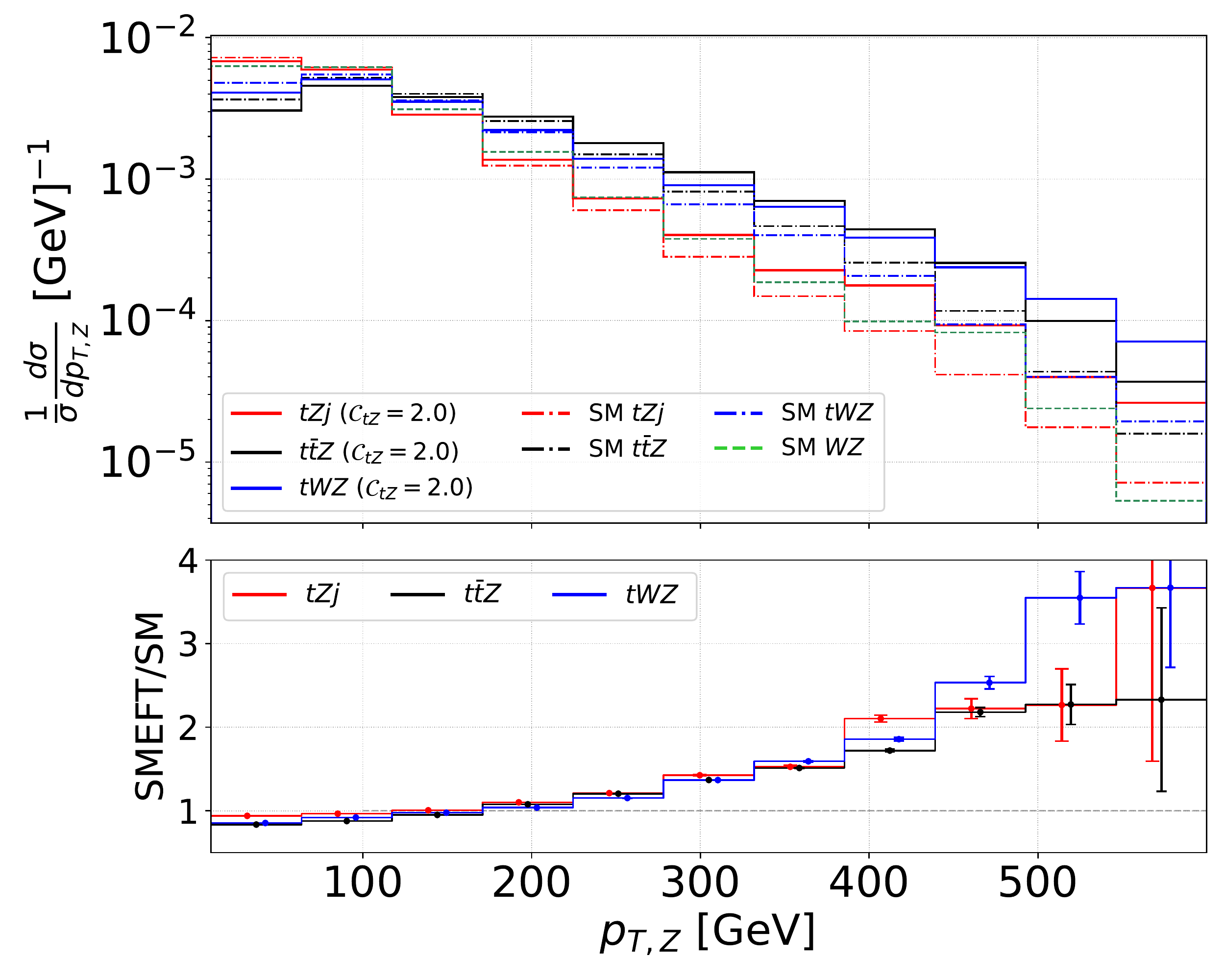}\includegraphics[scale=0.23]{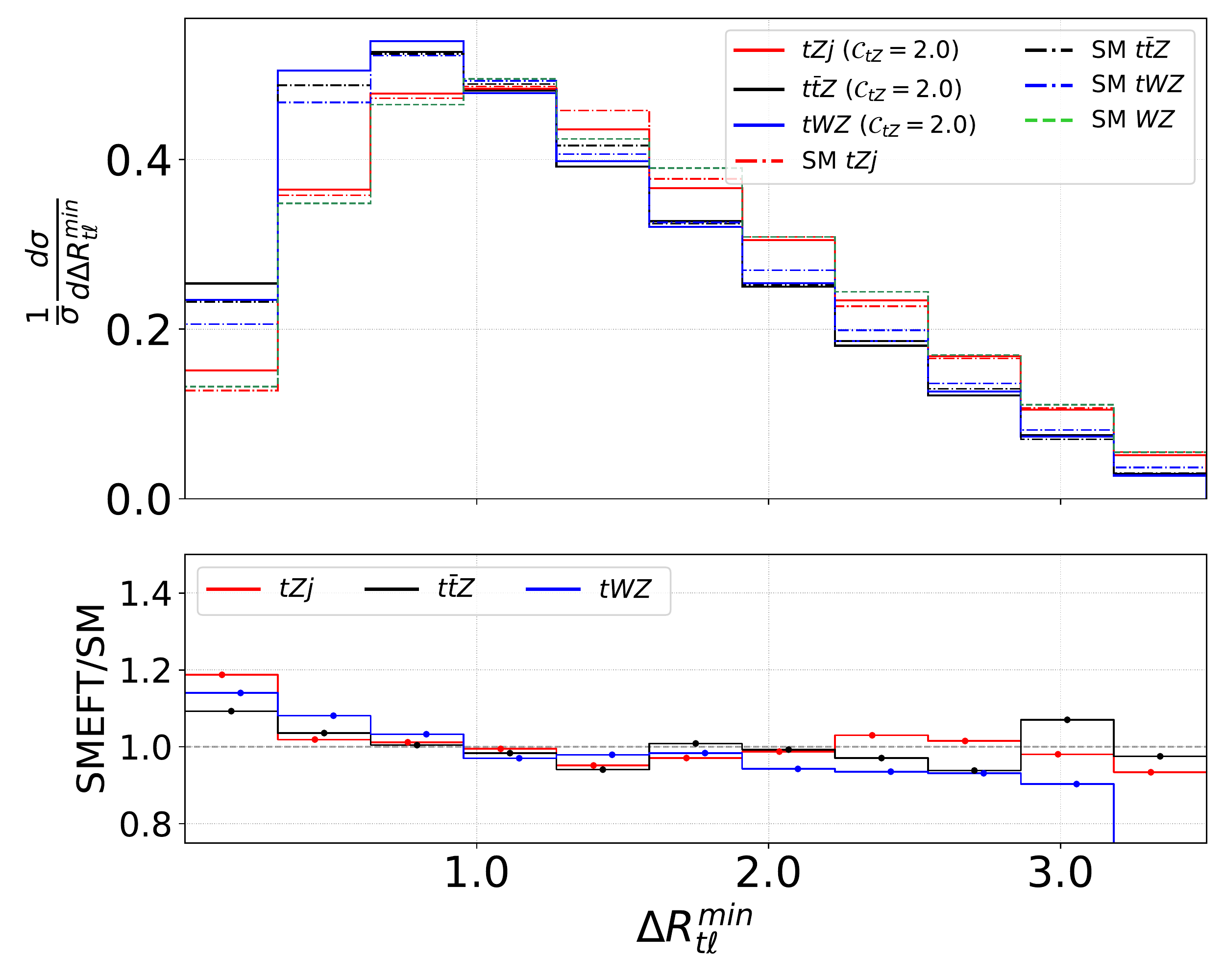}
    \caption{Distributions for the scalar sum of the transverse momenta of all visible final state objects $H_{T}$~(left), transverse momentum of the $Z$ boson $p_{T,Z}$~(center), and minimum $\Delta R$ separation between the top and lepton pair $\Delta R_{t\ell}^{\mathrm{min}}$~(right). The distributions correspond to SMEFT $tZj$~(red solid), $t\bar{t}Z$~(black solid) and $tWZ$~(blue solid) processes with $\mathcal{C}_{tZ}=2.0$. SM distributions for $tZj$~(red dashed), $t\bar{t}Z$~(black dashed), $tWZ$~(blue dashed) and $WZ+\mathrm{jets}$~(green dashed) are also shown. The distributions are presented at the detector level assuming $\sqrt{s}=13~$TeV at the LHC.}
    \label{fig:tzj_OtZ_distributions}
\end{figure*}

\begin{figure*}[!ht]
    \centering
    \includegraphics[scale=0.27]{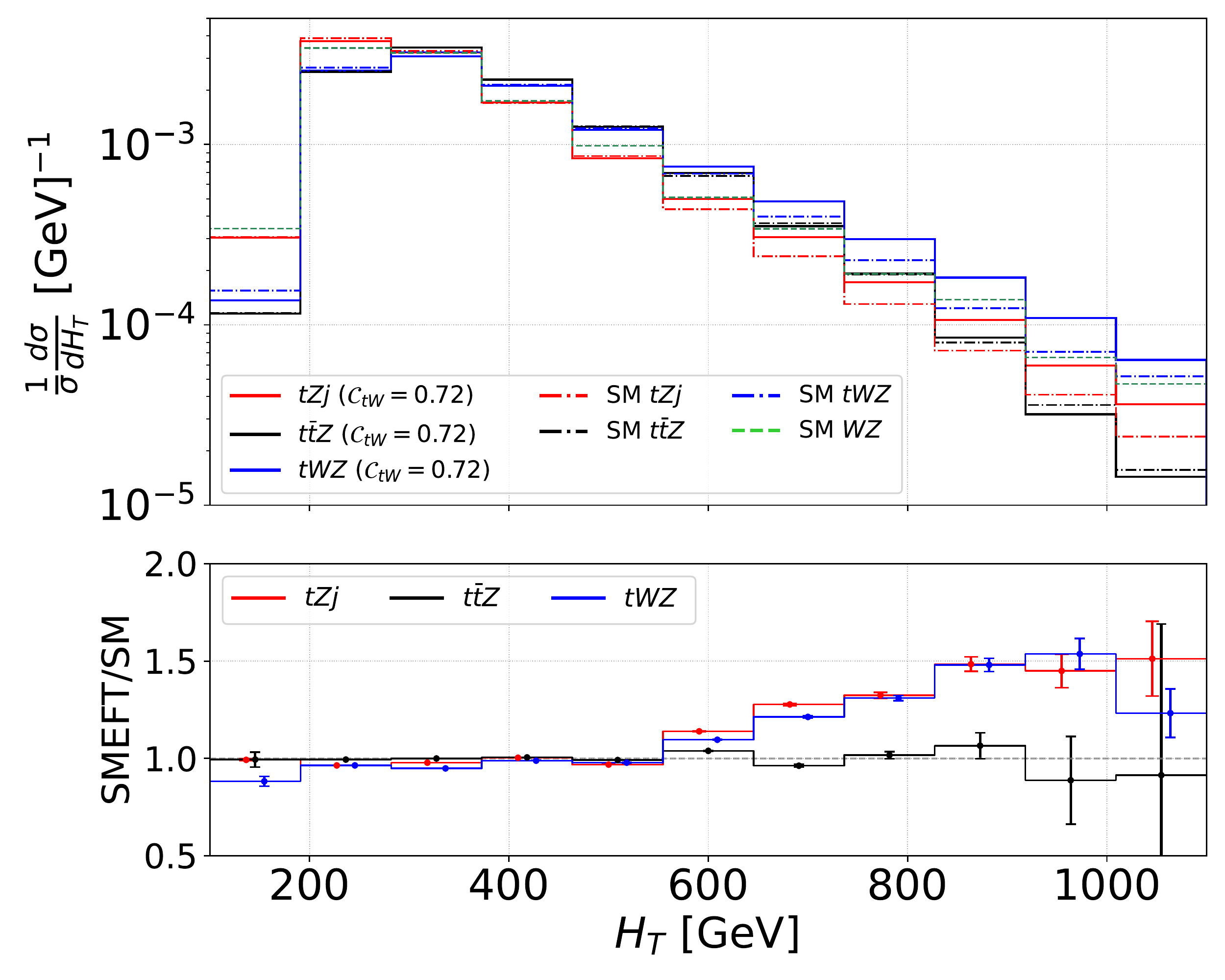}\hspace{2.0cm}\includegraphics[scale=0.27]{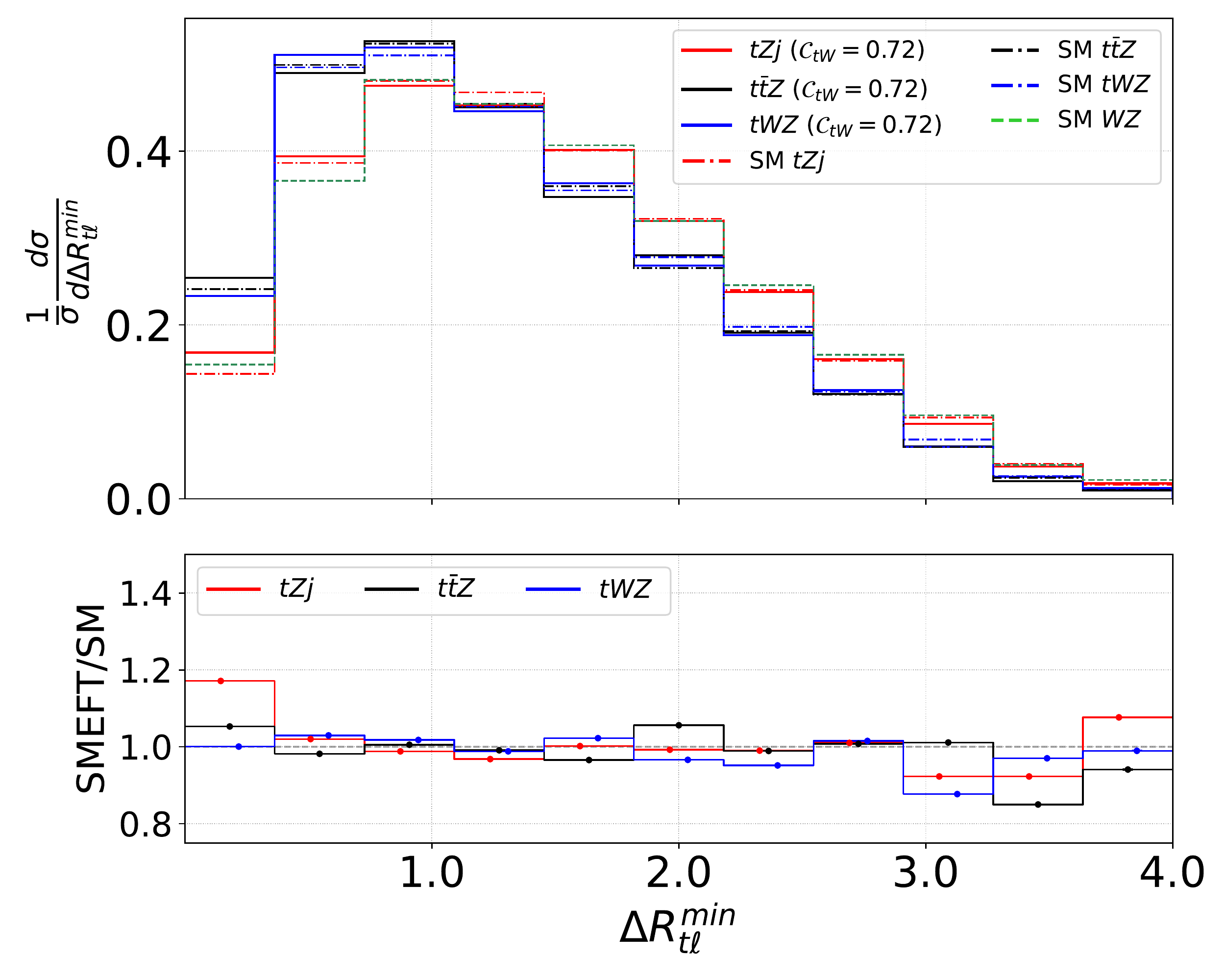}\\
    \includegraphics[scale=0.27]{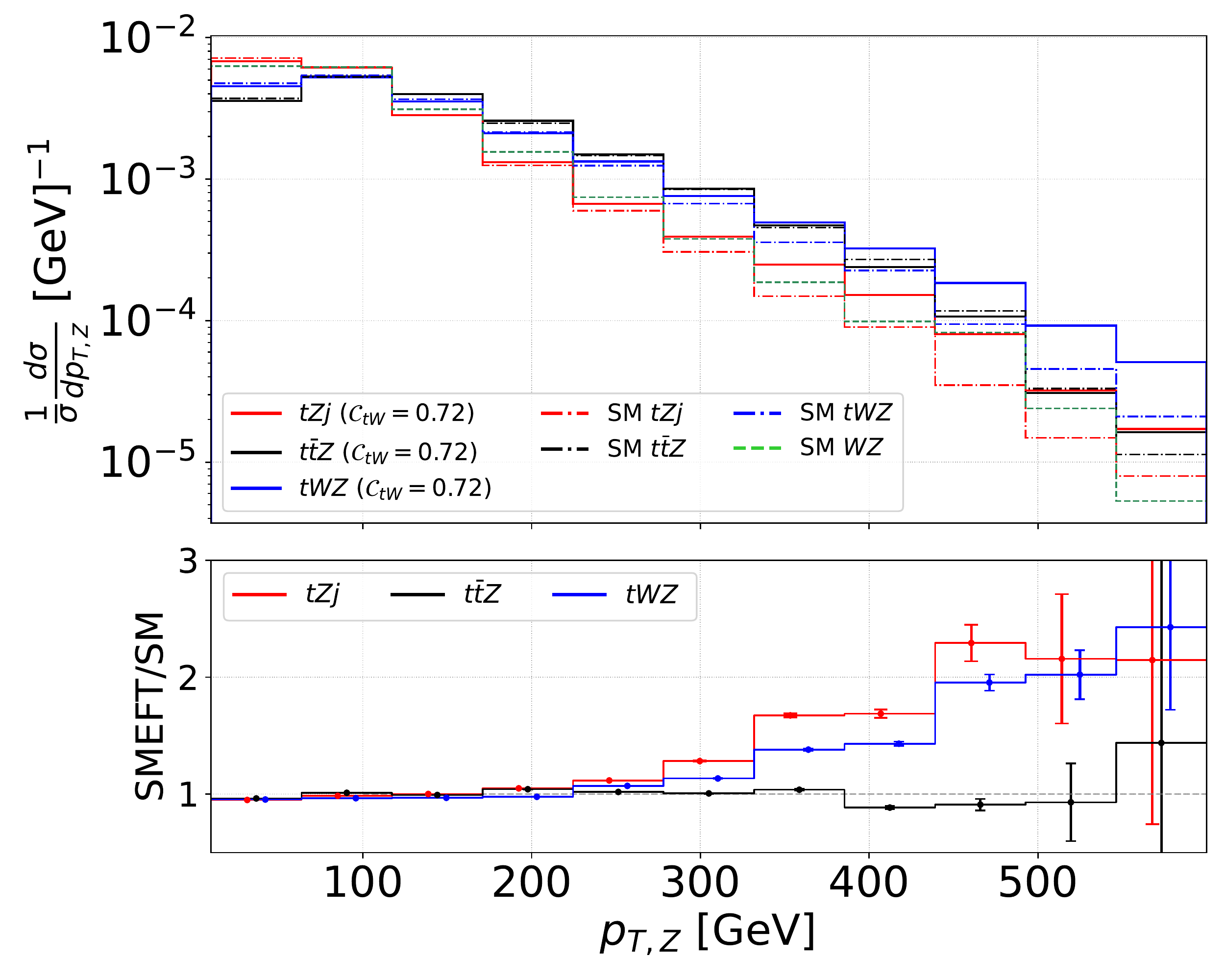}\hspace{2.0cm}\includegraphics[scale=0.27]{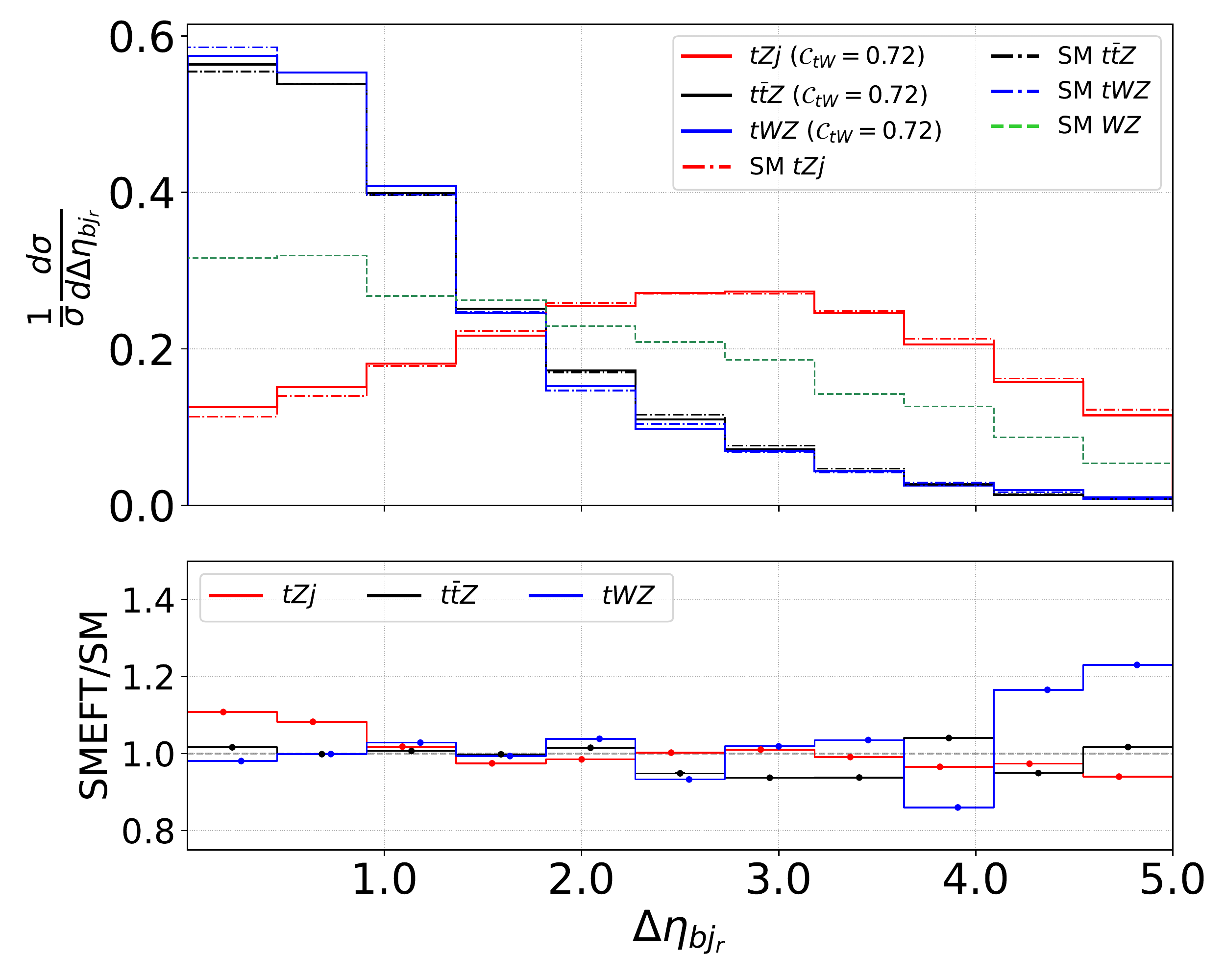}
    \caption{\textit{Top panels:} Distributions for the scalar sum of the transverse momenta of all visible final state objects $H_{T}$~(left), and minimum $\Delta R$ separation between the top and lepton pair $\Delta R_{t\ell}^{\mathrm{min}}$~(right). \textit{Bottom panels:} Distributions for transverse momentum of the $Z$ boson $p_{T,Z}$~(left), and difference between pseudorapidities of the $b$ tagged jet from top decay and the recoil jet $\Delta\eta_{bj_{r}}$~(right). The distributions correspond to SMEFT $tZj$~(red), $t\bar{t}Z$~(black solid) and $tWZ$~(blue solid) with $\mathcal{C}_{tW}=0.72$, SM $tZj$~(red dashed), $t\bar{t}Z$~(black dashed), $tWZ$~(blue dashed) and $WZ+\mathrm{jets}$~(green dashed). The results are shown at detector level for the LHC with $\sqrt{s}=13~$TeV.}
    \label{fig:tzj_OtW_distributions}
\end{figure*}

In this section we focus on the leptonic decay mode for $tZj$: $pp \to tZj \to (t \to \ell\nu b)(Z \to \ell\ell)j$. Other top electroweak processes, notably $t\bar{t}Z$ and $tWZ$ production, can also contribute to this final state. These processes would be affected by $\mathcal{O}_{tZ}$ and $\mathcal{O}_{tW}$, and we consider new physics modifications from SMEFT operators to them as part of our signal. The dominant background sources are SM $tZj$, $WZ +\ \mathrm{jets}$, $t\bar{t}Z$ and $tWZ$, while sub-dominant contributions can arise from $t\bar{t}\gamma$, $t\bar{t}h$ and $VVV~(V=W,Z)$. We ignore new physics modifications to $t\bar{t}\gamma$ and $t\bar{t}h$ since their relative production rates are considerably smaller compared to $tZj$ and $t\bar{t}Z$.

We select events containing exactly three isolated leptons, one $b$ tagged jet, and one or two light jets. The individual particles are required to pass the selection criteria in Eq.~(\ref{eqn:obj_selection_ttz}). Additionally, the leading~(sub-leading) lepton is required to have $p_{T} > 40$~GeV~(25~GeV).
We follow the strategy adopted in Sec.~\ref{sec:pp_ttz_intro} to reconstruct the $Z$ boson and identify the lepton $\ell_{W}$ associated with top decay. The unknown $\slashed{p}_{z}$ is computed by constraining the invariant mass of $\ell_{W}$ and the unobservable $\nu$ with the on-shell $W$ boson mass. Events which lead to zero solutions for $\slashed{p}_{z}$ are rejected. In cases with two solutions, we choose the solution that minimizes ${(m_{l_{t}\nu b} - m_{t})}^{2}$. The last missing piece of our event topology is the light jet $j_{r}$ that recoils against the $tZ$ system. We identify the leading jet that is not $b$ tagged as this jet.

We reconstruct various kinematic observables in the laboratory frame, and center of mass frames of the $W$, $t$ and $tZj_{r}$, to discriminate the new physics signal from background. The observables are listed below:
\be
\theta^{\star W}_{\alpha_W} \{\alpha_W = \ell_{W}, \nu\}, \theta^{\star t}_{\beta} \{\beta = \ell_{W},\nu,b\},\\
\theta^{\star tzj_{r}}_{\epsilon} \{\epsilon = \ell_{1},\ell_{2},\ell_{W},\nu,b,j_{r},t\},\\ 
p_{T,\zeta}, \eta_{\zeta}, \phi_{\zeta}, E_{\zeta}~\{\zeta = \epsilon,Z,tZ,tZj_{r}\}, \\
m_{k}~\{k = Z,t,tZ,tZj_{r}\}, m_{\ell_{1}\ell_2 \ell_{W}}, m_{jj}^{\mathrm{max}},\\
\Delta \phi_{\xi\rho}~\{\xi=\ell_{1},\ell_{2},Z; \rho=\ell_{W},b,j_{r}\}, \Delta \phi_{\ell_{W}j_{r}},\\ 
\Delta \phi_{\ell_{W} Z}^{tZj_{r}},\Delta \phi_{\ell\ell}^{\mathrm{max}}~\{\ell = \ell_{1},\ell_{2},\ell_{W}\}, \Delta R_{\ell b}^{\mathrm{min}}, \Delta R_{t\ell}^{\mathrm{min}}\\
m_{T,l_{W}},m_{T,tZ},p_{T, jj}^{\mathrm{max}},p_{T,jb}, p_{T,bj_{r}}, H_{T}.
\label{eqn:tzj_observables}
\ee
Here, $\alpha_{W}$ and $\beta$ denote the decay products of the $W$ and top respectively, $\ell_{1}$ and $\ell_{2}$ form the SFOS lepton pair that constitutes the $Z$ boson, $\theta^{\star i}_{k}$ is the angle between particle $k$ and the beam direction in the rest frame of particle $i$, $m_{\ell_{1}\ell_{2}\ell_{W}}$ is the invariant mass of the three leptons in the final state, and $\Delta \phi_{mn}^{tZj_{r}}$ represents the azimuthal angle difference between $m$ and $n$ in the rest frame of $tZj_{r}$. The other variables in Eq.~\ref{eqn:tzj_observables} have their usual meanings.

\begin{table}[!htb]
    \centering\scalebox{0.7}{
    \begin{tabular}{|c||c|c|c|} \hline
          & \multicolumn{3}{|c|}{$\mathcal{C}_{tZ}=2.0$}\\\cline{2-4} 
         Optimized & $H_{T}~$ $>$ & $p_{T,Z}~$ $>$ & $\Delta R_{t\ell}^{\mathrm{min}}$ $<$ \\ 
         cuts & - & 200~GeV & 3.0 \\ \hline 
         SMEFT $tZj$ & 3362 & 393  & 385 \\
         SMEFT $t\bar{t}Z$ & 3955 & 1204  & 1192 \\
         SMEFT $tWZ$ & 446 & 113  & 111 \\
         $tZj$ & 3190 & 278  & 270\\
         $t\bar{t}Z$ & 2924 & 664 & 654 \\
         $tWZ$ & 383 & 70.9 &  69.5\\
         $WZ$ & 6482  & 731 & 707 \\
         $t\bar{t}\gamma$ & 21.2 & 3.0 & 3.0\\ \hline
        Significance & 11.10 & 16.68 & 16.83  \\\hline \hline
         &  \multicolumn{3}{|c|}{$\mathcal{C}_{tZ}=1.5$}\\ \cline{2-4} 
         Optimized & $H_{T}~$ $>$ & $p_{T,Z}~$ $>$ & $\Delta R_{t\ell}^{\mathrm{min}}$ $<$ \\ 
         cuts & - & 250~GeV & 2.5 \\ \hline 
         SMEFT $tZj$ & 3312 & 197 & 183 \\
         SMEFT $t\bar{t}Z$ & 3515 & 587 & 562  \\
         SMEFT $tWZ$ & 415 & 59.9 & 57.0 \\
         $tZj$ & 3190 & 145 & 133\\
         $t\bar{t}Z$ & 2924 & 375  & 356 \\
         $tWZ$ & 383 & 40.4  & 38.2\\
         $WZ$ & 6482 & 380 & 344\\
         $t\bar{t}\gamma$ & 21.2 & 1.6 & 1.5\\ \hline
        Significance & 6.53 & 9.24 & 9.30 \\\hline \hline
         &  \multicolumn{3}{|c|}{$\mathcal{C}_{tZ}=1.0$}\\ \cline{2-4} 
         Optimized & $H_{T}~$ $>$ & $p_{T,Z}~$ $>$ & $\Delta R_{t\ell}^{\mathrm{min}}$ $<$ \\ 
         cuts & 450~GeV & 250~GeV & 2.75  \\ \hline 
         SMEFT $tZj$ & 613 & 165 &  158 \\
         SMEFT $t\bar{t}Z$ & 921 & 455 & 436 \\
         SMEFT $tWZ$ & 118 & 45.3 & 43.3 \\
         $tZj$ & 584 & 139 & 132\\
         $t\bar{t}Z$ & 783 & 348 & 331 \\
         $tWZ$ & 108 & 38.4 & 36.4 \\
         $WZ$ & 1497 & 367 & 337 \\
         $t\bar{t}\gamma$ & 4.0 & 1.5 & 1.4\\ \hline
        Significance & 3.24 & 4.68 & 5.00 \\\hline \hline
          & \multicolumn{3}{|c|}{$\mathcal{C}_{tZ}=0.5$}\\ \cline{2-4} 
         Optimized & $H_{T}~$ $>$ & $p_{T,Z}~$ $>$ & $\Delta R_{t\ell}^{\mathrm{min}}$ $<$ \\ 
         cuts & 450~GeV & 250~GeV & 0.75 \\ \hline 
         SMEFT $tZj$ & 578 & 138 & 55 \\
         SMEFT $t\bar{t}Z$ & 815 & 380 & 167  \\
         SMEFT $tWZ$ & 109 & 41.2 & 17.6 \\
         $tZj$ & 584 & 139 & 49.2 \\
         $t\bar{t}Z$ & 783 & 348 & 146 \\
         $tWZ$ & 108 & 38.5 & 16.8\\
         $WZ$ & 1497 & 367 & 120\\
         $t\bar{t}\gamma$ & 4.0 & 1.5 & 0.6 \\ \hline
        Significance & 0.49 & 1.06 & 1.51  \\\hline \hline
    \end{tabular}
    \begin{tabular}{|c|c|c|} \hline
         \multicolumn{3}{|c|}{$\mathcal{C}_{tZ}=-2.0$}\\\hline 
          $H_{T}~$ $>$ & $p_{T,Z}~$ $>$ & $\Delta R_{t\ell}^{\mathrm{min}}$ $<$ \\ 
         350~GeV & 250~GeV & 2.0 \\ \hline 
          1410 & 250 & 220 \\
          2329 & 771 & 693 \\
          256 & 76.9 & 68.6 \\
          1205 & 145 & 119\\
          1519 & 375 & 330 \\
          197 & 40.4 & 34.9\\
          2776 & 380 & 300 \\
          8.9 & 1.6 & 1.4\\ \hline
          14.22 & 17.51 & 17.76 \\\hline \hline
         \multicolumn{3}{|c|}{$\mathcal{C}_{tZ}=-1.5$}\\\hline 
         $H_{T}~$ $>$ & $p_{T,Z}~$ $>$ & $\Delta R_{t\ell}^{\mathrm{min}}$ $<$ \\ 
         350~GeV & 200~GeV & 2.5 \\ \hline 
         1324 & 351 & 325 \\
         1997 & 973 & 927 \\
         231 & 95.6 & 90.2 \\
         1205 & 273 & 249\\
         1519 & 653 & 616 \\
         197 &  69.8 & 65.5\\
         2776 & 727 & 655 \\
         8.9 & 3.0 & 2.9 \\ \hline
         8.35 & 10.20 & 10.33 \\\hline \hline
         \multicolumn{3}{|c|}{$\mathcal{C}_{tZ}=-1.0$}\\\hline 
         $H_{T}~$ $>$ & $p_{T,Z}~$ $>$ & $\Delta \eta_{bj_{r}}$ $<$ \\ 
         350~GeV & 200~GeV & 4.75 \\ \hline 
         1260 & 313 & 289 \\
         1696 & 774 & 771  \\
         207 & 80.6 & 80.1 \\
         1205 & 274 & 251 \\
         1519 & 653 & 648 \\
         197 & 69.8 & 69.2 \\
         2776 & 727 & 704\\
         8.9 & 3.0 & 2.9 \\ \hline
         3.20 & 4.11 & 4.20 \\\hline \hline
         \multicolumn{3}{|c|}{$\mathcal{C}_{tZ}=-0.5$}\\\hline 
         $H_{T}~$ $>$ & $p_{T,Z}~$ $>$ & $\Delta R_{t\ell}^{\mathrm{min}}$ $<$ \\ 
         450~GeV & 250~GeV & 0.75 \\ \hline 
         597 & 142 & 55 \\
         801 & 370 & 154 \\
         112 & 40.2 & 16.3 \\
         584 & 139 & 48.6\\
         783 & 348 & 136 \\
         108 & 38.4 & 16.0 \\
         1497 & 367 & 116 \\
         4.0 & 1.5 & 0.6 \\ \hline
        0.64 & 0.90 & 1.38 \\\hline \hline
    \end{tabular}}
\caption{Optimized selection cuts on $H_{T}$, $p_{T,Z}$ and $\Delta R_{t\ell}^{\mathrm{min}}$, applied successively, to maximize the signal significance $\sigma_{s}^{NP}$ for SMEFT signal benchmarks with $\{\mathcal{C}_{tZ} = \pm 2.0, \pm 1.5, \pm 1.0, \pm 0.5\}$ through searches in $pp \to tZj+ t\bar{t}Z + tWZ \to 3\ell + 1b + 1/2j $ channel at $\sqrt{s}=13~$TeV LHC with $\mathcal{L}=3~{\rm ab^{-1}}$. The optimized cuts, signal and background yields, and $\sigma_{s}^{NP}$ values are shown. No cuts are applied on $H_{T}$ in the signal regions optimized for $\mathcal{C}_{tZ}=2.0$ and 1.5.}
\label{tab:tzj_OtZ_cut_flow}
\end{table}

\begin{table}[!htb]
    \centering\scalebox{0.85}{
    \begin{tabular}{|c||c|c|} \hline
          & \multicolumn{2}{|c|}{$\mathcal{C}_{tW}=0.72$}\\\cline{2-3}
         Optimized & $H_{T}~$ $>$ & $\Delta R_{t\ell}^{\mathrm{min}}$ $<$ \\ 
         cuts & - & 3.0 \\ \hline 
         SMEFT $tZj$ & 3743 & 3578 \\
         SMEFT $t\bar{t}Z$ & 3697 & 3596 \\
         SMEFT $tWZ$ & 436 & 423 \\
         $tZj$ & 3185 & 3030 \\
         $t\bar{t}Z$ & 2935 & 2848\\
         $tWZ$ & 377 &  364 \\
         $WZ$ & 6483 & 6132 \\
         $t\bar{t}\gamma$ & 21.2 & 20.4 \\ \hline
        Significance & 12.1 & 12.2  \\\hline \hline
        &   \multicolumn{2}{|c|}{$\mathcal{C}_{tW}=0.48$}\\\cline{2-3} 
         Optimized & $H_{T}~$ $>$ &  $\Delta R_{t\ell}^{\mathrm{min}}$ $<$ \\ 
         cuts & 200~GeV & 3.75 \\ \hline 
         SMEFT $tZj$ & 3352 & 3331 \\
         SMEFT $t\bar{t}Z$ & 3414 & 3400 \\
         SMEFT $tWZ$ & 393 & 391 \\
         $tZj$ & 3038 & 3016 \\
         $t\bar{t}Z$ & 2881 & 2867 \\
         $tWZ$ & 368 & 366 \\
         $WZ$ & 6166 & 6096 \\
         $t\bar{t}\gamma$ & 20.1 & 20.0 \\ \hline
        Significance & 7.81 & 7.85 \\\hline \hline
          & \multicolumn{2}{|c|}{$\mathcal{C}_{tW}=0.24$}\\\cline{2-3} 
         Optimized & $H_{T}~>$ & $\Delta R_{t\ell}^{\mathrm{min}}$ $<$ \\ 
         cuts & 150~GeV & 3.25 \\ \hline 
         SMEFT $tZj$ & 3377 & 3292 \\
         SMEFT $t\bar{t}Z$ & 3212 & 3163 \\
         SMEFT $tWZ$ & 387 & 381\\
         $tZj$ & 3184 & 3099\\
         $t\bar{t}Z$ & 2935 & 2888\\
         $tWZ$ & 377 & 370 \\
         $WZ$ & 6479 & 6278 \\
         $t\bar{t}\gamma$ & 21.2 &  20.8 \\ \hline
        Significance & 4.21 & 4.24  \\\hline \hline
    \end{tabular}
    \begin{tabular}{|c|c|} \hline
         \multicolumn{2}{|c|}{$\mathcal{C}_{tW}=-0.72$}\\\hline 
          $H_{T}~$ & $\Delta \eta_{bj_r}$ $<$ \\ 
          150~GeV &  4.0 \\ \hline 
           2972 &  2322 \\
           2330 &  2297 \\
           397 &  389 \\
           3184 &  2515 \\
           2935 &  2889 \\
           377 &  371\\
           6479 &  5825 \\
           21.2 &  20.9\\ \hline
           6.98 &  7.10 \\\hline \hline
         \multicolumn{2}{|c|}{$\mathcal{C}_{tW}=-0.48$}\\\hline 
          $H_{T}~>$ & $\Delta \eta_{bj_r}$ $<$ \\ 
           150~GeV & 5.0 \\ \hline 
           3002 & 2768 \\
           2541 & 2533 \\
           388 & 386 \\
           3184 & 2950 \\
           2935 & 2925 \\
           378 & 376 \\
           6479 & 6284 \\
           21.2 & 21.1 \\ \hline
           4.96 & 5.03 \\\hline \hline
         \multicolumn{2}{|c|}{$\mathcal{C}_{tW}=-0.24$}\\\hline 
         $H_{T}~$ $>$ & $\Delta \eta_{bj_{r}}$ $<$ \\ 
         150~GeV & 2.5 \\ \hline 
         3105 & 1364 \\
         2706 & 2410 \\
         378 & 338 \\
         3184 & 1406 \\
         2935 & 2627 \\
         377 & 339 \\
         6479 & 4314 \\
         21.2 & 18.9 \\ \hline
         2.69 & 2.78 \\\hline \hline           
    \end{tabular}}
\caption{Optimized selection cuts on $H_{T}$, and $\Delta R_{t\ell}^{\mathrm{min}}$ or $\Delta \eta_{bj_{r}}$, applied successively, to maximize the signal significance $\sigma_{s}$ for SMEFT signal benchmarks with $\{\mathcal{C}_{tW} = \pm 0.72, \pm 0.48, \pm 0.24, \pm 0.12\}$ through searches in $pp \to tZj+ t\bar{t}Z + tWZ \to 3\ell + 1b + 1/2j $ channel at $\sqrt{s}=13~$TeV LHC with $\mathcal{L}=3~{\rm ab^{-1}}$. The optimized cuts, signal and background yields, and $\sigma_{s}$ values are shown. No cuts are applied on $H_{T}$ in the signal region optimized for $\mathcal{C}_{tW} = 0.72$.}
\label{tab:tzj_OtW_cut_flow}
\end{table}

\begin{figure*}[!t]
    \centering
    \includegraphics[scale=0.24]{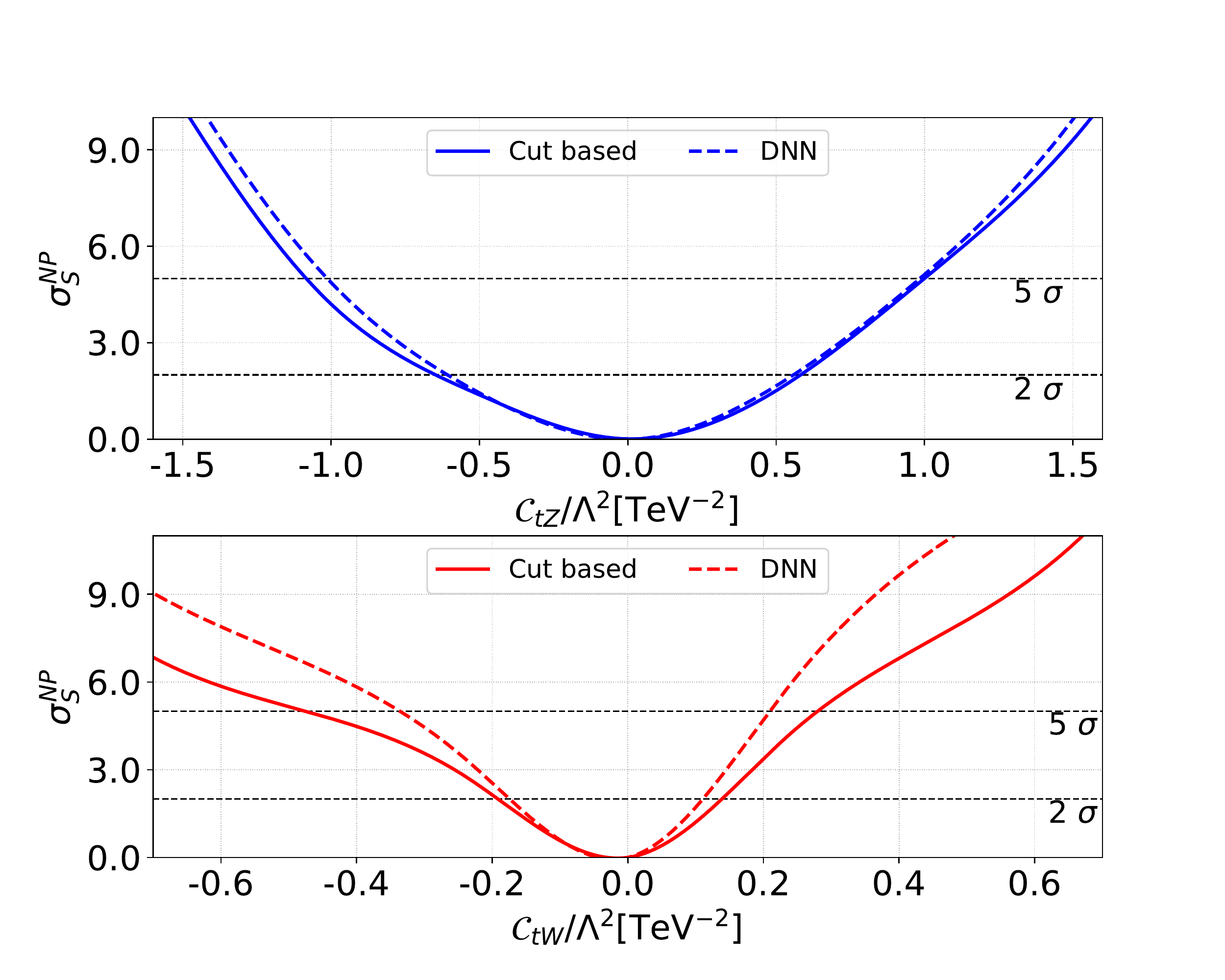}\includegraphics[scale=0.24]{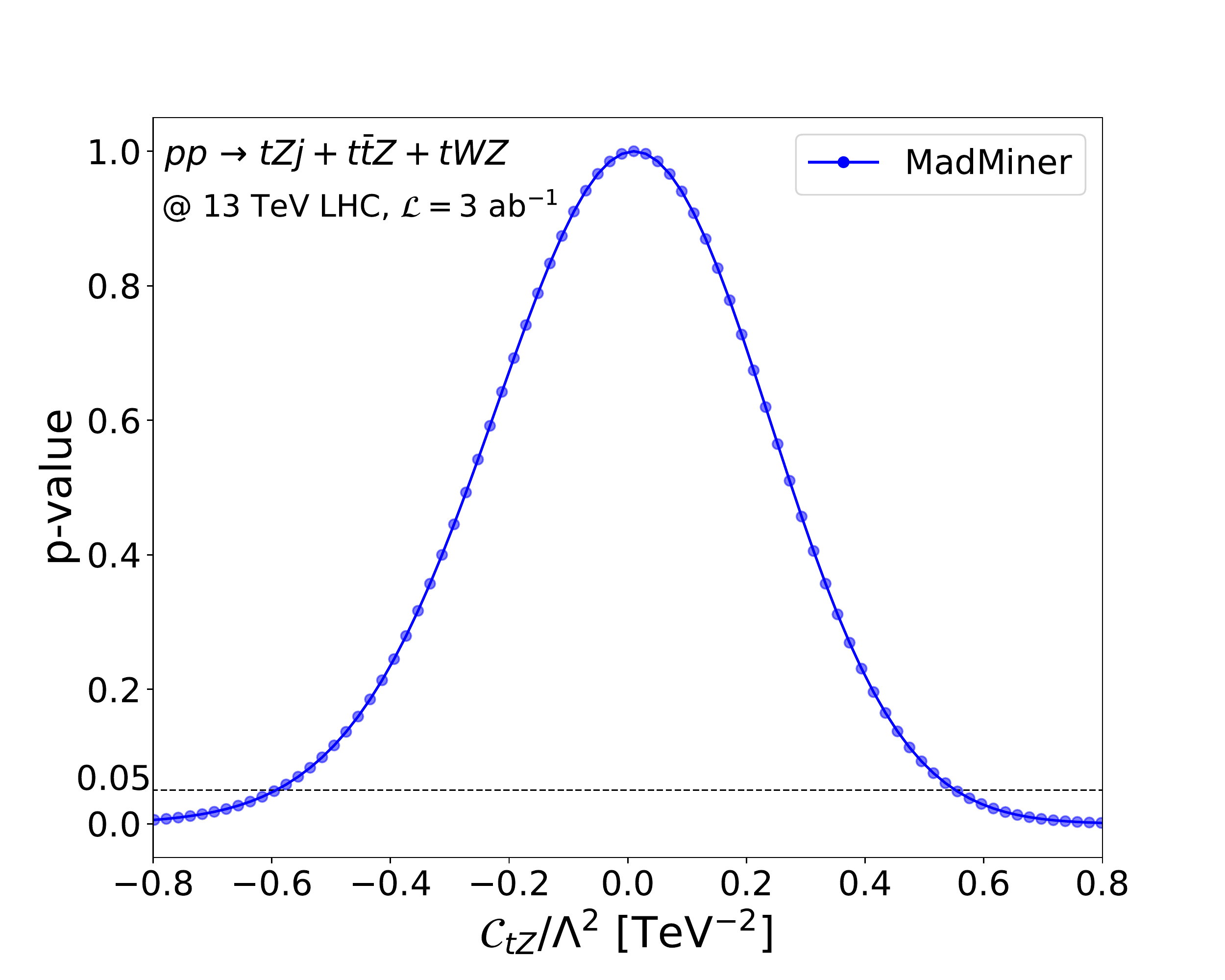}\includegraphics[scale=0.24]{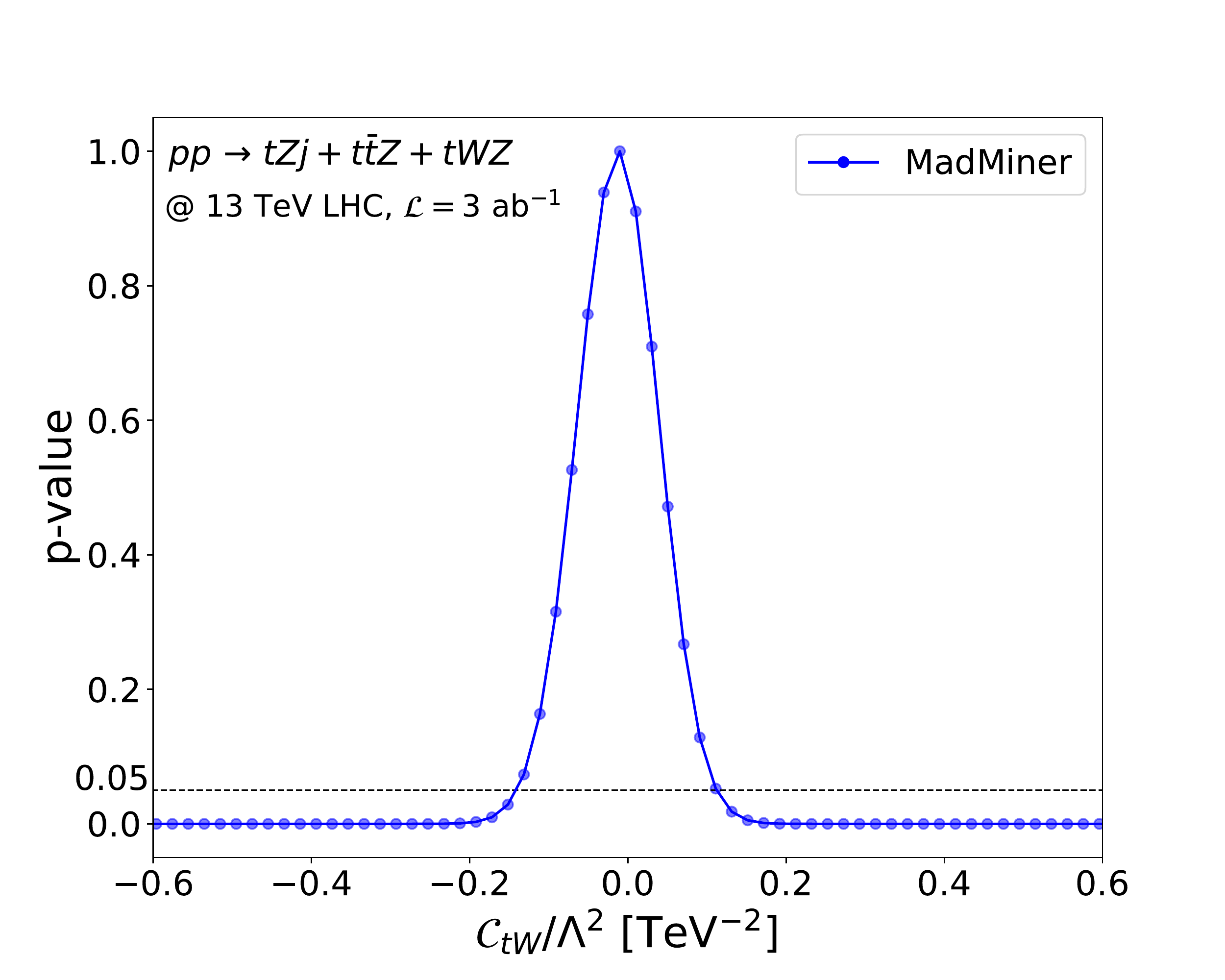}
    \caption{\textit{Left panel:} Projected sensitivity for $\mathcal{O}_{tZ}$~(top panel) and $\mathcal{O}_{tW}$~(bottom panel) from searches in $pp \to tZj + t\bar{t}Z + tWZ \to 3\ell + 1b +  1/2j$~(red) channels at the HL-LHC. The solid and dashed lines represent the projections from cut-based and DNN analysis, respectively. \textit{Central panel:} Projected sensitivity for $\mathcal{O}_{tZ}$ via searches in $pp \to tZj + t\bar{t}Z + tWZ \to 3\ell + 1b + 1/2j$ channel, from MadMiner. The vertical axis shows the p-values of the estimated negative log-likelihood ratio. The black-dashed line denotes a p-value of 0.05. \textit{Right panel:} Projected sensitivity for $\mathcal{O}_{tW}$ through searches in $pp \to tZj + t\bar{t}Z + tWZ \to 3\ell + 1b + 1/2j$ channel, from MadMiner. The color code is similar to that of the central panel. Results are presented for $\sqrt{s}=13~$TeV LHC with $\mathcal{L}=3~{\rm ab^{-1}}$.}
    \label{fig:tzj_limits}
\end{figure*}

We first perform a cut-based analysis to estimate the projected sensitivity for $\mathcal{C}_{tZ}$ by optimizing the selection cuts on $H_{T}$, $p_{T,Z}$ and $\Delta R_{tl}^{\mathrm{min}}$. Several other observable subsets from Eq.~(\ref{eqn:tzj_observables}) are also considered for the cut-based analysis; however, the combination of $\{H_{T},~p_{T,Z},~\Delta R_{tl}^{\mathrm{min}}\}$ leads to the best sensitivity. The optimization is performed for the 8 signal benchmarks considered in Sec.~\ref{sec:pp_ttz_intro}. We illustrate the distributions for $H_{T}$, $p_{T,Z}$ and $\Delta R_{tl}^{\mathrm{min}}$ at the detector level in Fig.~\ref{fig:tzj_OtZ_distributions}. The red, black and blue solid lines represent SMEFT $tZj$, $t\bar{t}Z$ and $tWZ$ events, respectively, for $\mathcal{C}_{tZ} = 2.0$, while the dashed lines represent the respective SM processes. The distributions for the $WZ+\mathrm{jets}$ background are presented as green dashed lines. In the bottom panel of the respective figures, the ratio SMEFT/SM is displayed. We observe that this ratio increases in the tails of the $H_{T}$ and $p_{T,Z}$ distributions for $tZj$, $t\bar{t}Z$, as well as for $tWZ$.
Based on the distributions for $H_{T}$ and $p_{T,Z}$ in Fig.~\ref{fig:tzj_OtZ_distributions}, the significance is likely to be enhanced in the boosted $Z$ regime. In addition, since the top quark recoils against the $Z$ boson, a boosted $Z$ would also imply a boosted top and $j_{r}$ system. Consequently, since the azimuthal angle difference between the decay products of the top is inversely correlated with its boost, going to low $\Delta R_{t\ell}^{\mathrm{min}}$ would also be effective in discriminating the SMEFT signal; generally, we expect that the final state lepton with the smallest $\Delta R$ separation from the top would be $\ell_{W}$. This effect leads to the ratio SMEFT/SM becoming greater than 1 in Fig.~\ref{fig:tzj_OtZ_distributions} for $\Delta R_{t\ell}^{\mathrm{min}} \lesssim 0.7$. We further note that for $p_{T,Z} \gtrsim 500~$GeV, the SMEFT contributions are larger than SM by at least a factor of 2. However, this high $p_{T}$ region is also marred by relatively larger statistical fluctuations as illustrated by the error bars in the bottom panel of the respective figure. In Table~\ref{tab:tzj_OtZ_cut_flow}, we present optimized cuts on $H_{T}$, $p_{T,Z}$, and $\Delta R_{tl}^{\mathrm{min}}$ that maximize $\sigma_{S}^{NP}$ in the present search channel for $\mathcal{C}_{tZ} = \pm 2.0, \pm 1.5, \pm 1.0,$ and $\pm 0.5$. As discussed previously, the optimized signal regions feature a strong lower cut on $H_{T}$ and/or $p_{T,Z}$, as well as an upper cut on $\Delta R_{t\ell}^{\mathrm{min}}$. We observe that $\sigma_{S}^{NP} = 1.51$~(1.38) for $\mathcal{C}_{tZ} = 0.5$~(-0.5), improving to 9.30~(10.33) for $\mathcal{C}_{tZ} = 1.5$~(-1.5). The variation of $\sigma_{S}^{NP}$ with $\mathcal{C}_{tZ}$ is presented in Fig.~\ref{fig:tzj_limits} as the blue solid line. We observe that the projected  $2\sigma$ sensitivity for $\mathcal{O}_{tZ}$ from searches in the $pp \to tZj + t\bar{t}Z + tWZ \to 3\ell + 1b + 1/2j$ channel at the HL-LHC reaches up to $-0.65 \lesssim \mathcal{C}_{tZ} \lesssim 0.58$.

A similar strategy is followed to estimate the sensitivity for $\mathcal{O}_{tW}$. Here again, we consider 6 signal benchmarks similar to that in Sec.~\ref{sec:pp_ttz_intro}. The cut-based optimization is performed for several subsets of observables from Eq.~(\ref{eqn:tzj_observables}). The strongest sensitivity is obtained for the subset $\{H_{T}, \Delta R_{t\ell}^{\mathrm{min}}, \Delta \eta_{bj_{r}}\}$. Comparable sensitivity is observed for the subset $\{p_{T,Z}, \Delta R_{t\ell}^{\mathrm{min}}, \Delta \eta_{bj_{r}}\}$.
We illustrate their distributions at the detector-level for SMEFT $tZj, t\bar{t}Z$ and $tWZ$ with $\mathcal{C}_{tW}=0.72$, and their SM counterparts, in Fig.~\ref{fig:tzj_OtW_distributions}. The color codes are adopted from Fig.~\ref{fig:tzj_OtZ_distributions}. As before, the SMEFT/SM ratio remains $\sim 1$ throughout for $t\bar{t}Z$. On the other hand, both $tZj$ and $tWZ$ have their kinematics affected by the new physics, with SMEFT/SM $\gtrsim 1$ in the tails of the $H_{T}$ and $p_{T,Z}$ distributions, as well as at smaller values of $\Delta R_{t\ell}^{min}$. Simultaneously, the tail of the $\Delta \eta_{bj_{r}}$ distribution for $WZ+\mathrm{jets}$ falls slowly when compared to that for $t\bar{t}Z$, suggesting that the rapidity gap between the $b$ jet and the recoiling light flavor jet can be used to reduce the $WZ+\mathrm{jets}$ background.
We optimize the selection cuts on our selected observables in order to maximize the signal significance. The cut flows for our final selections are shown in Table~\ref{tab:tzj_OtW_cut_flow}. We note that after adjusting cuts on the other observables, notably $H_T$, the $Z$ transverse momentum does not provide any additional sensitivity. Thus, we omit $p_{T,Z}$ from Table~\ref{tab:tzj_OtW_cut_flow} and display the sensitivity after cuts on $H_T$, $\Delta R^{\mathrm{min}}_{t\ell}$ and $\Delta \eta_{b j_r}$. As for the $3\ell + 2b\ + \geq 2j$ channel in the previous subsection, kinematic cuts alone provide limited increase in sensitivity to $\mathcal{O}_{tW}$.

Having performed the cut-based optimization, we next turn our attention to a multivariate DNN analysis. To begin, we follow a strategy similar to that in Sec.~\ref{sec:pp_ttz_intro} with similar training hyperparameters. The training dataset is constituted of pure EFT $tZj$ and $t\bar{t}Z$ events, as well as SM $tZj$, $t\bar{t}Z$, and $WZ+\mathrm{jets}$ events. Pure EFT $tWZ$ events are not included in the training dataset due to sub-dominant rates. The pure EFT events are assigned a score of 1 while SM events are given a score of 0. We train the DNN as a classification model. The test dataset includes SMEFT $tZj$, $t\bar{t}Z$, $tWZ$, and SM $tZj$, $t\bar{t}Z$, $tWZ$, $WZ+\mathrm{jets}$ and $t\bar{t}\gamma$.
As before, in the training dataset new physics effects are included only at the production level, and not in decay. However, new physics modifications are included in both production and top decay in the test dataset. The goal of the DNN model is to improve the difference between SMEFT events and their respective SM counterparts. The projected sensitivity obtained through this strategy is more or less comparable to that from cut-based optimization in the same channel~(Tables~\ref{tab:tzj_OtZ_cut_flow} and~\ref{tab:tzj_OtW_cut_flow}). The lack of improvement in the performance of the DNN when compared to the cut-based analysis could be partly ascribed to the imperfections in the training dataset as discussed in Sec.~\ref{sec:pp_ttz_intro}. Secondly, the DNN has to simultaneously learn the distinct features emerging from the presence of EFT operators in two different signal processes $viz$ $tZj$ and $t\bar{t}Z$, the two of which comprise the dominant contribution to the NP signal. Using this observation to our advantage, we follow a slightly different approach in the DNN analysis. Instead of training a single network, we train two distinct DNNs,

\begin{itemize}
    \item{$NN_{tZj}$}: trained to discriminate SMEFT $tZj$ events from SM backgrounds; training dataset includes pure EFT $tZj$, and SM $tZj$, $t\bar{t}Z$ and $WZ+\mathrm{jets}$ events. We note that it does not include pure EFT $t\bar{t}Z$ events.  
    \item{$NN_{t\bar{t}Z}$}: trained to discriminate SMEFT $t\bar{t}Z$ events from SM backgrounds; the model is trained on pure EFT $t\bar{t}Z$ events, and SM $tZj$, $t\bar{t}Z$ and $WZ+\mathrm{jets}$ events only. 
\end{itemize}

To quantify the gain in using networks targeting the influence of $tZj$ ad $t\bar{t}Z$ separately, we compute the F1 score,
\begin{equation}
    \mathrm{F1} = \frac{2\cdot(R*P)}{R+P} 
\end{equation}
where the recall $R$ is the fraction of signal events that are predicted by the model to be signal-like, and the precision $P$ is the fraction of true signal events among all events that are predicted to be signal-like. A higher F1 score indicates better classifier performance.
For the purpose of illustrating the benefit of multiple networks, we calculate the recall and precision of each network using a cutoff of 0.5, i.e.~taking events with a score of $\geq 0.5\ (< 0.5)$ to be classified as signal (background) by a given network. With this definition, the F1 scores for $NN_{tZj}$ and $NN_{t\bar{t}Z}$ are in the range of $\sim 0.2~$-$~0.3$ and $\sim 0.3~$-$~0.4$ across all signal benchmarks corresponding to different values of $\mathcal{C}_{tZ}$. By comparison, when training a single DNN to separate all SMEFT events from SM backgrounds, the F1 scores are smaller, typically in the range of $\sim 0.1~$-$~0.2$. This demonstrates the improvement that can be obtained with networks dedicated to capturing the effect of SMEFT operators in separate processes.

Then, the trained DNN models $NN_{tZj}$ and $NN_{t\bar{t}Z}$ are applied to the test dataset comprised of SMEFT $tZj$, $t\bar{t}Z$, $tWZ$ events, and all SM backgrounds $viz$ $tZj$, $t\bar{t}Z$, $tWZ$, $WZ+\mathrm{jets}$ and $t\bar{t}\gamma$. Rather than specifying a particular DNN score cutoff as for the F1 score comparison above, for our final analysis we identify the DNN scores $\alpha^{\prime} = \{\alpha_{tZj}, \alpha_{t\bar{t}Z}\}$ which maximize the signal significance $\sigma_{S}^{NP\star}$, 

\begin{equation}
    \sigma_{S}^{NP\star}(\alpha^{\prime}) = \frac{S_{SMEFT}^{\star}(\alpha^{\prime}) - S_{SM}^{\star}(\alpha^{\prime})}{\sqrt{B^{\star}(\alpha^{\prime})}},
    \label{eq:tzj_DNN_significance}
\end{equation}
where 
\begin{equation}
\begin{split}
    S_{SMEFT}^{\star}(\alpha^{\prime}) = \sum_{i} S_{SMEFT}^{i \star}(\alpha^{\prime}), \\ 
    S_{SM}^{\star}(\alpha^{\prime}) = \sum_{i} S_{SM}^{i \star}(\alpha^{\prime}),~ B^{\star}(\alpha^{\prime}) = \sum_{k} S_{SM}^{k \star}(\alpha^{\prime}); \\ \{i=tZj,t\bar{t}Z,tWZ\},\{k=i,t\bar{t}\gamma, WZ+\mathrm{jets}\}.
    \label{eq:tzj_DNN_1}
\end{split}
\end{equation}
In Eq.~(\ref{eq:tzj_DNN_1}), $S_{SMEFT}^{i \star}(\alpha^{\prime})$ and $S_{SM}^{k \star}(\alpha^{\prime})$ are computed in the following way, 

\begin{equation}
    \begin{split}
    S_{SMEFT}^{i \star}(\alpha^{\prime}) = {NN}_{tZj}^{\alpha_{tZj}}(S_{SMEFT}^{i\star}) +  {NN}_{t\bar{t}Z}^{\alpha_{t\bar{t}Z}}(S_{SMEFT}^{i\star}) \\
    S_{SM}^{k \star}(\alpha) = 
    {NN}_{tZk}^{\alpha_{tZj}}(S_{SM}^{k\star})+{NN}_{t\bar{t}Z}^{\alpha_{t\bar{t}Z}}(S_{SM}^{k\star}),
    \label{eq:tzj_DNN_2}
    \end{split}
\end{equation}
where, ${NN}_{tZj}^{\alpha_{tZj}}(S_{SMEFT}^{i\star})$ and ${NN}_{tZj}^{\alpha_{tZj}}(S_{SM}^{k\star})$ represent the number of events at the HL-LHC for the SMEFT process $i$~($i=tZj$, $t\bar{t}Z$, $tWZ$) and SM process $k$~($k=i,t\bar{t}\gamma, WZ+\mathrm{jets}$), respectively, with DNN output score greater than $\alpha_{tZj}$ after being passed through $NN_{tZj}$. Similarly, ${NN}_{t\bar{t}Z}^{\alpha_{t\bar{t}Z}}(S_{SMEFT}^{i\star})$ and ${NN}_{t\bar{t}Z}^{\alpha_{t\bar{t}Z}}(S_{SM}^{k\star})$ denote the respective event rates with DNN score greater than $\alpha_{t\bar{t}Z}$ when subjected to $NN_{t\bar{t}Z}$. The SMEFT signal yield for process $i$, $S_{SMEFT}^{i\star}(\alpha^{\prime} = \{\alpha_{tZj}, \alpha_{t\bar{t}Z}\})$~(c.f. Eq.~(\ref{eq:tzj_DNN_2})), is computed by adding the event rates for SMEFT process $i$ with DNN score $\geq \alpha_{tZj}$ and $\geq \alpha_{t\bar{t}Z}$ when subjected to $NN_{tZj}$ and $NN_{t\bar{t}Z}$, respectively. This is equivalent to summing the contributions from two signal regions, each of which is trained with distinct neural networks, $NN_{tZj}$ and $NN_{t\bar{t}Z}$. The total signal yields for the SM processes $k$ are also computed in a similar manner. We then employ Eq.~(\ref{eq:tzj_DNN_significance}) to compute the signal significance $\sigma_{S}^{NP\star}$.
In Table~\ref{tab:tzj_OtZ_DNN}, we present the event rates $S^{i\star}_{SMEFT}$ and $S^{k\star}_{SM}$ at the HL-LHC for the optimized DNN score $\alpha^{\prime}$ that maximizes the signal significance $\sigma_{S}^{NP\star}$ for 8 signal benchmarks for $\mathcal{C}_{tZ}$ considered in Sec.~\ref{sec:pp_ttz_intro}. Separate $NN_{tZj}$ and $NN_{t\bar{t}Z}$ models are trained at each of these benchmarks. We follow an analogous strategy for $\mathcal{C}_{tW}$ where we consider 6 signal benchmarks similar to that in Sec.~\ref{sec:pp_ttz_intro}. The optimized signal and background event rates along with signal significance values are presented in Table~\ref{tab:tzj_OtW_DNN}. 

\begin{table}[!t]
\centering\scalebox{0.83}{
\begin{tabular}{|c|c|c|c|c|c|c|c|c|c|c|c|} \hline 
    \multirow{2}{*}{$\mathcal{C}_{tZ}$} & \multirow{2}{*}{$NN$} & \multicolumn{3}{c|}{$S_{SMEFT}^{\star}~(\alpha)$} & \multicolumn{5}{c|}{$S_{SM}^{\star}(\alpha)$} & \multirow{2}{*}{$\alpha^{\prime}$}  & \multirow{2}{*}{$\sigma_{S}^{NP\star}$} \\ \cline{3-10}
    & & $tZj$ & $t\bar{t}Z$ & $tWZ$ & $tZj$ & $t\bar{t}Z$ & $tWZ$ & $WZ$ & $t\bar{t}\gamma$ &  & \\ \hline
     \multirow{2}{*}{2.0} & $NN_{tZj}$ & 287 & 140 & 16.0 & 177 & 65.8 & 9.9 & 244 & 0.4 & 0.63 & \multirow{2}{*}{18.3}  \\   
       & $NN_{t\bar{t}Z}$ & 143 & 1557 & 153 & 97.8 & 926 & 107.6 & 838 & 5.1 & 0.46 &  \\ \hline
     \multirow{2}{*}{1.5} & $NN_{tZj}$ & 171 & 66.2 & 9.0 & 114 & 42.0 & 6.3 & 165 & 0.2 & 0.7 & \multirow{2}{*}{9.94}  \\   
       & $NN_{t\bar{t}Z}$ & 200 & 1810 & 191 & 167 & 1380 & 162 & 1316 & 8.1 & 0.4 &  \\ \hline
     \multirow{2}{*}{1.0} & $NN_{tZj}$ & 184 & 79.0 & 9.5 & 156 & 56.9 & 8.7 & 220 & 0.33 & 0.65 & \multirow{2}{*}{5.12}  \\   
       & $NN_{t\bar{t}Z}$ & 86.1 & 892 & 92.1 & 72.6 & 733 & 85.3 & 678 & 3.8 & 0.49 &  \\ \hline
     \multirow{2}{*}{0.5} & $NN_{tZj}$ & 201 & 67.2 & 8.7 & 184 & 55.7 & 7.6 & 264 & 0.2 & 0.64 & \multirow{2}{*}{1.66}  \\   
       & $NN_{t\bar{t}Z}$ & 63.8 & 794 & 91.0 & 62.6 & 750.5 & 87.2 & 784.7 & 4.5 & 0.49 &  \\ \hline
     \multirow{2}{*}{-0.5} & $NN_{tZj}$ & 619 & 214 & 27.5 & 601 & 202 & 26.7 & 819 & 1.3 & 0.47 & \multirow{2}{*}{1.44}  \\   
       & $NN_{t\bar{t}Z}$ & 280 & 1770 & 209 & 274 & 1700 & 208 & 1907 & 9.9 & 0.37 &  \\ \hline
     \multirow{2}{*}{-1.0} & $NN_{tZj}$ & 3272 & 3195 & 391 & 3189 & 2914 & 380 & 6420 & 21.1 & 0.01 & \multirow{2}{*}{4.87}  \\   
       & $NN_{t\bar{t}Z}$ & 417 & 2202 & 254 & 388 & 1965 & 242 & 2385 & 11.9 & 0.32 &  \\ \hline
     \multirow{2}{*}{-1.5} & $NN_{tZj}$ & 282 & 153 & 17.2 & 203 & 94.7 & 12.1 & 255 & 0.5 & 0.59 & \multirow{2}{*}{11.21}  \\   
       & $NN_{t\bar{t}Z}$ & 152 & 1272 & 130 & 119 & 906 & 103 & 884 & 4.1 & 0.45 &  \\ \hline
     \multirow{2}{*}{-2.0} & $NN_{tZj}$ & 3464 & 4075 & 449 & 3192 & 2927 & 383 & 6483 & 21.2 & 0.0 & \multirow{2}{*}{19.47}  \\   
       & $NN_{t\bar{t}Z}$ & 982 & 3584 & 386 & 818 & 2521 & 322 & 3634 & 17.0 & 0.2 &  \\ \hline
    \end{tabular}}
    \caption{Signal significance $\sigma_{S}^{NP\star}(\alpha)$ from DNN analysis in $pp \to tZj + t\bar{t}Z + tWZ \to 3\ell + 1b + 1/2j $ channel for $\{\mathcal{C}_{tZ} = \pm 2.0, \pm 1.5, \pm 1.0, \pm 0.5\}$ at $\sqrt{s}=13~$TeV LHC with $\mathcal{L}=3~{\rm ab^{-1}}$. The signal rates for SMEFT $tZj$, $t\bar{t}Z$ and $tWZ$ processes, background rates for SM $tZj$, $t\bar{t}Z$, $tWZ$, $t\bar{t}\gamma$ and $WZ+\mathrm{jets}$, and the optimal DNN scores are presented.}
    \label{tab:tzj_OtZ_DNN}
\end{table}

\begin{table}[!htb]
\centering\scalebox{0.83}{
\begin{tabular}{|c|c|c|c|c|c|c|c|c|c|c|c|} \hline 
    \multirow{2}{*}{$\mathcal{C}_{tW}$} & \multirow{2}{*}{$NN$} & \multicolumn{3}{c|}{$S_{SMEFT}^{\star}~(\alpha)$} & \multicolumn{5}{c|}{$S_{SM}^{\star}(\alpha)$} & \multirow{2}{*}{$\alpha^{\prime}$}  & \multirow{2}{*}{$\sigma_{S}^{NP\star}$} \\ \cline{3-10}
    & & $tZj$ & $t\bar{t}Z$ & $tWZ$ & $tZj$ & $t\bar{t}Z$ & $tWZ$ & $WZ$ & $t\bar{t}\gamma$ &  & \\ \hline
     \multirow{2}{*}{0.72} & $NN_{tZj}$ & 3610 & 1840 & 245 & 3075 & 1471 & 211 & 5161 & 10.7 & 0.2 & \multirow{2}{*}{14.2}  \\   
       & $NN_{t\bar{t}Z}$ & 831 & 3252 & 379 & 661 & 2568 & 328 & 3327 & 17.9 & 0.2 &  \\ \hline
     \multirow{2}{*}{0.48} & $NN_{tZj}$ & 3506 & 3467 & 402 & 3178 & 2934 & 377 & 6479 & 21.2 & 0.02 & \multirow{2}{*}{10.99}  \\   
       & $NN_{t\bar{t}Z}$ & 3512 & 3469 & 403 & 3185 & 2935 & 377 & 6484 & 21.2 & 0.0 &  \\ \hline
     \multirow{2}{*}{0.24} & $NN_{tZj}$ & 3345 & 3171 & 383 & 3150 & 2891 & 373 & 6398 & 20.9 & 0.06 & \multirow{2}{*}{5.93}  \\   
       & $NN_{t\bar{t}Z}$ & 1889 & 3115 & 374 & 1757 & 2842 & 365 & 5183 & 20.4 & 0.05 &  \\ \hline
     \multirow{2}{*}{-0.24} & $NN_{tZj}$ & 3022 & 1424 & 223 & 3097 & 1555 & 222 & 5252 & 11.6 & 0.2 & \multirow{2}{*}{3.38}  \\   
       & $NN_{t\bar{t}Z}$ & 925 & 2510 & 347 & 965 & 2719 & 348 & 4015 & 19.3 & 0.15 &  \\ \hline
     \multirow{2}{*}{-0.48} & $NN_{tZj}$ & 3002 & 2536 & 387 & 3182 & 2931 & 21.2 & 6473 & 21.2 & 0.04 & \multirow{2}{*}{6.7}  \\   
       & $NN_{t\bar{t}Z}$ & 1538 & 2451 & 372 & 1625 & 2822 & 363 & 5042 & 20.0 & 0.06 &  \\ \hline
     \multirow{2}{*}{-0.72} & $NN_{tZj}$ & 2829 & 2025 & 348 & 3025 & 2556 & 332 & 5848 & 18.6 & 0.1 & \multirow{2}{*}{9.31}  \\   
       & $NN_{t\bar{t}Z}$ & 1707 & 2266 & 386 & 1811 & 2854 & 366 & 5208 & 20.4 & 0.04 &  \\ \hline
    \end{tabular}}
    \caption{Signal significance $\sigma_{S}$ from DNN analysis in $pp \to tZj+ t\bar{t}Z + tWZ \to 3\ell + 1b + 1/2j $ channel for $\{\mathcal{C}_{tW} = \pm 0.72, \pm 0.48, \pm 0.24\}$ at $\sqrt{s}=13~$TeV LHC with $\mathcal{L}=3~{\rm ab^{-1}}$. The signal rates for SMEFT $tZj$, $t\bar{t}Z$ and $tWZ$ processes, background rates for SM $tZj$, $t\bar{t}Z$, $tWZ$, $t\bar{t}\gamma$ and $WZ+\mathrm{jets}$ are presented. The optimal DNN score $\alpha$ and corresponding signal significance~$\sigma_{S}^{NP}$ are also shown.}
    \label{tab:tzj_OtW_DNN}
\end{table}

With the DNN analysis, we observe a signal significance of $\sigma_{S}^{NP\star}=18.3~$(19.47) at $2\sigma$ for $\mathcal{C}_{tZ}=2.0~(-2.0)$ which is $\sim 10\%$ higher compared to that from cut-based optimization~($\sigma_{S^{NP}} = 16.8$~(17.76)). The margin of improvement reduces as we move towards smaller values of $\mathcal{C}_{tZ}$. For example, at $\mathcal{C}_{tZ}=0.5~(-0.5)$, the DNN leads to $\sigma_{S}^{NP\star} = 1.66~(1.44)$ which is roughly $5\%$ higher than their cut-based counterparts~($\sigma_{S}^{NP} = 1.51~(1.38)$). The improvement from the DNN is more apparent in the scenario where $\mathcal{O}_{tW}$ is the NP operator. For $\mathcal{C}_{tW}=0.72~(-0.72)$, we observe $\sigma_{S}^{NP\star} = 14.2~(9.31)$ which is roughly $\gtrsim 10\%$ higher than the cut-based results. Even at smaller values of $\mathcal{C}_{tW} = 0.24~(-0.24)$, we observe an improvement of $\gtrsim 20\%$ over the cut-based results.

We interpolate $\sigma_{S}^{NP\star}$ as a function of $\mathcal{C}_{tZ}$ using the results from Table~\ref{tab:tzj_OtZ_DNN}. The results are illustrated in the left panel of Fig.~\ref{fig:tzj_limits} as blue dashed lines. In the same figure, we illustrate the obtainable significance $\sigma_{S}^{NP\star}$ as a function of $\mathcal{C}_{tW}$ as red dashed lines. Our results indicate that $\mathcal{C}_{tZ}$ and $\mathcal{C}_{tW}$ could be probed up to $-0.61 \lesssim \mathcal{C}_{tZ} \lesssim 0.55$ and $-0.16\lesssim \mathcal{C}_{tW} \lesssim 0.12$ at $2\sigma$, respectively, through searches in the $pp \to tZj + t\bar{t}Z + tWZ \to 3\ell + 1b + 1/2 j$ channel at the HL-LHC.

Before concluding the present section, we also employ \texttt{MadMiner} to estimate the projected sensitivities for $\mathcal{O}_{tZ}$ and $\mathcal{O}_{tW}$ through searches in the $pp \to tZj + t\bar{t}Z + tWZ \to 3\ell + 1b + 1/2j$ channel at the HL-LHC. Unlike the DNN training dataset, where new physics effects from $\mathcal{O}_{tW}$ are considered only at the production level, the MC event samples for $tZj$, $t\bar{t}Z$, and $tWZ$, used in \texttt{MadMiner}, include new physics modifications to top decay as well. We include $10^{6}$ event samples for each of these processes as well as the $WZ+\mathrm{jets}$ background, and reconstruct all kinematic observables in Eq.~(\ref{eqn:tzj_observables}). Similar to the analysis in Sec.~\ref{sec:pp_ttz_intro}, we assume a quartic ansatz for $\mathcal{C}_{tW}$ and a squared ansatz for $\mathcal{C}_{tZ}$ in the morphing setup.  The network structure is also similar to that in Sec.~\ref{sec:pp_ttz_intro}. In the present case, we perform the training using the \texttt{RASCAL} algorithm~\cite{Brehmer:2018eca} over 150 epochs. The hyperparameter for the \texttt{RASCAL} loss function is set to 1. We consider the \texttt{RASCAL} algorithm in the present section since it leads to better projected sensitivities compared to that from \texttt{ALICES} algorithm.  Similar to \texttt{ALICES}, the \texttt{RASCAL} loss functional is defined using the joint likelihood ratio $r(x,z|\theta,\theta_{SM})$ as well as the joint scores $t(x,z|\theta_{0})$, and its minimizing function is the intractable event likelihood ratio $r(x|\theta,\theta_{SM})$. The other network hyperparameters are selected as in Sec.~\ref{sec:pp_ttz_intro}.

\begin{figure}[!t]
    \centering
    \includegraphics[scale=0.3]{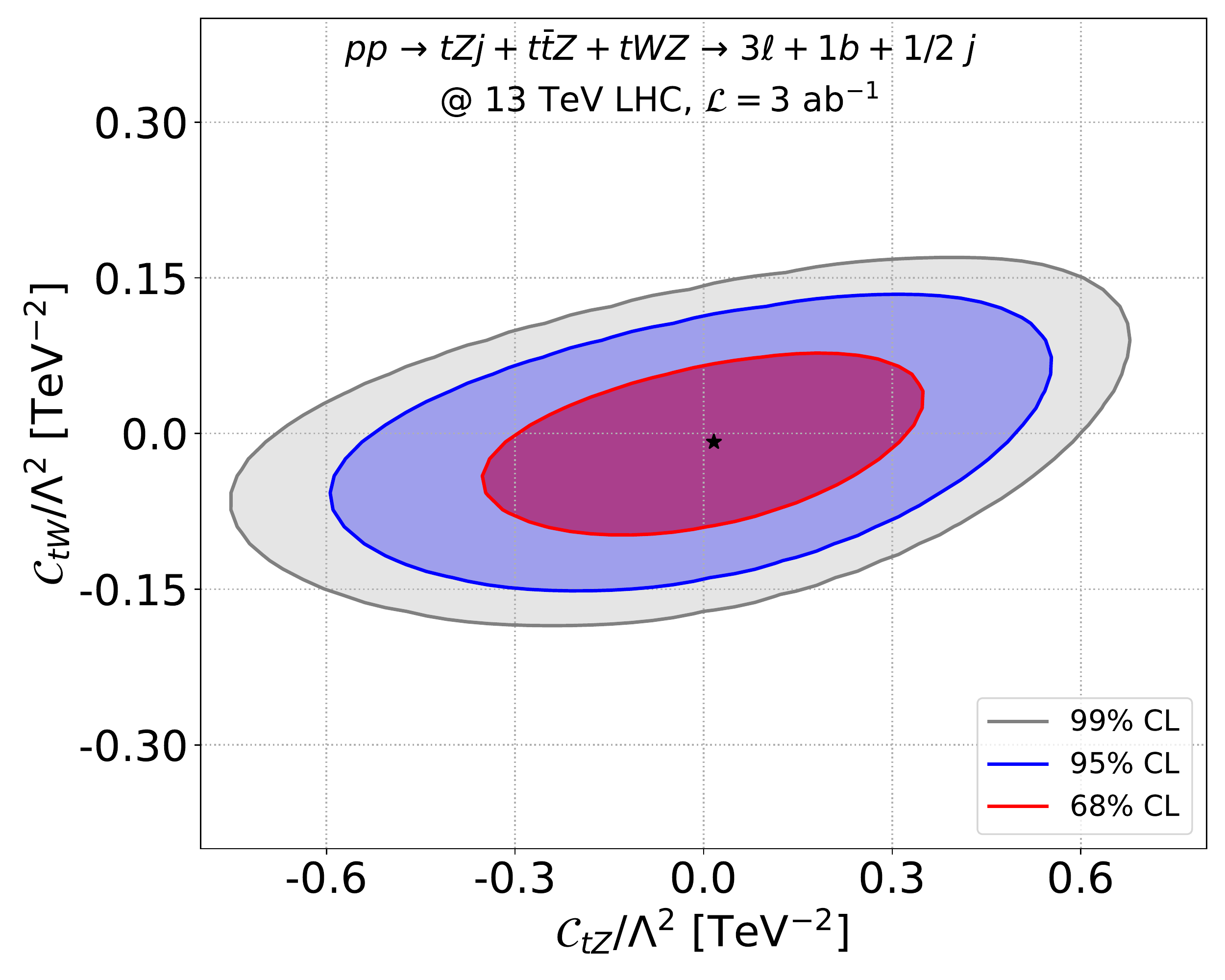}
    \caption{Projected sensitivity in the $\{\mathcal{C}_{tZ},\mathcal{C}_{tW}\}$ plane from searches in the $pp \to tZj + t\bar{t}Z + tWZ \to 3\ell + 1b +  1/2j $ channel at the 13~TeV LHC with $\mathcal{L}=3~{\rm ab^{-1}}$.}
    \label{fig:tzj_2d_madminer}
\end{figure}

We present the projection contours from searches in the $pp \to tZj + t\bar{t}Z + tWZ \to 3\ell + 1b + 1/2j$ channel at the HL-LHC using \texttt{MadMiner} in the $\{\mathcal{C}_{tZ},\mathcal{C}_{tW}\}$ plane in Fig.~\ref{fig:tzj_2d_madminer}. The sensitivities at $68\%$, $95\%$ and $99\%$ CL are illustrated as red, blue and grey shaded regions, respectively. Since the estimated likelihood ratio is a function of both $\theta = \{\mathcal{C}_{tW}, \mathcal{C}_{tZ}\}$, in order to set 1d limits for $\mathcal{C}_{tW}$ or $\mathcal{C}_{tZ}$, we profile over the other. We present the distribution of p-values as a function of $\mathcal{C}_{tZ}$ and $\mathcal{C}_{tW}$ in central and right panels of Fig.~\ref{fig:tzj_limits}, respectively. We observe that $\mathcal{C}_{tZ}$ could be probed up to $-0.59 \lesssim \mathcal{C}_{tZ} \lesssim 0.55$ while the projected sensitivity for $\mathcal{O}_{tW}$ reaches up to $-0.14 \lesssim \mathcal{C}_{tW} \lesssim 0.11$ at $95\%$ CL.  

\begin{figure}[!htb]
    \centering
    \includegraphics[scale=0.37]{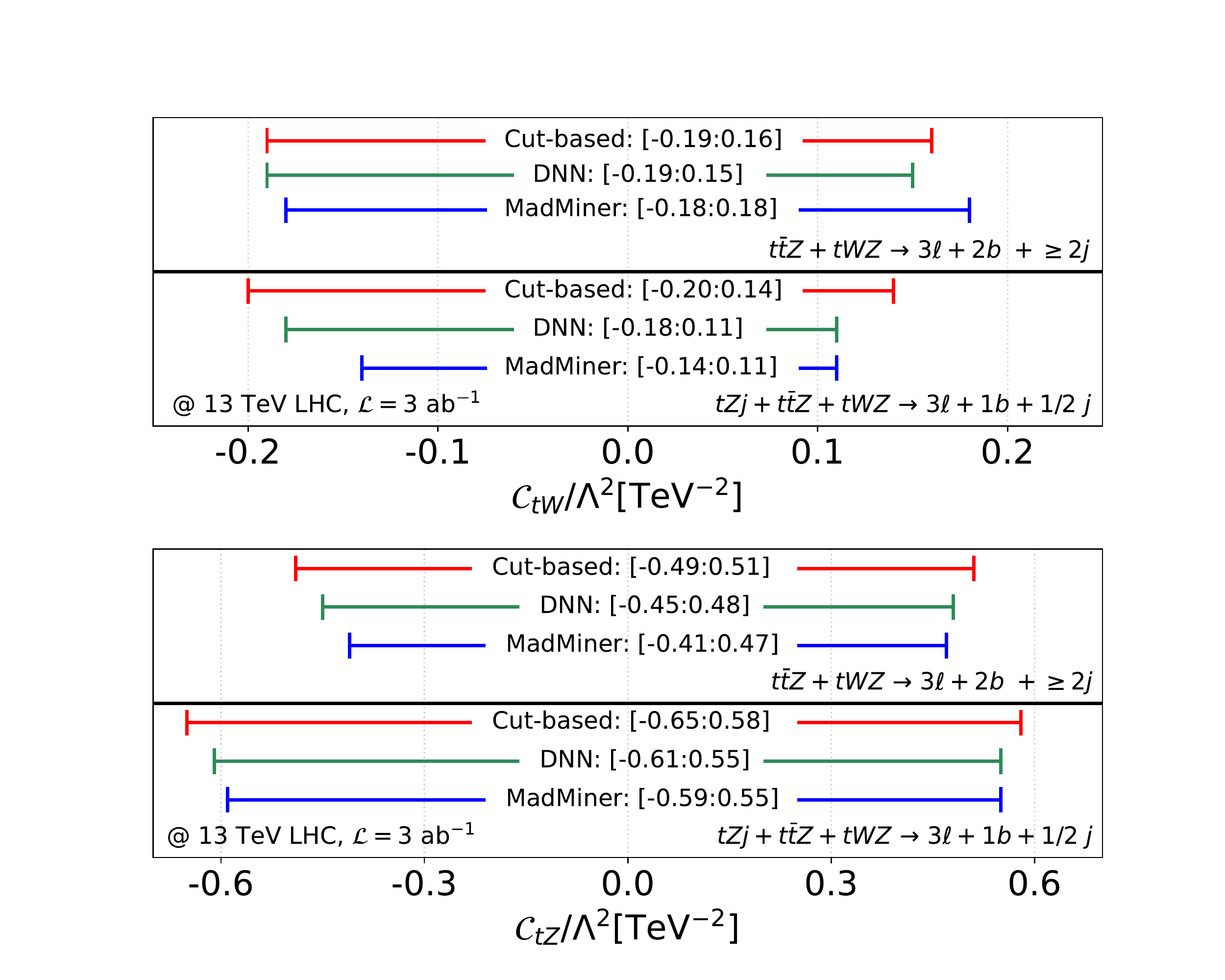}
    \caption{Projected sensitivity for $\mathcal{C}_{tW}$~(\textit{upper panel}) and $\mathcal{C}_{tZ}$~(\textit{lower panel}) from searches inI  the $pp \to t\bar{t}Z + tWZ \to 3\ell + 2b\ + \geq 2j$ and $pp \to tZj + t\bar{t}Z + tWZ \to 3\ell + 1b + 1/2j$ channels at $\sqrt{s}=13~\mathrm{TeV}$ LHC with $\mathcal{L}=3~\mathrm{ab^{-1}}$, using cut-and-count analysis, DNN analysis and \texttt{MadMiner}.}
    \label{fig:summary_plot}
\end{figure}

The 1d projection limits for $\mathcal{C}_{tZ}$ and $\mathcal{C}_{tW}$ from searches in the $pp \to tZj + t\bar{t}Z + tWZ \to 3\ell + 1b + 1/2j$ channel at the HL-LHC using the cut-and-count methodology and machine learning techniques are summarized in the bottom panels of the subfigures in Fig.~\ref{fig:summary_plot}. In the $\mathcal{C}_{tZ}$ scenario, the optimized selection cuts on $p_{T,Z}$, $H_{T}$ and $\Delta R_{t\ell}^{min}$~($vid$ Table~\ref{tab:tzj_OtZ_cut_flow}), lead to a considerable improvement~($\gtrsim 15\%$) in signal significance over rate-only measurements for all signal benchmarks considered in the present study. As discussed previously, the cut-based study yields a projected sensitivity of $-0.65 \lesssim \mathcal{C}_{tZ} \lesssim 0.58$ at $2\sigma$ uncertainty. The DNN and \texttt{MadMiner} analyses lead to a further improvement of roughly $5\%$ in the projected sensitivities. The aforesaid observations are indicative of the potent sensitivity of differential measurements in the $pp \to tZj + t\bar{t}Z + tWZ \to 3\ell + 1b + 1/2j$ channel to $\mathcal{C}_{tZ}$ and $\mathcal{C}_{tW}$, and emphasize the necessity of including them along with rate measurements at the high luminosity run of the LHC. Our results also illustrate that the machine learning techniques are relatively more efficient in extracting the new physics information from differential measurements in the signal channel considered in this section.

For comparison, we also summarize the analogous limits arising from the $t\bar{t}Z$ analyses of Sec.~\ref{sec:pp_ttz_intro} in the top panels of Fig.~\ref{fig:summary_plot}. Notably, the sensitivity to $\mathcal{O}_{tW}$ only improves with the use of kinematic information for the $tZj$ channel. The reason that $t\bar{t}Z$ production is less sensitive to this operator is that the dominant diagrams for this process, consisting of top pair production with a single electroweak vertex for $Z$ emission, are unaffected by $\mathcal{O}_{tW}$. Thus, $\mathcal{O}_{tW}$ only affects $t\bar{t}Z$ through its influence on the top quark decay. For $tZj$, by contrast, $\mathcal{O}_{tW}$ affects production as well as decay, so its effects show up in all of the kinematic distributions more readily.

\section{Conclusions}
\label{sec:conclusions}

While top production in association with electroweak bosons is beginning to be observed experimentally, the HL-LHC offers the possibility of measuring these processes with sufficient statistics to leverage kinematic information in the final state. New physics can affect top electroweak production even if any BSM states are too heavy to observe directly. In this work, we analyzed the projected sensitivities for the dimension 6 SMEFT electroweak dipole operators $\mathcal{O}_{tZ}$ and $\mathcal{O}_{tW}$ at leading order at the HL-LHC through searches in $pp \to t\bar{t}Z + tWZ \to 3\ell + 2b\ + \geq 2j$ and $pp \to tZj + t\bar{t}Z + tWZ \to 3\ell + 1b + 1/2j$ channels. We considered a comprehensive set of observables in both channels, and combined rate information with differential cross-section measurements to boost the projected sensitivities. Both conventional cut-and-count techniques and machine-learning based multivariate analysis techniques have been used. In the latter category, we adopted Deep Neural Networks and \texttt{MadMiner}. 

We considered new physics contributions from $\mathcal{O}_{tZ}$ at the production level, and from $\mathcal{O}_{tW}$ at the production level as well as top decay. Furthermore, non-linear pure SMEFT $\mathcal{O}(\Lambda^{-4})$ contributions in addition to $\mathcal{O}(\Lambda^{-2})$ interference contributions have been considered. We optimized a cut-based analysis using different subsets of observables, identifying those which lead to the strongest sensitivity. We also identified the most important observables that steer the sensitivity to $\mathcal{C}_{tZ}$ and $\mathcal{C}_{tW}$ obtained through a multivariate DNN technique.

In the $pp \to t\bar{t}Z + tWZ \to 3\ell + 2b\ + \geq 2j$ channel, sensitivity to $\mathcal{C}_{tZ}$ improved with the utilization of kinematic measurements on top of rate information for all signal benchmarks considered in this work. In the aforesaid channel, $\mathcal{C}_{tZ}$ can be probed at the HL-LHC up to $-0.49 \leq \mathcal{C}_{tZ} \leq 0.51$ at $2\sigma$ using cut-and-count techniques. The potential reach for $\mathcal{C}_{tZ}$ improves further upon adopting the ML-based techniques $viz$ DNN and \texttt{MadMiner}. The strongest sensitivity is obtained with \texttt{MadMiner}, which can probe $\mathcal{C}_{tZ}$ up to $-0.41 \leq \mathcal{C}_{tZ} \leq 0.47$ at $95\%$ CL. In the case of $\mathcal{C}_{tW}$, the projected sensitivity does not increase significantly after the inclusion of kinematic information due to the smaller sensitivity of $t\bar{t}Z$ to $\mathcal{C}_{tW}$.

The $pp \to tZj + t\bar{t}Z + tWZ \to 3\ell + 1b + 1/2j$ channel exhibits a relatively weaker sensitivity to $\mathcal{C}_{tZ}$. Using cut-based techniques, the projected sensitivity to $\mathcal{C}_{tZ}$ at the HL-LHC stands at $-0.65 \leq \mathcal{C}_{tZ} \leq 0.58$ at $2\sigma$. The DNN and \texttt{MadMiner} analyses lead to $\sim 5\%$ improvement in the projected limits, $-0.61 \leq \mathcal{C}_{tZ} \leq 0.55$~(DNN) and $-0.59 \leq \mathcal{C}_{tZ} \leq 0.55$~(\texttt{MadMiner}). Compared to $pp \to t\bar{t}Z + tWZ \to 3\ell + 2b\ + \geq 2j$ channel, relatively stronger limits are found on $\mathcal{C}_{tW}$ in the $pp \to tZj + t\bar{t}Z + tWZ \to 3\ell + 1b + 1/2j$ channel. Here, the inclusion of kinematic information leads to a noticeable improvement in the projected sensitivity to $\mathcal{C}_{tW}$ with the use of deep learning. While the cut-based technique resulted in only a marginal improvement over rate-only measurements, both the DNN and \texttt{MadMiner} analyses led to a $\gtrsim 10\%$ improvement. The strongest sensitivity to $\mathcal{C}_{tW}$ is exhibited by \texttt{MadMiner} in the $pp \to tZj + t\bar{t}Z + tWZ \to 3\ell + 1b + 1/2j$ channel wherein $\mathcal{C}_{tW}$ can be probed up to  $-0.14 \leq \mathcal{C}_{tW} \leq 0.11$ at $95\%$ CL at the HL-LHC. Among the two channels considered in this work, the $pp \to t\bar{t}Z + tWZ \to 3\ell + 2b\ + \geq 2j$ channel has a stronger sensitivity to $\mathcal{C}_{tZ}$ while the $pp \to tZj + t\bar{t}Z + tWZ \to 3\ell + 1b + 1/2j$ channel is more sensitive to $\mathcal{C}_{tW}$, with both ultimately displaying considerable improvements upon the inclusion of differential measurements.

The inclusion of next to leading order effects in the signal and background processes while developing search strategies may help to further improve the projected reach since both production rates and differential distributions in $t\bar{t}Z$ and $tZj$ are susceptible to modifications from higher order effects~\cite{Degrande:2018fog,Goldouzian:2020ekx}. Furthermore, combination with other relevant decay channels for $t\bar{t}Z$ and $tZj$ at the HL-LHC may lead to potential improvements.

The high statistics afforded by the HL-LHC will enable detailed study of kinematics in a variety of processes which are currently relatively rare, notably those involving electroweak couplings. Machine learning techniques can enhance our ability to maximize the sensitivity to new physics from complex final states. This work demonstrates the application of such methods to top production in association with neutral gauge bosons, finding success in improved limits on higher-dimensional operators that parametrize new physics. It is likely that there are further opportunities to exploit kinematic information in electroweak physics at the HL-LHC.

\section{Acknowledgements}
We thank Giacomo Magni and Juan Rojo for helpful discussions. R.K.B. and A.I. thank the U.S. Department of Energy for the financial support, under grant number DE-SC0016013. Some of the computing for this project was performed at the High Performance Computing Center at Oklahoma State University, supported in part through the National Science Foundation grant OAC-1531128.

\bibliography{ref}
\end{document}